\documentclass[aps,prd,showpacs,letterpaper,reprint,superscriptaddress,notitlepage,nofootinbib,floatfix]{revtex4-2}

\usepackage[caption=false]{subfig}
\usepackage{graphicx}
\usepackage{bm}
\usepackage{amssymb}
\usepackage{amsmath}
\usepackage{amsfonts}
\usepackage[utf8]{inputenc}
\usepackage{booktabs}
\usepackage{gensymb}
\usepackage[colorlinks=true,allcolors=blue]{hyperref}
\usepackage{microtype}
\usepackage{multirow, makecell}
\usepackage[capitalise]{cleveref}
\usepackage[utf8]{inputenc}
\usepackage{listings}
\usepackage{siunitx}
\usepackage{xspace}
\usepackage{comment}
\usepackage{floatrow}
\usepackage{ragged2e}
\usepackage{xspace}
\usepackage{multirow}
\usepackage{dcolumn}
\usepackage{bm}
\usepackage[dvipsnames]{xcolor}
\usepackage{gensymb}
\usepackage[colorlinks=true,allcolors=blue]{hyperref}
\usepackage{siunitx}

\DeclareSIUnit \s {\second}
\DeclareSIUnit \ns {\nano\second}
\DeclareSIUnit \mus {\micro\second}
\DeclareSIUnit \ms {\milli\second}
\DeclareSIUnit \MB {\mega\byte}
\DeclareSIUnit \GB {\giga\byte}
\DeclareSIUnit \TB {\tera\byte}
\DeclareSIUnit \PB {\peta\byte}
\DeclareSIUnit \Mbps {\mega\bit/\s}
\DeclareSIUnit \Gbps {\giga\bit/\s}
\DeclareSIUnit \Tbps {\tera\bit/\s}
\DeclareSIUnit \Pbps {\peta\bit/\s}
\DeclareSIUnit \kton {\kilo\tonne} 
\DeclareSIUnit \kt {\kilo\tonne}
\DeclareSIUnit \Mt {\mega\tonne}
\DeclareSIUnit \eV {\electronvolt}
\DeclareSIUnit \keV {\kilo\electronvolt}
\DeclareSIUnit \MeV {\mega\electronvolt}
\DeclareSIUnit \GeV {\giga\electronvolt}
\DeclareSIUnit \PeV {\peta\electronvolt}
\DeclareSIUnit \EeV {\exa\electronvolt}
\DeclareSIUnit \m {\meter}
\DeclareSIUnit \cm {\centi\meter}
\DeclareSIUnit \in {\inchcommand}
\DeclareSIUnit \km {\kilo\meter}
\DeclareSIUnit \kV {\kilo\volt}
\DeclareSIUnit \kW {\kilo\watt}
\DeclareSIUnit \MW {\mega\watt}
\DeclareSIUnit \MHz {\mega\hertz}
\DeclareSIUnit \mrad {\milli\radian}
\DeclareSIUnit \year {years}
\DeclareSIUnit \POT {POT}
\DeclareSIUnit \sig {$\sigma$}
\DeclareSIUnit\parsec{pc}
\DeclareSIUnit\lightyear{ly}
\DeclareSIUnit\foot{ft}
\DeclareSIUnit\ft{ft}
\DeclareSIUnit \ppb{ppb}
\DeclareSIUnit \ppt{ppt}
\DeclareSIUnit \samples{S}
\DeclareSIUnit \pe{PE}
\DeclareSIUnit \mppe {\meter/\pe}
\DeclareSIUnit \mb{mb}

\usepackage{xcolor, soul}
\sethlcolor{blue}

\newcommand{\reffig}[1]{Fig.~\ref{#1}}

\usepackage{mathtools}
\newcommand{\cosz}{\ensuremath{\cos(\theta_\mathrm{z})}\xspace}
\newcommand{\numu}{\ensuremath{\nu_\mu}\xspace}

\newcommand{\nue}{\ensuremath{\nu_e}\xspace}

\newcommand{\nutau}{\ensuremath{\nu_\tau}\xspace}

\usepackage{array}
\usepackage{booktabs}
\usepackage[dvipsnames]{xcolor}
\usepackage{tikz}
\usepackage{rotating}

\newcolumntype{M}[1]{>{\centering\arraybackslash}m{#1}}

\begin{document}

\title{Searching for eV-scale sterile neutrinos with eight years of atmospheric neutrinos at the IceCube neutrino telescope}

\affiliation{III. Physikalisches Institut, RWTH Aachen University, D-52056 Aachen, Germany}
\affiliation{Department of Physics, University of Adelaide, Adelaide, 5005, Australia}
\affiliation{Dept. of Physics and Astronomy, University of Alaska Anchorage, 3211 Providence Dr., Anchorage, AK 99508, USA}
\affiliation{Dept. of Physics, University of Texas at Arlington, 502 Yates St., Science Hall Rm 108, Box 19059, Arlington, TX 76019, USA}
\affiliation{CTSPS, Clark-Atlanta University, Atlanta, GA 30314, USA}
\affiliation{School of Physics and Center for Relativistic Astrophysics, Georgia Institute of Technology, Atlanta, GA 30332, USA}
\affiliation{Dept. of Physics, Southern University, Baton Rouge, LA 70813, USA}
\affiliation{Dept. of Physics, University of California, Berkeley, CA 94720, USA}
\affiliation{Lawrence Berkeley National Laboratory, Berkeley, CA 94720, USA}
\affiliation{Institut f{\"u}r Physik, Humboldt-Universit{\"a}t zu Berlin, D-12489 Berlin, Germany}
\affiliation{Fakult{\"a}t f{\"u}r Physik {\&} Astronomie, Ruhr-Universit{\"a}t Bochum, D-44780 Bochum, Germany}
\affiliation{Universit{\'e} Libre de Bruxelles, Science Faculty CP230, B-1050 Brussels, Belgium}
\affiliation{Vrije Universiteit Brussel (VUB), Dienst ELEM, B-1050 Brussels, Belgium}
\affiliation{Dept. of Physics, Massachusetts Institute of Technology, Cambridge, MA 02139, USA}
\affiliation{Dept. of Physics and Institute for Global Prominent Research, Chiba University, Chiba 263-8522, Japan}
\affiliation{Department of Physics, Loyola University Chicago, Chicago, IL 60660, USA}
\affiliation{Dept. of Physics and Astronomy, University of Canterbury, Private Bag 4800, Christchurch, New Zealand}
\affiliation{Dept. of Physics, University of Maryland, College Park, MD 20742, USA}
\affiliation{Dept. of Astronomy, Ohio State University, Columbus, OH 43210, USA}
\affiliation{Dept. of Physics and Center for Cosmology and Astro-Particle Physics, Ohio State University, Columbus, OH 43210, USA}
\affiliation{Niels Bohr Institute, University of Copenhagen, DK-2100 Copenhagen, Denmark}
\affiliation{Dept. of Physics, TU Dortmund University, D-44221 Dortmund, Germany}
\affiliation{Dept. of Physics and Astronomy, Michigan State University, East Lansing, MI 48824, USA}
\affiliation{Dept. of Physics, University of Alberta, Edmonton, Alberta, Canada T6G 2E1}
\affiliation{Erlangen Centre for Astroparticle Physics, Friedrich-Alexander-Universit{\"a}t Erlangen-N{\"u}rnberg, D-91058 Erlangen, Germany}
\affiliation{Physik-department, Technische Universit{\"a}t M{\"u}nchen, D-85748 Garching, Germany}
\affiliation{D{\'e}partement de physique nucl{\'e}aire et corpusculaire, Universit{\'e} de Gen{\`e}ve, CH-1211 Gen{\`e}ve, Switzerland}
\affiliation{Dept. of Physics and Astronomy, University of Gent, B-9000 Gent, Belgium}
\affiliation{Dept. of Physics and Astronomy, University of California, Irvine, CA 92697, USA}
\affiliation{Karlsruhe Institute of Technology, Institut f{\"u}r Kernphysik, D-76021 Karlsruhe, Germany}
\affiliation{Dept. of Physics and Astronomy, University of Kansas, Lawrence, KS 66045, USA}
\affiliation{SNOLAB, 1039 Regional Road 24, Creighton Mine 9, Lively, ON, Canada P3Y 1N2}
\affiliation{Department of Physics and Astronomy, UCLA, Los Angeles, CA 90095, USA}
\affiliation{Department of Physics, Mercer University, Macon, GA 31207-0001, USA}
\affiliation{Dept. of Astronomy, University of Wisconsin, Madison, WI 53706, USA}
\affiliation{Dept. of Physics and Wisconsin IceCube Particle Astrophysics Center, University of Wisconsin, Madison, WI 53706, USA}
\affiliation{Institute of Physics, University of Mainz, Staudinger Weg 7, D-55099 Mainz, Germany}
\affiliation{Department of Physics, Marquette University, Milwaukee, WI, 53201, USA}
\affiliation{Institut f{\"u}r Kernphysik, Westf{\"a}lische Wilhelms-Universit{\"a}t M{\"u}nster, D-48149 M{\"u}nster, Germany}
\affiliation{Bartol Research Institute and Dept. of Physics and Astronomy, University of Delaware, Newark, DE 19716, USA}
\affiliation{Dept. of Physics, Yale University, New Haven, CT 06520, USA}
\affiliation{Dept. of Physics, University of Oxford, Parks Road, Oxford OX1 3PU, UK}
\affiliation{Dept. of Physics, Drexel University, 3141 Chestnut Street, Philadelphia, PA 19104, USA}
\affiliation{Physics Department, South Dakota School of Mines and Technology, Rapid City, SD 57701, USA}
\affiliation{Dept. of Physics, University of Wisconsin, River Falls, WI 54022, USA}
\affiliation{Dept. of Physics and Astronomy, University of Rochester, Rochester, NY 14627, USA}
\affiliation{Oskar Klein Centre and Dept. of Physics, Stockholm University, SE-10691 Stockholm, Sweden}
\affiliation{Dept. of Physics and Astronomy, Stony Brook University, Stony Brook, NY 11794-3800, USA}
\affiliation{Dept. of Physics, Sungkyunkwan University, Suwon 16419, Korea}
\affiliation{Institute of Basic Science, Sungkyunkwan University, Suwon 16419, Korea}
\affiliation{Dept. of Physics and Astronomy, University of Alabama, Tuscaloosa, AL 35487, USA}
\affiliation{Dept. of Astronomy and Astrophysics, Pennsylvania State University, University Park, PA 16802, USA}
\affiliation{Dept. of Physics, Pennsylvania State University, University Park, PA 16802, USA}
\affiliation{Dept. of Physics and Astronomy, Uppsala University, Box 516, S-75120 Uppsala, Sweden}
\affiliation{Dept. of Physics, University of Wuppertal, D-42119 Wuppertal, Germany}
\affiliation{DESY, D-15738 Zeuthen, Germany}

\author{M. G. Aartsen}
\affiliation{Dept. of Physics and Astronomy, University of Canterbury, Private Bag 4800, Christchurch, New Zealand}
\author{R. Abbasi}
\affiliation{Department of Physics, Loyola University Chicago, Chicago, IL 60660, USA}
\author{M. Ackermann}
\affiliation{DESY, D-15738 Zeuthen, Germany}
\author{J. Adams}
\affiliation{Dept. of Physics and Astronomy, University of Canterbury, Private Bag 4800, Christchurch, New Zealand}
\author{J. A. Aguilar}
\affiliation{Universit{\'e} Libre de Bruxelles, Science Faculty CP230, B-1050 Brussels, Belgium}
\author{M. Ahlers}
\affiliation{Niels Bohr Institute, University of Copenhagen, DK-2100 Copenhagen, Denmark}
\author{M. Ahrens}
\affiliation{Oskar Klein Centre and Dept. of Physics, Stockholm University, SE-10691 Stockholm, Sweden}
\author{C. Alispach}
\affiliation{D{\'e}partement de physique nucl{\'e}aire et corpusculaire, Universit{\'e} de Gen{\`e}ve, CH-1211 Gen{\`e}ve, Switzerland}
\author{N. M. Amin}
\affiliation{Bartol Research Institute and Dept. of Physics and Astronomy, University of Delaware, Newark, DE 19716, USA}
\author{K. Andeen}
\affiliation{Department of Physics, Marquette University, Milwaukee, WI, 53201, USA}
% ...
\date{\today}

\collaboration{IceCube Collaboration}
\noaffiliation

\date{\today}

\begin{abstract}
We report in detail on searches for eV-scale sterile neutrinos, in the context of a 3+1 model, using eight years of data from the IceCube neutrino telescope.
By analyzing the reconstructed energies and zenith angles of 305,735 atmospheric $\nu_\mu$ and $\bar{\nu}_\mu$ events we construct confidence intervals in two analysis spaces: $\sin^2 (2\theta_{24})$ vs. $\Delta m^2_{41}$ under the conservative assumption $\theta_{34}=0$; and $\sin^2(2\theta_{24})$ vs. $\sin^2 (2\theta_{34})$ given sufficiently large $\Delta m^2_{41}$ that fast oscillation features are unresolvable.
Detailed discussions of the event selection, systematic uncertainties, and fitting procedures are presented.
No strong evidence for sterile neutrinos is found, and the best-fit likelihood is consistent with the no sterile neutrino hypothesis with a p-value of 8\% in the first analysis space and 19\% in the second.
\end{abstract}

\maketitle
\tableofcontents

\section{Introduction~\label{sec:introduction}}

Anomalies in short-baseline oscillation experiments studying neutrinos from pion decay-at-rest~\cite{aguilar:lsnd}, meson decay-in-flight beams~\cite{aguilar:miniboone}, and nuclear reactors~\cite{mention:reactor_anomaly} have produced a string of experimental observations that suggest unexpected neutrino flavor transformation at short baselines.
These observations are anomalies under the well-established three massive neutrino framework, but can be accommodated, to some extent, by addition of a new heavy neutrino mass state $\nu_4$.
For consistency with constraints from invisible $Z$-boson decay~\cite{ALEPH:2005ab} and existing unitarity constraints on the three Standard Model neutrinos~\cite{Parke:2015goa}, the new state is mostly composed of a sterile flavor $\nu_s$ which does not participate in Standard Model electroweak interactions.
The sterile neutrino hypothesis is the minimal explanation of the anomalous observations. It has motivated a worldwide program to search for new particle states with mass-squared differences between $\SI{0.1}\eV^2$ and $\SI{10}\eV^2$~\cite{Abazajian:2012ys}.  Notable other explanations include, for example, phenomenology that modifies the vacuum oscillation probability relevant to short-baseline neutrino experiments~\cite{Murayama:2000hm,Strumia:2002fw,Barenboim:2002ah, GonzalezGarcia:2003jq,Barger:2003xm,Sorel:2003hf,Barenboim:2004wu, Zurek:2004vd, Kaplan:2004dq, Pas:2005rb,deGouvea:2006qd,Schwetz:2007cd, Farzan:2008zv,Hollenberg:2009ws,Nelson:2010hz,Akhmedov:2010vy,Diaz:2010ft,Bai:2015ztj, Giunti:2015mwa,Papoulias:2016edm, Moss:2017pur,Carena:2017qhd,Moulai:2019gpi,Barenboim:2019hso}, modifications of neutrino propagation in matter~\cite{Liao:2016reh, Liao:2018mbg,Asaadi:2017bhx,Doring:2018cob,Denton:2018dqq}, or production of new particles in the beam or in the detector and its surroundings~\cite{Gninenko:2009ks,Gninenko:2010pr,Dib:2011jh,McKeen:2010rx,Masip:2012ke, Masip:2011qb,Gninenko:2012rw,Magill:2018jla,Bertuzzo:2018itn,Ballett:2018ynz,Ballett:2019pyw,Fischer:2019fbw,Dentler:2019dhz,deGouvea:2019qre}.

The simplest sterile neutrino model, called the ``3+1'' model, introduces a single mass eigenstate $\nu_4$ that is heavier than the three flavor states mostly composed of active neutrinos ($\nu_1$, $\nu_2$, $\nu_3$) by a fixed difference  $\Delta m_{41} = m_4 – m_1$.
Three flavor neutrino mixing is described by the well-known $3\times3$ Pontecorvo-Maki-Nakagawa-Sakata (PMNS) matrix~\cite{pontekorvo1957mesonium, maki1962remarks}.
In the 3+1 model, an extended $4\times4$ PMNS matrix $U^{4\times4}$ includes additional elements $U_{e4}$, $U_{\mu4}$, and $U_{\tau4}$ to account for the heavy neutrino fraction of the three active flavors. This extended PMNS matrix can be parameterized as
\begin{equation}
    U^{4 \times 4} = R_{34}R_{24}R_{14} U_{\rm PMNS},
    \label{eq:pmns}
\end{equation}
where $U_PMNS$ is block diagonal between the first three and the forth component, and the new matrix elements are expressed in terms of three mixing angles $\theta_{14}$,\,$\theta_{24}$,\,$\theta_{34}$, and two observable $CP$-violating phases, $\delta_{14},\,\delta_{24}$~\cite{Diaz:2019fwt}.
The 3+1 model is widely used as a benchmark for experimental datasets to examine whether they show evidence for a sterile neutrino. Extensions to this model have been proposed such as adding more neutrino mass states~\cite{Diaz:2019fwt}, allowing the heavier mass states to decay~\cite{PalomaresRuiz:2005vf,Moss:2017pur,Dentler:2019dhz,deGouvea:2019qre}, or introducing secret neutrino interactions~\cite{Gninenko:2009ks,Gninenko:2010pr,Masip:2012ke,Radionov:2013mca,Blennow:2016jkn,Ballett:2018ynz,Bertuzzo:2018itn,Arguelles:2018mtc,Liao:2018mbg,Denton:2018dqq,Ballett:2019pyw, Fischer:2019fbw, Jones:2019tow}; these more complex models are not considered further in this work.

In terrestrial experiments with low-energy neutrinos ($<\SI{100}\GeV$) and short-baselines ($\leq\SI{1}\km$), neutrino oscillations involving the heavier mass state proceed as in vacuum, parameterized by a single effective mixing angle determining the oscillation amplitude, and the value of $\Delta m_{41}^2$, which controls the oscillation wavelength.
Although mixing generally depends on all of the model parameters, results of vacuum-like neutrino oscillation experiments are often presented in a two-dimensional parameter space of one mixing angle and one mass-squared difference, where the relationship between the effective mixing angle and the rotation mixing angles -- $\theta_{14}$, $\theta_{24}$, and $\theta_{34}$ -- depends on oscillation channel.

In the 3+1 model, the disappearance and appearance oscillation probabilities are related.
A consistent interpretation of all data sets requires that non-zero oscillations be observed in $\nu_\mu\rightarrow\nu_e$, $\nu_\mu\rightarrow\nu_\mu$, and $\nu_e\rightarrow\nu_e$ channels.
At the present time this is not the case.
Flavor change consistent with the presence of an eV-scale new neutrino mass state has been observed in some $\nu_\mu\rightarrow\nu_e$ appearance experiments~\cite{athanassopoulos1996evidence,Aguilar-Arevalo:2018gpe}.
A general deficit of antineutrinos observed from nuclear reactors can be interpreted as a finite $\nu_e\rightarrow\nu_e$ disappearance signature~\cite{mention:reactor_anomaly}, although the complexities of  modeling the reactor antineutrino flux normalization and shape remain controversial, with much ongoing theoretical and experimental work~\cite{An:2016srz,An:2017osx,Ashenfelter:2015uxt,Hayes:2013wra,Hayes:2015yka,Hayes:2017res}.  
The complementary $\nu_\mu\rightarrow\nu_\mu$ channel has been studied by various experiments, but no anomalous flavor change has been observed~\cite{adamson2011active,adamson2012improved, adamson2011search,adamson2016search,stockdale1984limits,abe2015limits,mahn2012dual,cheng2012dual,stockdale1984limits}.


Global fits to world data prefer the 3+1 model over a model with no sterile neutrinos by more than $5\sigma$~\cite{Diaz:2019fwt}.  This is despite the fact that the apparent observation of flavor change in the $\nu_\mu\rightarrow\nu_e$ channel, apparent non-observation of flavor change in the $\nu_\mu\rightarrow\nu_\mu$ channel, and present knowledge of the allowed magnitude of $\nu_e\rightarrow\nu_e$ disappearance remain difficult to reconcile under the 3+1 model.
Furthermore, the removal of no single data set relieves this tension to an acceptable level~\cite{maltoni_michele_2018_1287015}.
Continued study of the $\nu_\mu$ disappearance channel with increased sample size and systematically controlled experimental datasets therefore represents a critical aspect of reaching a conclusive statement about eV-scale sterile neutrinos.

One of the strongest constraints on sterile-neutrino-induced $\nu_\mu$ and $\bar{\nu}_\mu$ disappearance is from the study of high-energy atmospheric neutrinos observed by IceCube~\cite{TheIceCube:2016oqi}.
The IceCube Neutrino Observatory~\cite{Aartsen:2016nxy} is a gigaton ice-Cherenkov detector located near the South Pole.
It is comprised of 5160 digital optical modules~\cite{Abbasi:2008aa}, or DOMs, which are self-sufficient detection units made of photomultiplier tubes enclosed in pressure housings, deployed on 86 vertically orientated ``strings" extending between $\SI{2450}\meter$ and $\SI{1450}\meter$ below the surface of the Antarctic ice sheet~\cite{Aartsen:2013rt,icecube_instrumentation}.
These modules detect the Cherenkov light emitted by charged particles created in high-energy neutrino interactions.
Most of IceCube's detected neutrinos are produced in cosmic-ray air showers and span an energy range from approximately $\SI{10}\GeV$ to $\SI{1}\PeV$, with the peak detected flux around $\SI{1}\TeV$~\cite{icecube_instrumentation}. High-energy muons produced in the cosmic-ray air showers can penetrate the Antarctic ice sheet and dominate the downwards-going event rate in IceCube. Therefore to select neutrino events, it is common to only look at events originating below the horizon (upward-going).

\begin{figure*}[t]
\begin{center}
\includegraphics[width=0.95\columnwidth]{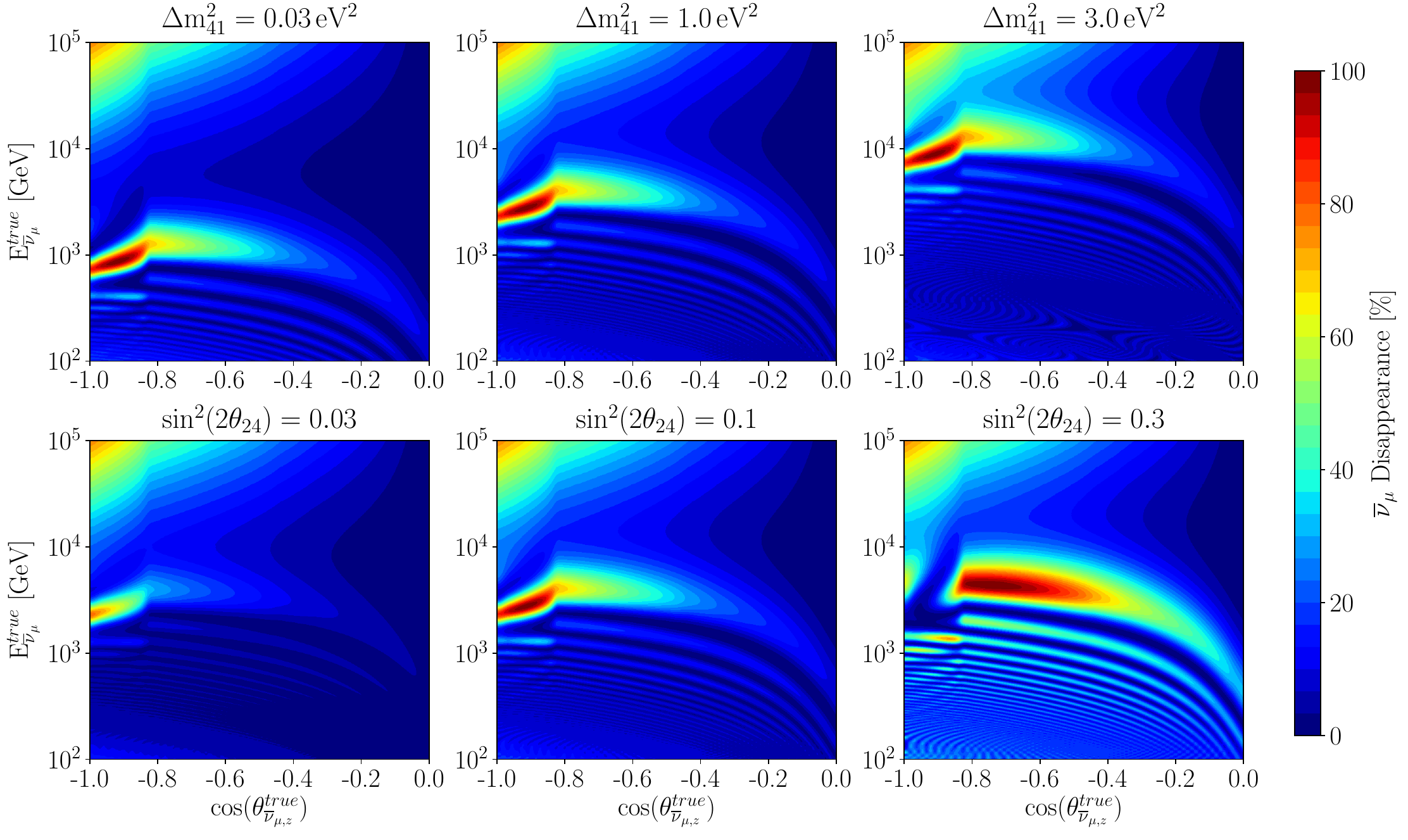}
\caption{
\textbf{\textit{Disappearance probability calculated using NuSQuIDS~\cite{nusquids} for several sterile neutrino parameters.}}
The $\bar{\nu}_\mu$ disappearance probability in terms of true neutrino energy and cosine of the zenith is shown for several sterile neutrino parameters.
Top row has fixed $\sin^2(2\theta_{24}) = 0.1$ and increasing mass-squared differences from left to right.
Bottom row has fixed $\Delta m^2_{41}=\SI{1}\eV^2$ and increasing mixing from left to right. There are visible discontinuities between inner to outer core ($\cos(\theta_\nu^{\rm true}) = -0.98$) and outer core to mantle ($\cos(\theta_z^{\rm true}) = -0.83$).  Note that the peak of the IceCube flux is at around 1~TeV.}
\label{fig::true_signal}
\end{center}
\end{figure*}

Sterile neutrinos in IceCube would give rise to a suite of oscillation effects - not only simple vacuum-like oscillations, but also matter-enhanced resonant effects induced as high-energy neutrinos cross the core of the Earth~\cite{Choubey:2007ji,Barger:2011rc,esmaili2012constraining,esmaili2013restricting,lindner2016sterile}.
These resonances can lead to a dramatic magnification of the $\nu_\mu$ disappearance signature within the IceCube atmospheric neutrino sample for mass-squared differences between $\SI{0.1}\eV^2$. and $\SI{10}\eV^2$.  For example, order-of-magnitude enhancements in the disappearance probability occur at the peak energy of 1 TeV for plausible values of $\Delta m_{41}^2$ and $\theta_{24}$.
Some examples are shown in Fig.~\ref{fig::true_signal}.

IceCube has previously searched for matter-enhanced signatures of sterile neutrinos using one year of high-energy atmospheric neutrino data~\cite{aartsen2016searches}.
The one-year sample contained 20,145 upward-going muon tracks in the approximate energy range $\SI{400}\GeV$ to $\SI{20}\TeV$, with a non-neutrino induced background of less than 0.01\%.
The study found no evidence for $\nu_\mu$ disappearance and placed a constraint in an unexplored area of the $\sin^2 (2\theta_{24})$-$\Delta m^2_{41}$ parameter space. 
The resulting upper limit on $\theta_{24}$ was constructed assuming that all other heavy-neutrino related mixing angles are zero.
The value of $\theta_{14}$ does not affect the upper limit, while the choice of $\theta_{34}=0$ yields a conservative upper limit on $\sin^2( 2\theta_{24})$~\cite{Collin:2016aqd} when $\theta_{34} < 25\degree$~\cite{Lindner:2015iaa}.
The statistical uncertainties in that analysis were at or below 5\% per bin, mandating a comparable level of control of systematic uncertainties in the Antarctic ice model, atmospheric neutrino flux, and detector response.

For values of $|\Delta m_{41}^2 L/E| \gg 1 $, sterile neutrino oscillations are rapid in energy and become unresolvable within detector resolution of approximately $\log_{10}(E/\SI{1}\TeV) \approx 0.3$~\cite{Aartsen:2013vja}, leading instead to a deficit that is approximately independent of $\Delta m_{41}^2$, $L$ and $E$.
In this regime, the presence of flavor-violating mixing makes it possible to search for signatures of sterile neutrinos either as modification of standard neutrino oscillations~\cite{Razzaque:2012tp} or anomalous flavor conversion~\cite{Blennow:2018hto}, proportional to the matter density traversed.
IceCube has previously searched for this effect using its low-energy dataset, examining 5,118 total events collected over three years in the energy range of $\SI{6.3}\GeV$ to $\SI{56}\GeV$~\cite{aartsen2017search}.
Because of their low energies, event reconstruction is more challenging in this sample and backgrounds are difficult to reduce.
To mitigate background contamination, the DeepCore sub-array~\cite{Collaboration:2011ym,icecube_instrumentation} was used for event selection and reconstruction with the remainder of the IceCube array serving as a veto against atmospheric muon backgrounds.
No evidence of atmospheric $\nu_\mu$ disappearance was observed, leading to a limit expressed in terms of the mixing matrix elements $|U_{\mu 4}|^2=\sin^2  \theta_{24}$ and $|U_{\tau 4}|^2=\sin^2 \theta_{34}\cos^2 \theta_{34}$.


This article presents an update of the search for sterile neutrinos with IceCube in both resonant (``Analysis I'') and large-$\Delta m^2_{41}$ (``Analysis II'') scenarios using the IceCube 8-year high-energy neutrino dataset. The data comprise of 305,735 $\nu_\mu$ events with reconstructed energies between $\SI{500}\GeV$ and $\SI{10}\TeV$.
Relative to earlier analyses, we use an improved high-efficiency event selection and significantly updated detector model and calibration.
The increased sample size over IceCube's previously published event selection~\cite{Aartsen:2015rwa} has mandated a substantial overhaul of the systematic uncertainty treatment related to the glacial ice, detector response, incident neutrino flux, and neutrino interactions in order to achieve systematic control at the few-percent level.
This paper aims to provide a comprehensive explanation of the search for sterile neutrinos with the IceCube 8-year high-energy neutrino data set presented in Ref. ~\cite{MEOWSPRL}.
Further information can be found in Refs.~\cite{SpencersThesis,jones2015sterile,delgado2015new}.

\section{Analysis overview~\label{sec:overview}}

The main results presented in this paper are two independent sets of frequentist confidence intervals applied in distinct analysis sub-spaces, which we will refer to as Analysis I and Analysis II.  The two analysis spaces are constructed such that the only parameter point shared by both is the ``no sterile neutrinos'' hypothesis.
For Analysis I, a Bayesian model comparison is also constructed, as reported in Ref.~\cite{MEOWSPRL}.
This paper provides detailed information on both frequentist and Bayesian analysis techniques and results.
These two statistical approaches aim to answer different, though often related, questions.
The Bayesian approach informs us about the likelihood of the model given the observed data and has the advantage of a unified interpretation of systematic and statistical uncertainties, whereas the frequentist approach allows us to construct intervals that encompass the regions of parameters that best match the data, enabling direct comparison with other published confidence intervals.

The results of Analysis I are given in terms of $\Delta m^2_{41}$ and $\sin^2(2\theta_{24})$, with $|U_{\tau 4}|^2$ (or equivalently, $\theta_{34}$) set to zero. 
As with the previous IceCube sterile neutrino search, this choice is conservative since $\theta_{34}=0$ minimizes sensitivity to the effects of non-zero $\theta_{24}$~\cite{lindner2016sterile,Collin:2016aqd}.
Neither analysis has sensitivity to $|U_{e4}|^2$ so it is set to zero throughout; similarly the new $CP$-phases are set to zero as they affect the results only marginally.
Analysis II applies to the regime where sterile-neutrino-driven oscillations are fully averaged within the energy resolution of the detector, which is the case for $\Delta m^2_{41} \gtrsim \SI{20}\eV^2$ given our energy range and resolution. For the purpose of  calculation we have fixed $\Delta m^2_{41}=\SI{50}\eV^2$.  The results of Analysis II are intervals in the two rotation angles $\sin^2(2 \theta_{34})$ and $\sin^2(2 \theta_{24})$. 

The expected at-detector neutrino flux is calculated at each hypothesis point in the physics parameter space.
This involves simulating the neutrino oscillations and absorption across the Earth; determining the interaction point in the ice or rock; producing and propagating final-state particles; modeling the detector response; emulating the online triggering and at-Pole event selection; performing event reconstruction; and applying the high-level event selection.
The signal expectations at each parameter point can then be used to generate pseudo-experiments for construction of frequentist intervals, or to compute the Bayesian evidence when constructing the Bayesian hypothesis test.

Events are selected using a new high-efficiency and high-purity event selection described in detail in Sec.~\ref{sec:EvtSel}.
All data in the sample have been reprocessed with the most up-to-date IceCube calibration protocols described in Ref.~\cite{IceCube:2020nwx} and only include events where all 86 strings of the IceCube array were fully functional. 
The total data set contains 305,735 events collected over a livetime of 7.634 years, starting on May 13$^{\mathrm{th}}$, 2011 and ending on May 19$^{\mathrm{th}}$, 2019.
The energy proxy and directional reconstructions are calculated using the latest versions of internal IceCube event reconstruction software packages, similar to those used in Ref.~\cite{Aartsen:2015rwa}.   The expected angular resolution $\sigma_{\cos \theta_z}$ varies between 0.005 and 0.015 as a function of energy, and the energy resolution is approximately $\sigma_{\log_{10} ( E_\mu)}\sim 0.5$~\cite{weaver2015evidence}.
The data are divided into 260 bins in reconstructed muon energy and the cosine of the zenith angle, \cosz.
The reconstructed energy is logarithmically binned in steps of 0.10, from $\SI{500}\GeV$ to $\SI{9976}\GeV$ (13 bins).
The \cosz is binned linearly in steps of 0.05, from $-1.0$ to 0.0 (20 bins).

We perform frequentist parameter estimation using a maximum-likelihood approach. In this work, the likelihood function, which describes the probability of the observed data given a specified physics model, is defined as:
\begin{equation}
\mathcal{L}(\vec{\Theta},\vec{\eta}) =  \prod_{i=1}^{\mathrm{N_{bins}}} \mathcal{L}_{\rm eff}(\mu_i(\vec{\Theta},\vec{\eta}),\sigma_i(\vec{\Theta},\vec{\eta}); x_i),
\label{eq::likelihood}
\end{equation}
where $x_i$ is the number of observed events in the bin; $\mu_i(\vec{\Theta},\vec{\eta})$ and $\sigma_i(\vec{\Theta},\vec{\eta})$ are the expected number of events and its corresponding Monte Carlo (MC) statistical uncertainty in the same bin; and $\vec{\Theta}$ and $\vec{\eta}$ are the set of physics and systematic nuisance parameters respectively. 
The bin-wise likelihood function $\mathcal{L}_{\rm eff}$ is a modified version of the Poisson likelihood that accounts for Monte Carlo statistical uncertainties, first introduced in Ref.~\cite{Arguelles:2019izp}.  Using this protocol, the effects of finite Monte Carlo statistics on the analysis results become negligible.

For the frequentist analysis, we use the profile likelikehood technique to account for systematic uncertainties.
The profile likelihood is defined as the constrained optimization of the likelihood,
\begin{equation}
\mathcal{L}_{\rm profile}(\vec{\Theta}) = {\rm max}_{\vec\eta} \mathcal{L}(\vec{\Theta},\vec{\eta}) \Pi(\vec{\eta}),
\label{eq::profile_likelihood}
\end{equation}
where the constraints from external information on the nuisance parameters are encoded in the function:
\begin{equation}
\Pi(\vec{\eta}) =  \prod_{j=1}^{\mathrm{N_{syst.}}} \Pi(\eta_j).
\label{eq::prior}
\end{equation}
The penalty terms for each nuisance parameter are Gaussian functions with central values and standard deviations given in Table.~\ref{table::Priors}.
The minimization is performed with the limited-memory BFGS-B algorithm~\cite{BFGS}, which is aware of box constraints on the parameters. 

In order to construct confidence regions for our parameters of interest we use the following test statistic:
\begin{eqnarray}
\mathrm{TS}(\vec{\Theta}) &=& - 2\Delta \log \mathcal{L}_{\rm profile}(\vec{\Theta}) \\
&=& - 2\left[\log \mathcal{L}_{\rm profile}\vec{\Theta} - \log \mathcal{L}_{\rm profile}(\hat{\vec{\Theta}})\right],
~\label{eq:LLHDiff}
\end{eqnarray}
where $\hat{\vec{\Theta}}$ is the best-fit point that maximizes $\mathcal{L}_{\rm profile}$.
We construct frequentist confidence regions using the Neyman construction~\cite{10.2307/91337} with the Feldman-Cousins ordering scheme~\cite{Feldman:1997qc}.
Based on validations at several points in the parameter space with Monte Carlo ensembles we find that the test-statistic distribution (Eq.~\ref{eq:LLHDiff}) follows Wilks' theorem faithfully~\cite{wilks1938}, so we use this to draw the final contours.

We have also performed a Bayesian model selection analysis reported in Ref.~\cite{MEOWSPRL}.
In order to avoid dependence on the physics parameter priors we compare each physics parameter point to the no sterile neutrino hypothesis. 
We compute the Bayesian evidence $\mathcal{E}$ at each parameter point $\vec{\eta}$. The evidence of a model with prior $\Pi_{\vec\Theta}(\vec{\tilde\Theta}) = \delta(\vec\Theta- \vec{\tilde\Theta})$ is given by~\cite{Gariazzo:2019xhx}:
\begin{equation}
    \mathcal{E}(\vec{\Theta}) = \int d\vec{\eta}~ \mathcal{L}(\vec{\Theta},\vec{\eta}) \Pi(\vec{\eta}),
\end{equation}
where the priors on the nuisance parameters are the constraints used in the frequentist analysis given in Table.~\ref{table::Priors} and the integral is evaluated using the \texttt{MultiNest} algorithm~\cite{Feroz:2013hea}.
The ratio between the evidence for a given model parameter point and for the null hypothesis is known as the Bayes factor, and quantifies the preference for the alternative model over the null.
Following usual practices, whenever appropriate, we assign a qualitative statement to the model comparison using Jeffreys scale~\cite{jeffreys1998theory}.
In this scale, strong preference for the alternative model over the null is stated when there is 95\% certainty of the alternative when both hypothesis have \textit{a priori} equal likelihoods.

\section{Signal prediction}

A critical aspect of this analysis is calculation of the expected detected muon distribution for each physics parameter point.
We now describe the method for computing the “central” Monte Carlo model, i.e., with no systematic variations applied.
This involves prediction of the atmospheric neutrino flux, calculation of expected oscillated flux at IceCube, creation of a weighted ensemble of final-state particles, propagation of these particles through the detector model, and event reconstruction.
The first two of these steps will be performed for each point in the physics parameter space, which is projected onto a grid defined as:
\begin{itemize}
\item $\Delta m^2_{41}$ from $\SI{0.01}\eV^2$ to $\SI{100}\eV^2$ logarithmically in steps of 0.05 (80 bins);
\item $\sin^2(2\theta_{24})$ from 0.001 to 1.0 logarithmically in steps of 0.05 (60 bins);
\item $\sin^2(2\theta_{34})$ from $10^{-2.2}$ to 1.0 logarithmically in steps of 0.05 (44 bins).
\end{itemize}

\subsection{Atmospheric and astrophysical neutrino flux predictions}

The neutrino flux is assumed to be composed of atmospheric and astrophysical neutrinos.
The atmospheric neutrino flux is divided into two components: the conventional flux produced by the decay of pions, kaons, and muons; and the ``prompt'' flux produced by the decay of charmed hadrons.
The astrophysical neutrino component, first observed by IceCube~\cite{aartsen2014observation}, is still of unknown origin.
Its angular and energy distribution are compatible with an isotropic arrival directions and a power-law spectrum~\cite{schneider2019characterization}.

The conventional component is computed using the Matrix Cascade Equation ({\tt MCEq}) package~\cite{fedynitch2015calculation,fedynitch2012influence}.
{\tt MCEq} solves the atmospheric shower cascade equations numerically, and takes as inputs the cosmic-ray model, hadronic interaction model, and atmospheric density profile, which are scanned here as continuous nuisance parameters.
For our central flux we use the Hillas-Gaisser 2012 H3a~\cite{gaisser2013cosmic2} primary cosmic-ray model.  The hadronic interactions involved in the development of the extensive air showers are modeled using the {\tt Sibyll2.3c}~\cite{riehn2017hadronic} model.
The atmospheric density profile, required to predict the matter density through which the air showers will develop, is extracted from {\tt AIRS} satellite data~\cite{AIRS}; further details are provided in Sec.~\ref{sec::atm_denstiy_unce}.
The month-by-month temperature profiles for each year are approximated using the 2011 temperature profile. 
Using 2011 {\tt AIRS} data, we compute the atmospheric flux for each month to account for seasonal variations and then construct a weighted sum over monthly livetime of the multi-year data sample.
To ensure that the effects of annual variability of systematic climate change were not so large as to invalidate this approximation, a simulation set was generated assuming an especially hot year with two Septembers and no January.
The observed change in time-integrated neutrino flux was found to be comfortably within neutrino flux systematic uncertainties.

Since re-interaction is not competitive with decay for the prompt atmospheric neutrino flux component, it is unaffected by the atmospheric density variations and is approximately isotropic. 
The prompt flux normalization, however, is not well known, carrying uncertainties arising from the charm quark mass and lack of hadronic data in the very forward direction from collider experiments. 
In this analysis, we set the prompt $\nu_\mu$ spectrum to the BERSS~\cite{bhattacharya2015perturbative} calculation.  This prompt flux is sufficiently sub-leading within our energy range that we do not include independent nuisance parameters to characterize its uncertainty, but rather allow any discrepancy with the central model to be absorbed by the nuisance parameters associated with the astrophysical flux.

The astrophysical neutrino flux is assumed to be isotropic, following an unbroken single power-law energy spectrum, with a central spectral index of $\gamma = -2.5$ and normalization obtained from astrophysical neutrino measurements performed by IceCube in various channels and energy ranges~\cite{schneider2019characterization,Aartsen:2020aqd}. We also assume an astrophysical neutrino to antineutrino ratio of 1:1 and uniform distribution over flavors.  Systematic uncertainties on the atmospheric and astrophysical fluxes are described in detail in Sec.~\ref{ch::systematics}.

\subsection{Oscillation prediction}

Neutrino oscillation probabilities at IceCube are functions of energy and zenith angle, with the latter affecting both the distance of travel and the matter density profile traversed. The oscillation probability of high-energy atmospheric neutrinos crossing the Earth is non-trivial to calculate for several reasons: 1) the oscillation is significantly influenced by matter effects, especially in the vicinity of resonances; 2) the matter density and composition varies as a function of position in the Earth; 3) all four neutrino states participate in the oscillation; 4) absorption competes with oscillation, implying that the evolution cannot be exactly described by Schr{\"o}dinger's equation.
For these reasons, the neutrino oscillation probability is not solvable analytically. Instead, it is calculated numerically using the {\tt nuSQuIDS} software package~\cite{nusquids}.  

The \texttt{nuSQuIDS} package is built using the Simple Quantum Integro-Differential equation Solver (\texttt{SQuIDS}) framework~\cite{squids}.
The neutrino flavor density matrix is decomposed in terms of $SU(N)$ generators plus the identity, and an open-system Liouville–Von-Neumann equation is solved numerically, seeded with the initial atmospheric neutrino flux at a height of $\SI{20}\km$ above Earth.
The terms included in the evolution include effects deriving from neutrino mass (vacuum effects), matter effects on oscillations, and absorption due to charged- and neutral-current interactions in the Earth.
Neutrino and antineutrino fluxes are propagated through the Earth in all four flavors.
Additional subleading effects are also included in \texttt{nuSQuIDS},  including the production of secondary neutrinos in charged-current $\nu_\tau$ interactions, a process known as $\tau$-regeneration~\cite{Halzen:1998be}.
For sterile neutrinos in the mass range of interest, both vacuum-like and resonant oscillation effects are generally observed; see~\cite{Akhmedov:1988kd,Krastev:1989ix,Chizhov:1998ug,Chizhov:1999az,Akhmedov:1999va,Petcov:2016iiu} for an extended discussion. 

Our oscillation predictions consistently include the three-flavor active neutrino oscillations, with neutrino mixing parameters set to the current global best-fit values for normal ordering given in Ref.~\cite{pdg}.
In practice both analyses are insensitive to all active neutrino mixing parameters and mass differences, since for $E_\nu > \SI{100}\GeV$ the active neutrino oscillation probability is insignificant over the Earth diameter.
Fig.~\ref{fig::true_signal} shows some example oscillograms that illustrate the \texttt{nuSQuIDS} predictions, expressed as transition probabilities between the initial and final flux, $\Phi_{\mathrm{initial}}$ and $\Phi_{\mathrm{final}}$, namely as $1-\Phi_{\mathrm{final}}(E,\cosz)/\Phi_{\mathrm{initial}}(E,\cosz)$.

Neutrino absorption in the Earth is computed by \texttt{nuSQuIDS} using neutrino-nucleon isoscalar deep-inelastic cross section calculation given in Ref.~\cite{cooper2011high}.
The Earth density is assumed to be spherically symmetric with density profile given by the preliminary reference Earth model (PREM)~\cite{dziewonski1981preliminary}.
Past versions of this analysis associated a systematic uncertainty with the density profile of the Earth.
However, this was found to be sub-dominant to other sources of systematic uncertainty and to the per-bin statistical precision, so the effect is no longer included here.

After propagation to the vicinity of the IceCube detector, the final-state density matrix is projected into flavor space to yield the energy- and zenith-dependent flux of each neutrino flavor. 
This information is used along with the doubly-differential deep-inelastic neutrino scattering cross sections to weight pre-generated Monte-Carlo events.

\subsection{Neutrino interaction cross section}

In the energy range of this analysis the only relevant neutrino interaction is neutrino-nucleon deep-inelastic scattering~\cite{Formaggio:2013kya}. 
We use the calculation reported in Ref.~\cite{cooper2011high} for neutrinos and antineutrinos. 
The neutrino cross section is used both in calculating the Earth opacity to high-energy neutrinos and in determining the interaction rate.
The uncertainties in the cross section reported in Ref.~\cite{cooper2011high} imply that the latter effect is negligible with respect to the uncertainty in the atmospheric neutrino fluxes.
The effect of the cross section uncertainty on the Earth opacity has been recently discussed in Ref.~\cite{Vincent:2017svp} and is small when considering the effect on the total rate. 
However, since we are now searching for 1\%-level distortions of the angular distribution, we incorporate an uncertainty contribution for the Earth opacity.
This is implemented by computing the spectrum-dependent absorption strength for each neutrino flux component, namely atmospheric conventional, prompt, and astrophysical, given re-scaled cross sections, given the uncertainties reported in Ref.~\cite{cooper2011high}.
The resulting absorption distributions are used to generate a continuous parameterization of the systematic uncertainty due to Earth opacity using the {\tt PHOTOSPLINE}~\cite{Whitehorn:2013nh} interpolation package.

\subsection{Detector simulation}

Monte Carlo samples are constructed and employed in fits using a final-state reweighting technique. Events are generated using a reference flux and propagated through the standard IceCube Monte Carlo simulation chain, to be re-weighted to a physical flux for analysis.
Cherenkov light is simulated directly through layered IceCube ice models which include the effects of absorption and scattering of light on dust and other impurities, and the response of the IceCube DOMs is simulated using the techniques described in Ref.~\cite{Aartsen:2016nxy}.
The optical effects of refrozen ice immediately surrounding the IceCube strings are parameterized, as is the optical anisotropy of the ice, and the tilt (non-planarity) of the glacier~\cite{chirkin2013evidence}.

Secondary particles are injected into the target volume encompassing the detector according to a reference energy spectrum and a continuous doubly differential cross section.
We consider a range of injected primary $\nu_\mu$ energies from $\SI{100}\GeV$ to $\SI{1}\PeV$ from zenith angle 80$^\circ$ (10$^\circ$ above the horizon) to 180$^\circ$ (upward-going).
The injected energies are sampled using an $E^{-2}$ power-law energy spectrum and the arrival directions are distributed isotropically in azimuth and \cosz.
The interaction is assigned by randomly selecting a point within a cylindrical volume centered on IceCube, whose axis is aligned with the trajectory of the incoming particle, with an injection radius of $\SI{800}\m$.
The cylinder length is set to be the 99.9\% muon range in ice plus two additional ``endcaps'', each with a length of $\SI{1200}\m$.  This procedure allows for efficient generation of representative Monte Carlo samples of $\nu_\mu$ interactions that deposit light in the IceCube detector.

For each event, the incident neutrino energy, final-state lepton energy and zenith, Bjorken $x$ and $y$ interaction variables,  probability of the neutrino interaction, and the properties associated with the injected point (total column depth and impact parameter) are recorded.
A full simulation set in this analysis contains $2\times 10^9$ such events each generated with independent seeds, yielding a number of events approximately equal to 500 years of detector data.
For each oscillation hypothesis and systematic uncertainty parameter configuration the event ensembles are re-weighted according to the final-state prediction for that model.

The charged final-state secondaries are propagated through the ice according to the expected ionization energy loss and stochastic losses~\cite{koehne2013proposal}, accounting for ionization, Bremsstrahlung, photonuclear processes, electron pair production, Landau–Pomeranchuk–Migdal  and Ter-Mikaelian effects, and Moli{\`e}re scattering using the {\tt PROPOSAL} package~\cite{koehne2013proposal}.
Along the track, photons are generated randomly according to the parameterization of the Cherenkov radiative emission from tracks (muons) or cascades (electromagnetic or hadronic showers).
Each photon is tracked as it propagates through ice until it is either absorbed or interacts with a DOM.
The photon propagation accounts for random scatters according to the ice model, described as depth- and wavelength-dependent scattering and absorption~\cite{Askebjer:1997ep,Askebjer:1994yn,Price:97,JGRD:JGRD12654}, and anisotropy~\cite{chirkin2013evidence} along a major and minor axis.
At each photon scattering point, the algorithm randomizes the new photon direction based on a scattering angle distribution parameterized by the mean scattering angle and scattering coefficient.
For each photon that strikes a DOM in the simulation, the detector response is modeled according to standard IceCube methods, which are outlined in Ref.~\cite{icecube_instrumentation}.
Simulated events are reconstructed using the same algorithms that are applied to real events, in the same manner as the previous generation of IceCube sterile neutrino searches~\cite{Aartsen:2015rwa,TheIceCube:2016oqi}.

\begin{figure}[tb]
\centering
\includegraphics[width=\columnwidth]{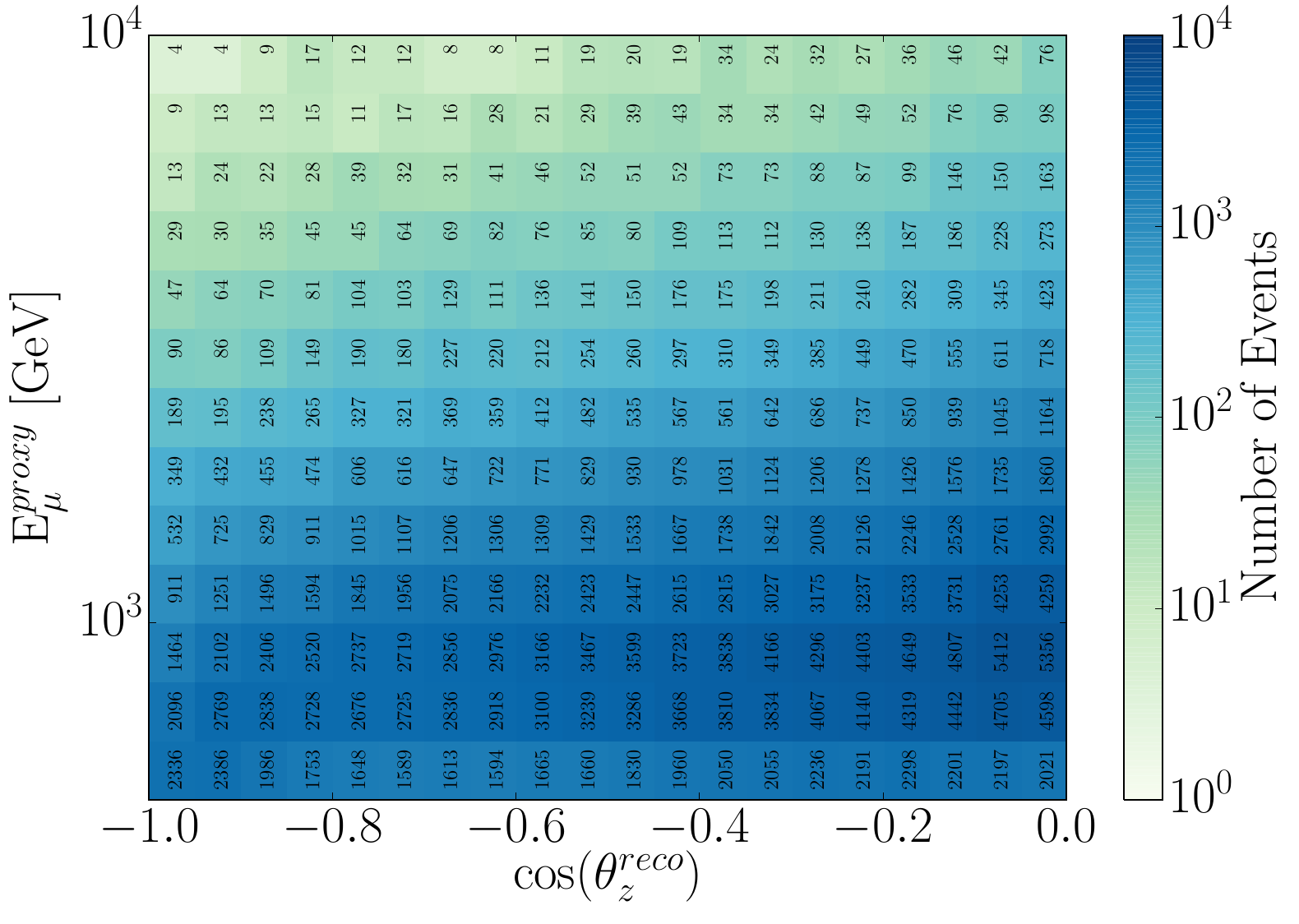}
\caption{\textbf{\textit{
Number of observed events per bin in the full eight-year dataset used in this work.}}
}
\label{fig::n_events}
\end{figure}

\section{Event selection\label{sec:EvtSel}}

\begin{figure}[t]  
    \centering
    \subfloat{\includegraphics[width=0.90\linewidth]{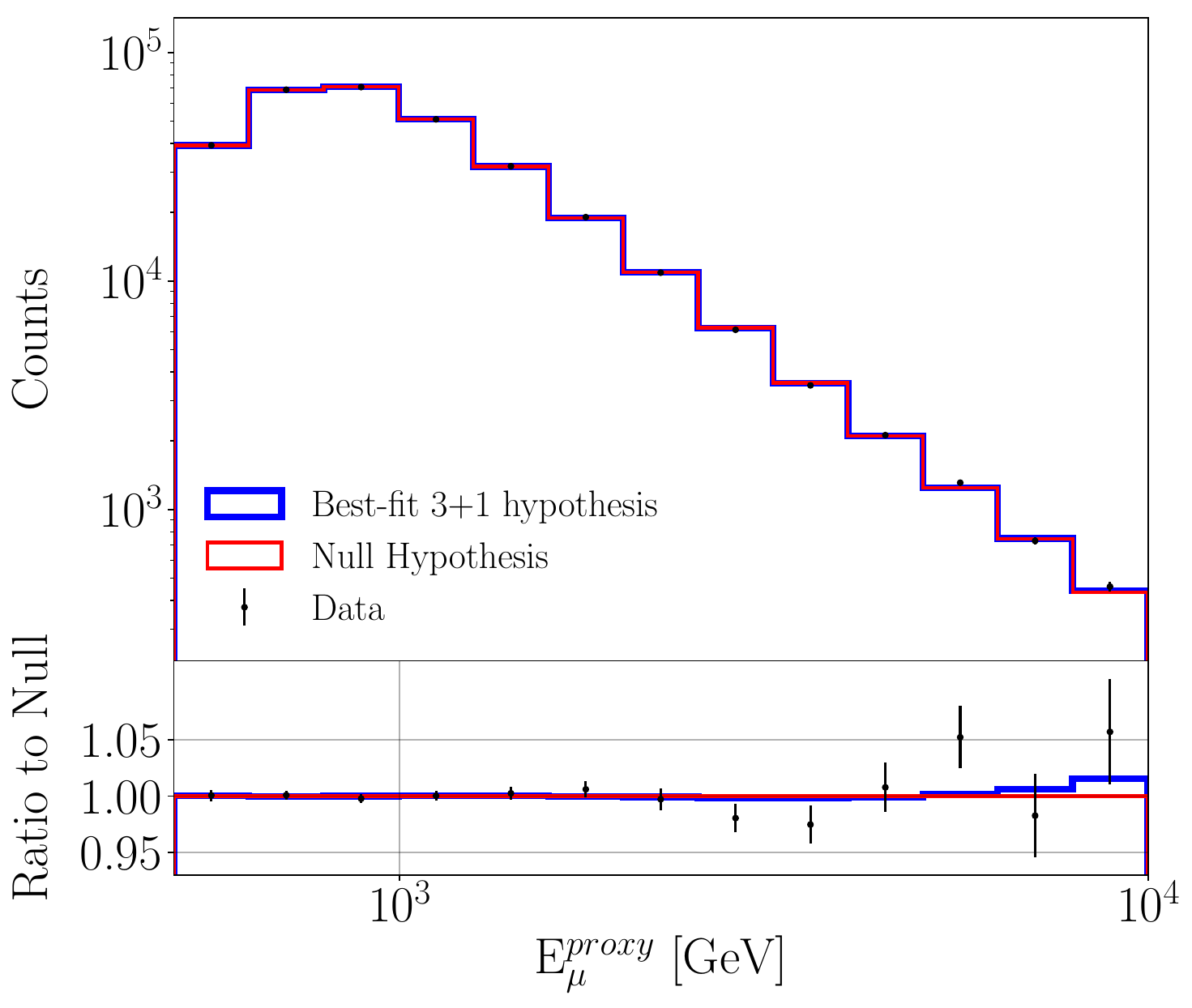}}
    
    \subfloat{\includegraphics[width=0.90\linewidth]{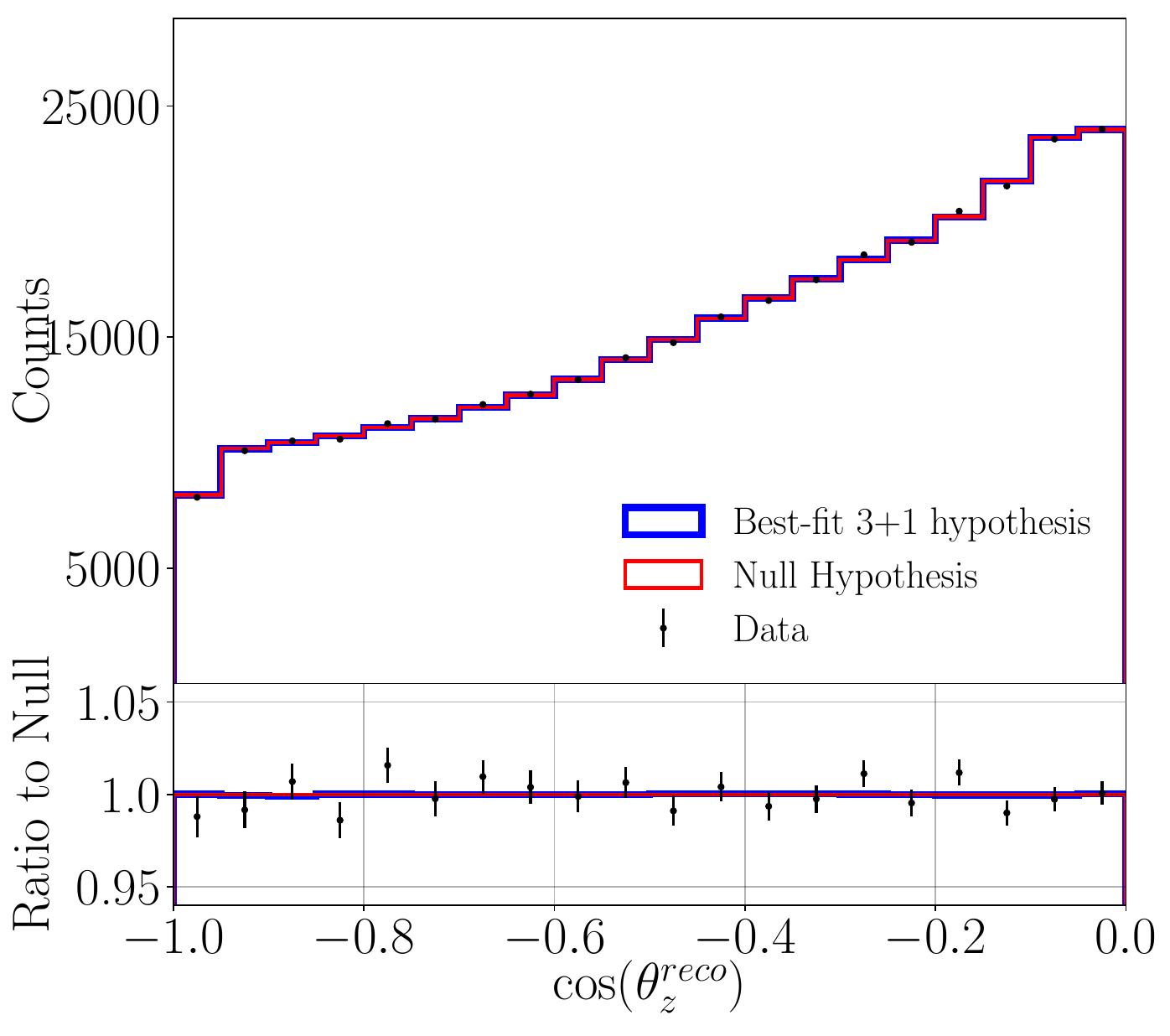}}
    \caption{\textbf{\textit{Reconstructed muon energy (top) and cosine of the zenith (bottom) distributions.}} Data points are shown in the blue histogram with the error bars that represent the statistical error.
    The solid blue and red lines show the best-fit sterile neutrino hypothesis and the null (no sterile neutrino) hypothesis, respectively, with nuisance parameters set to their best-fit values in each case.}
    \label{fig::Energybs1}
\end{figure}

Muons are identifiable in IceCube by the track-like nature of emitted Cherenkov light as they propagate through the ice.
The event selection defines the set of criteria used to reduce the background event contamination while maintaining a high-efficiency selection of atmospheric $\nu_\mu$ events. The primary background contributions comprise air-shower cosmic-ray (sometimes referred to herein as ``atmospheric'') muons, neutral-current neutrino interactrions, charged-current electron neutrino interactions, and charged-current tau neutrino interactions.
The event selection described in this section identifies 305,735 events,  shown distributed in reconstructed energy and \cosz in Fig.~\ref{fig::n_events}.
The energy and zenith distributions of the data are shown separately in Fig.~\ref{fig::Energybs1}.

Despite the $\SI{1.5}\km$ of overburden directly above IceCube, the detector is triggered at a rate of approximately $\SI{3}\kHz$~\cite{aartsen2016characterization} by downward-going muons produced in cosmic-ray air showers.
The simulation of cosmic-ray air showers is handled by the \texttt{CORSIKA} Monte Carlo package~\cite{heck1998corsika,heck2000extensive}.
Eight independent \texttt{CORSIKA} simulation sets containing $6\times10^8$ events are used to quantify the amount of cosmic-ray muon contamination in the event selection, covering a primary cosmic-ray energy from $6 \times 10^2~\si\GeV$ to $1 \times 10^{11}~\si\GeV$.
\texttt{CORSIKA} simulates the air showers to ground level, propagating the cosmic-ray muons through the firn and ice to a sampling surface around the detector.
The cosmic-ray muons are then weighted to an initial cosmic-ray flux, in this case HillasGaisser2012 H3a~\cite{gaisser2013cosmic2}.


\begin{figure}[t]  
 \begin{minipage}{\textwidth}
   \includegraphics[width=\textwidth]{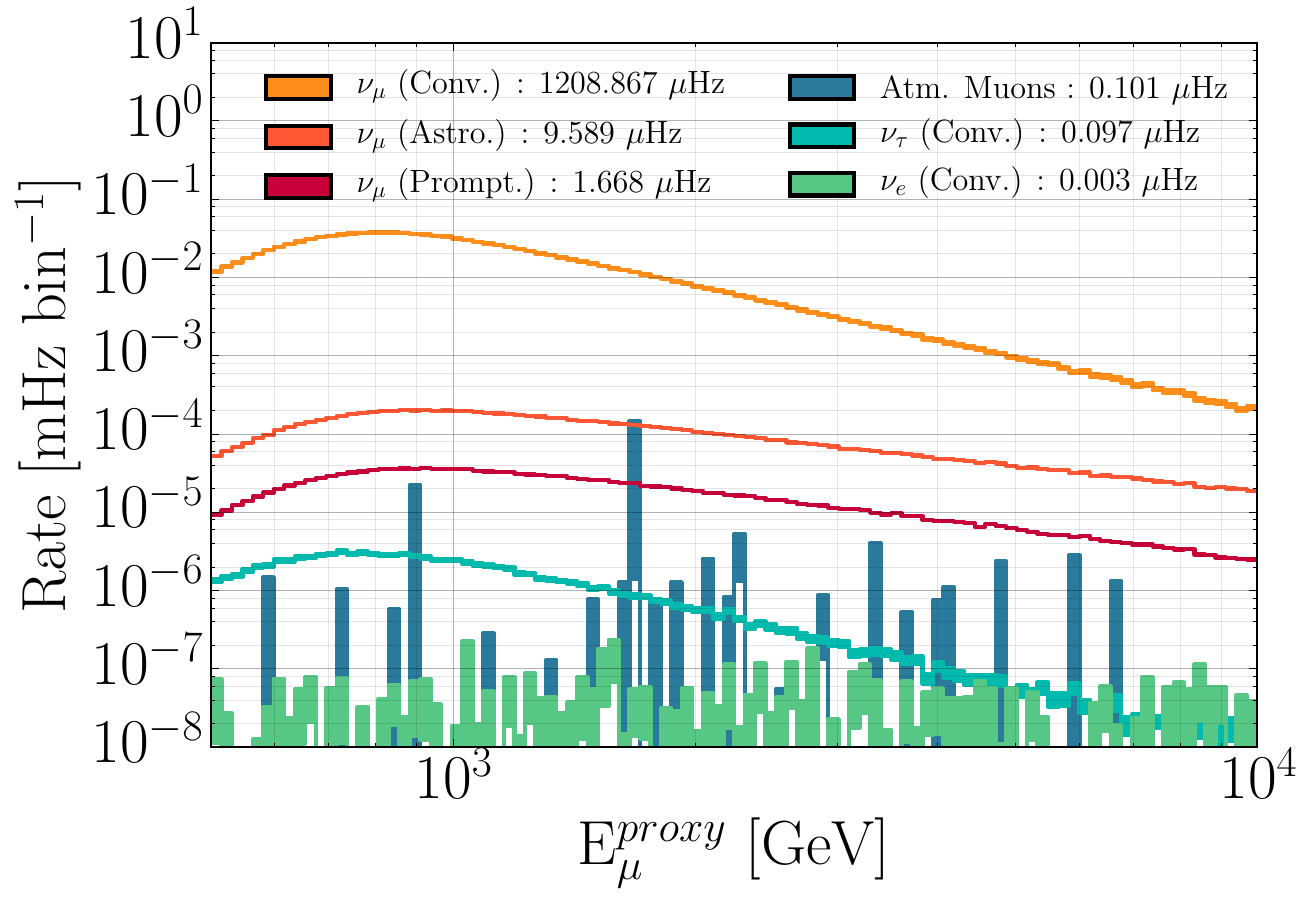}
  \end{minipage}\\
  \begin{minipage}{\textwidth}
   \includegraphics[width=\textwidth]{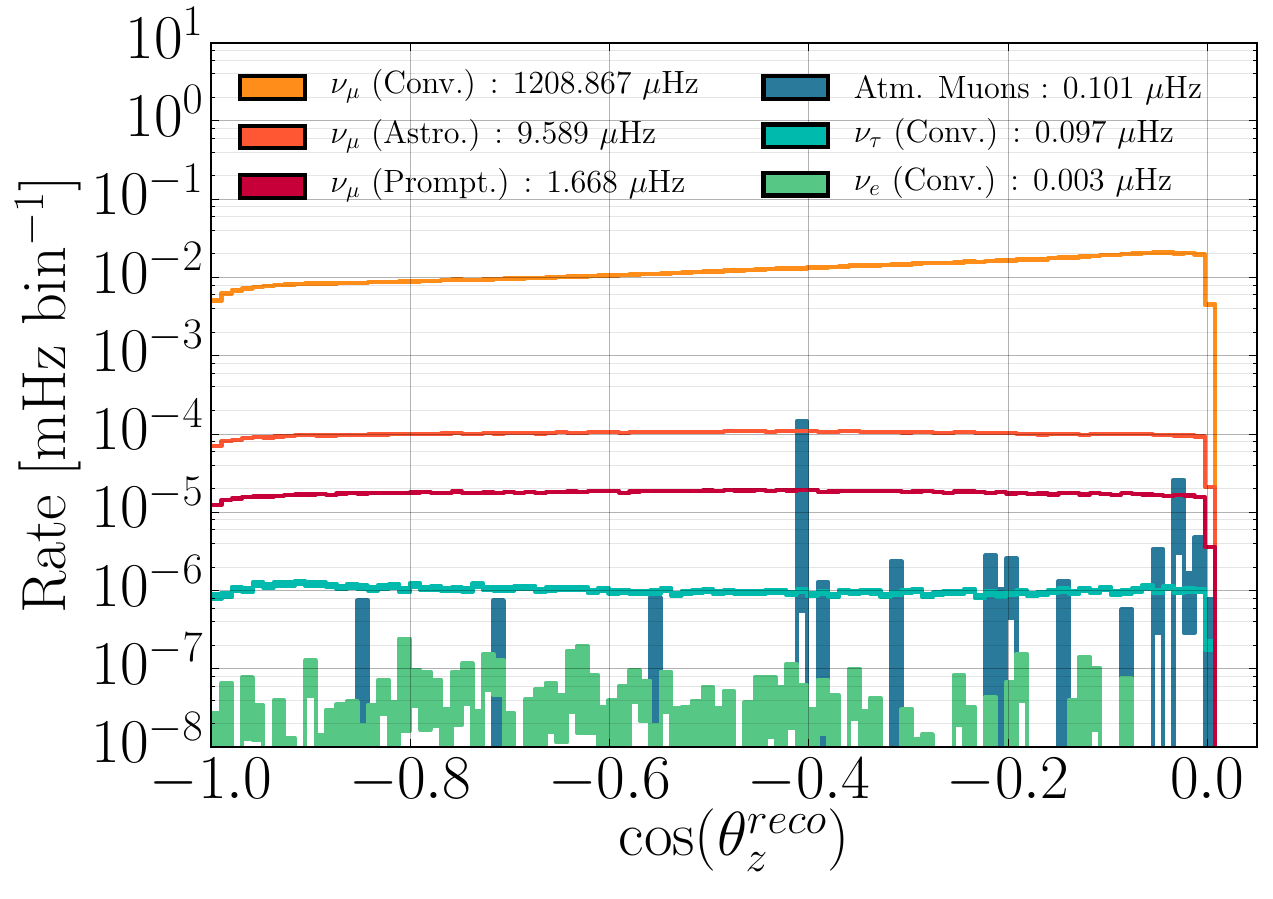}
  \end{minipage}
 \caption{
 \textbf{\textit{Expected composition of the energy and zenith distributions.}}
 Top: The reconstructed energy distribution for signal (conventional atmospheric, prompt atmospheric, and astrophysical $\numu$ flux) and backgrounds (atmospheric muons, \nutau, and \nue).
 Bottom: The corresponding reconstructed zenith direction.}
 \label{fig::final_dist}
\end{figure}    

\begin{figure}[t]
\centering
\includegraphics[width=\textwidth]{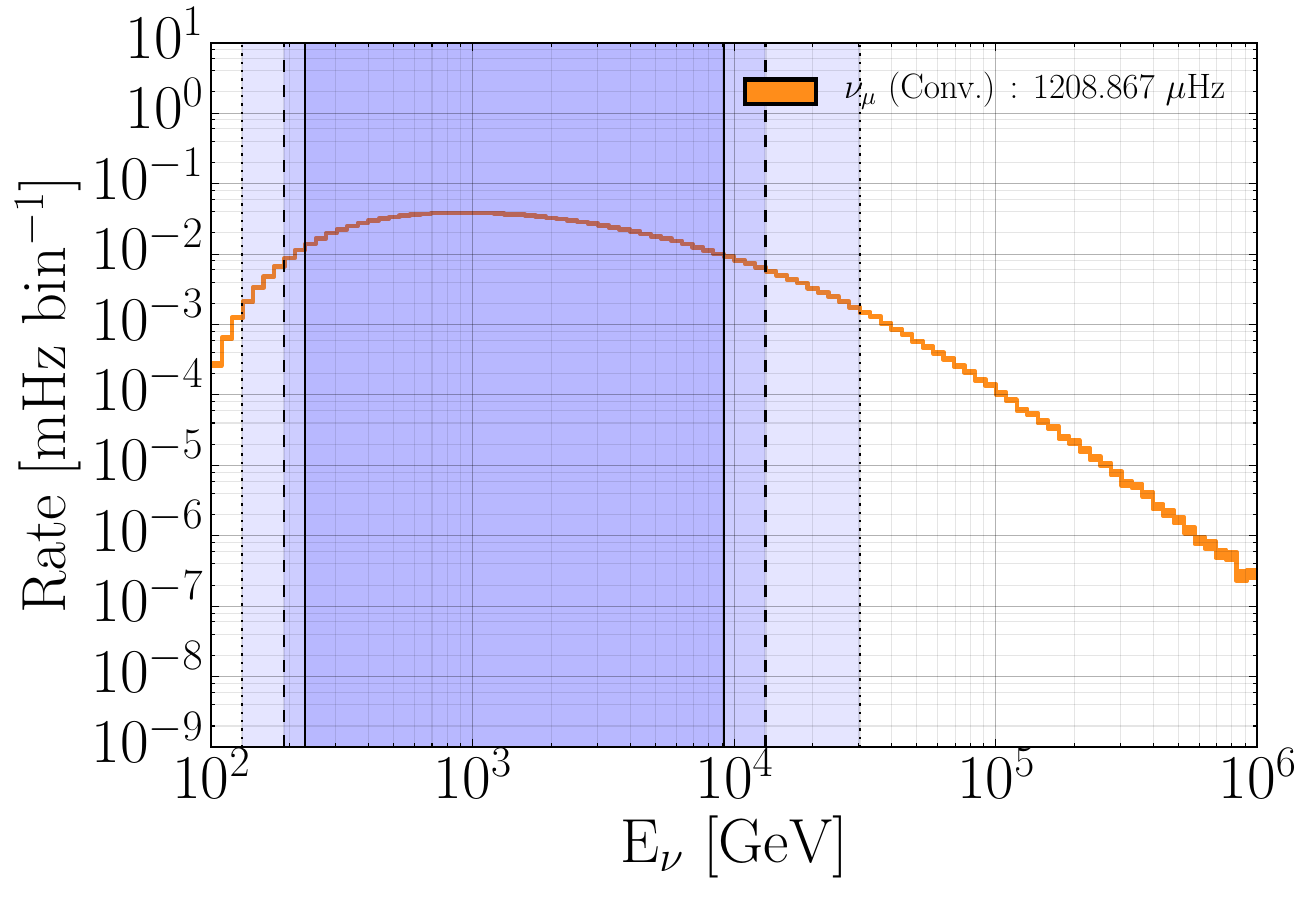}
\caption{
    \textbf{\textit{Predicted true conventional neutrino energy distribution.}}
    The distribution is shown as an orange histogram for the conventional atmospheric component for default nuisance parameters.
    The translucent regions indicate the area that contains 90\% (solid lines), 95\% (dashed), and 99\% (dotted) of the data.
}
\label{fig::final_dist_true}
\end{figure}

At the Earth's surface, the conventional $\nu_\mu$ flux dominates the neutrino flavor composition.
The sub-dominant electron and tau neutrino flavors represent a far lower fraction of the background than the cosmic-ray muons.
The topological signature of cascades, primarily caused by electron-neutrino and neutral-current neutrino interactions, is sufficiently different from the track-like topology that they are efficiently rejected.
Tau neutrinos can interact via charged-current interactions producing a tau lepton and a cascade-like shower.
When the tau lepton decays producing hadrons these events are also efficiently rejected.
However, the tau lepton can subsequently decay to a muon and flavor conserving neutrinos with a branching ratio of 17.39\,$\pm$\,0.04\%~\cite{pdg}.
While the signature of these events are track-like in nature, the $\nutau$-appearance probability from standard oscillations is small at TeV energies considering the first $\numu\rightarrow\nutau$ oscillation maximum occurs at approximately $\SI{25}\GeV$ for upward-going neutrinos.
The electron and tau neutrino backgrounds are accounted for in dedicated simulation sets, each with an effective livetime of approximately 250 years.

The event selection for this analysis is the union of two event filters, referred to hereafter as the \textit{Golden} and the \textit{Diamond} filters (Sec.~\ref{sec::golden} and~\ref{sec::diamond}).
If an event passes either one of these filters, it is included in our final sample.
For both filters, we first require that every event passes the online IceCube muon filter, which selects track-like events. We then pass the event through a series of precuts used to reduce the data and MC to a manageable level (Sec.~\ref{sec::precuts}).
An energy cut is placed at $\SI{500}\GeV$ reconstructed energy, since events below this energy are found to contribute minimally to the sensitivity of the analysis, and may be subject to additional low-energy systematic uncertainties.
We also place a cut to select upward-going events with \textit{i.e.} \cosz\,$\leq$\,0, above which the sample is most likely to have atmospheric muon background contamination, also with minimal impact on sensitivity.
Fig.~\ref{fig::final_dist} shows plots of the expected event rate distributions in reconstructed energy and reconstructed cosine zenith for the different event types generated using MC events passing the event selection.
We show the predicted true neutrino energy distribution of the conventional atmospheric neutrinos in the sample in Fig.~\ref{fig::final_dist_true}.
We find that greater than 90\% of our events originate from a neutrino with a true energy between $\SI{200}\GeV$ and $\SI{10}\TeV$. 
The observed zenith angle can be taken as the true zenith angle, $\theta^{reco}_z=\theta_z$, for practical purposes, since within our angular bins the difference in zenith angle between the reconstructed muon track and the MC truth is negligible ($<1^\circ$, discussed in more detail in Ref.~\cite{Aartsen:2019epb}).

\begin{table}[t]
\footnotesize
\centering
\resizebox{\columnwidth}{!}{
\begin{tabular}{|l| c c c c c|}
\hline
\textbf{Sub-Selection} & \numu & \nutau & \nue & $\mu$ &  Purity\\ [0.5ex]
\hline
\hline
Golden Filter & 154,970$\pm$393 & 16$\pm$4 & 1$\pm$1& 16$\pm$4 &$>$99.9\%\\
Diamond Filter & 295,416$\pm$543& 22$\pm$5 & 1$\pm$1&4$\pm$2 &$>$99.9\% \\
\hline
\end{tabular}}
\vskip 0.15cm
\centering

\caption{\textbf{\textit{Sub-sample event composition.}}
The expected number of events that pass the Golden, Diamond filters. The uncertainties are statistical only.}
\label{table::evtsel_expectation2}
\end{table}

\begin{table}[t]
\footnotesize

\centering

\begin{tabular}{|l| c |}
\hline
\textbf{Component} & \textbf{Full Sample Composition}\\
\hline
\hline
Conv. \numu & 315,214$\pm$561 \\
Astro \numu & 2,350$\pm$48 \\
Prompt. \numu & 481$\pm$22\\
All \nutau & 23$\pm$5  \\
All \nue &  1$\pm$1 \\
Atmospheric $\mu$ &  18$\pm$4 \\ 
Purity & $>$99.9\% \\ 
\hline

\end{tabular}

\caption{\textbf{\textit{Final event selection expected number of events.}}
The expected final sample composition.
The uncertainties are statistical only.}
\label{table::evtsel_expectation1}
\end{table}

\subsection{Precuts and low-Level reconstruction}\label{sec::precuts}

Before applying high-level event selections, a series of precuts are applied to reduce data volume and reject low-quality event candidates.  These precuts are:
\begin{enumerate}
\item If the reconstructed direction is above the horizon, $\cosz \geq 0.0$,  require that the total event charge (Qtot) is greater than 100 photoelectrons (PE) and the Average Charge Weighted Distance (AQWD) is less than $\SI{200}\mppe$.
The AQWD is defined as the average distance of the pulses produced by Cherenkov light in each PMT, weighted by the total charge of the event from the track hypothesis.

\begin{figure*}[t!]
    \centering
    \begin{minipage}{0.45\textwidth}
        \centering
        \includegraphics[width=0.9\linewidth]{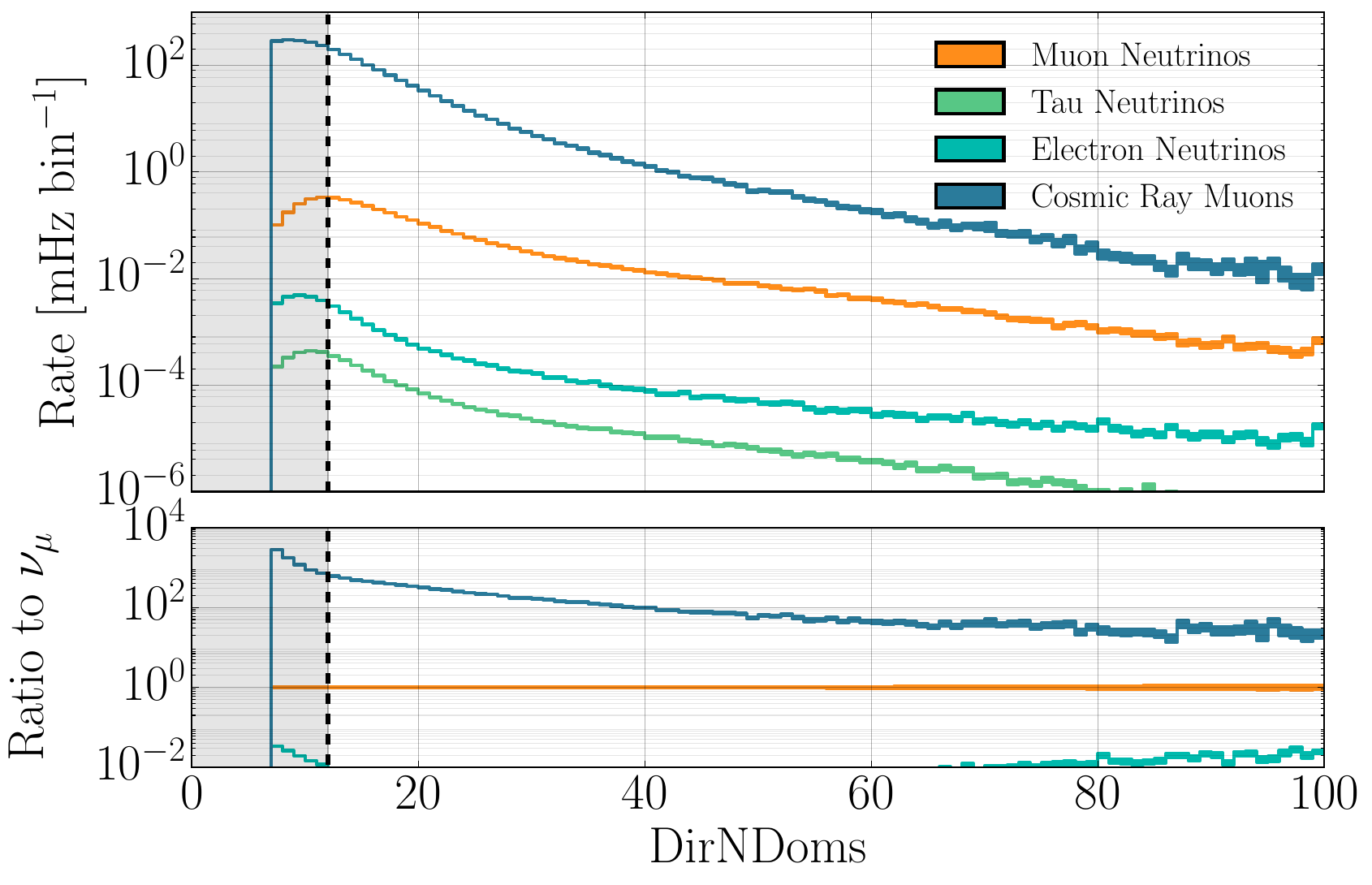}
    \end{minipage}%
    \begin{minipage}{0.45\textwidth}
        \centering
        \includegraphics[width=0.9\linewidth]{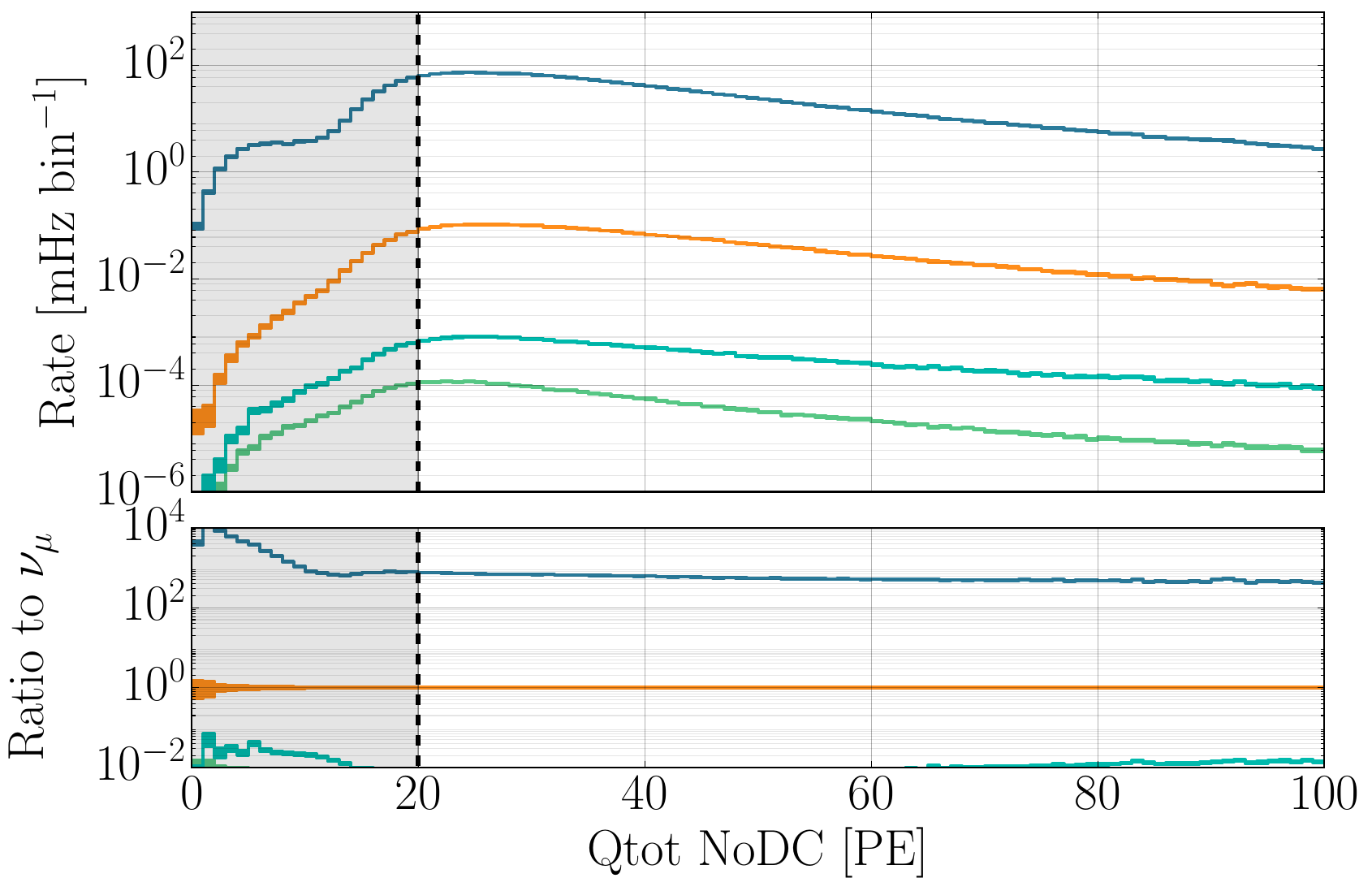}
    \end{minipage}
    \begin{minipage}{0.45\textwidth}
        \centering
        \includegraphics[width=0.9\linewidth]{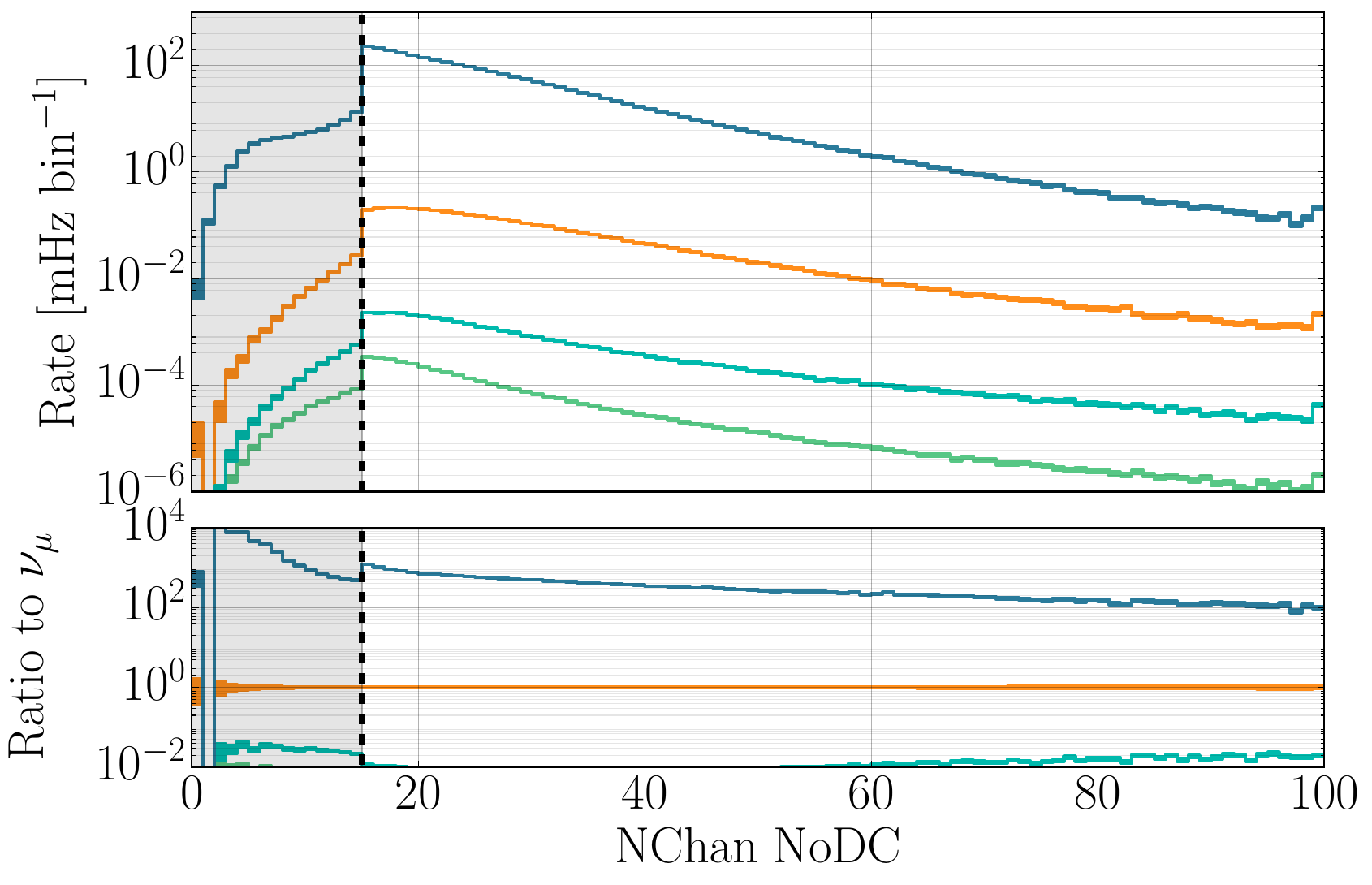}
    \end{minipage}%
    \begin{minipage}{0.45\textwidth}
        \centering
        \includegraphics[width=0.9\linewidth]{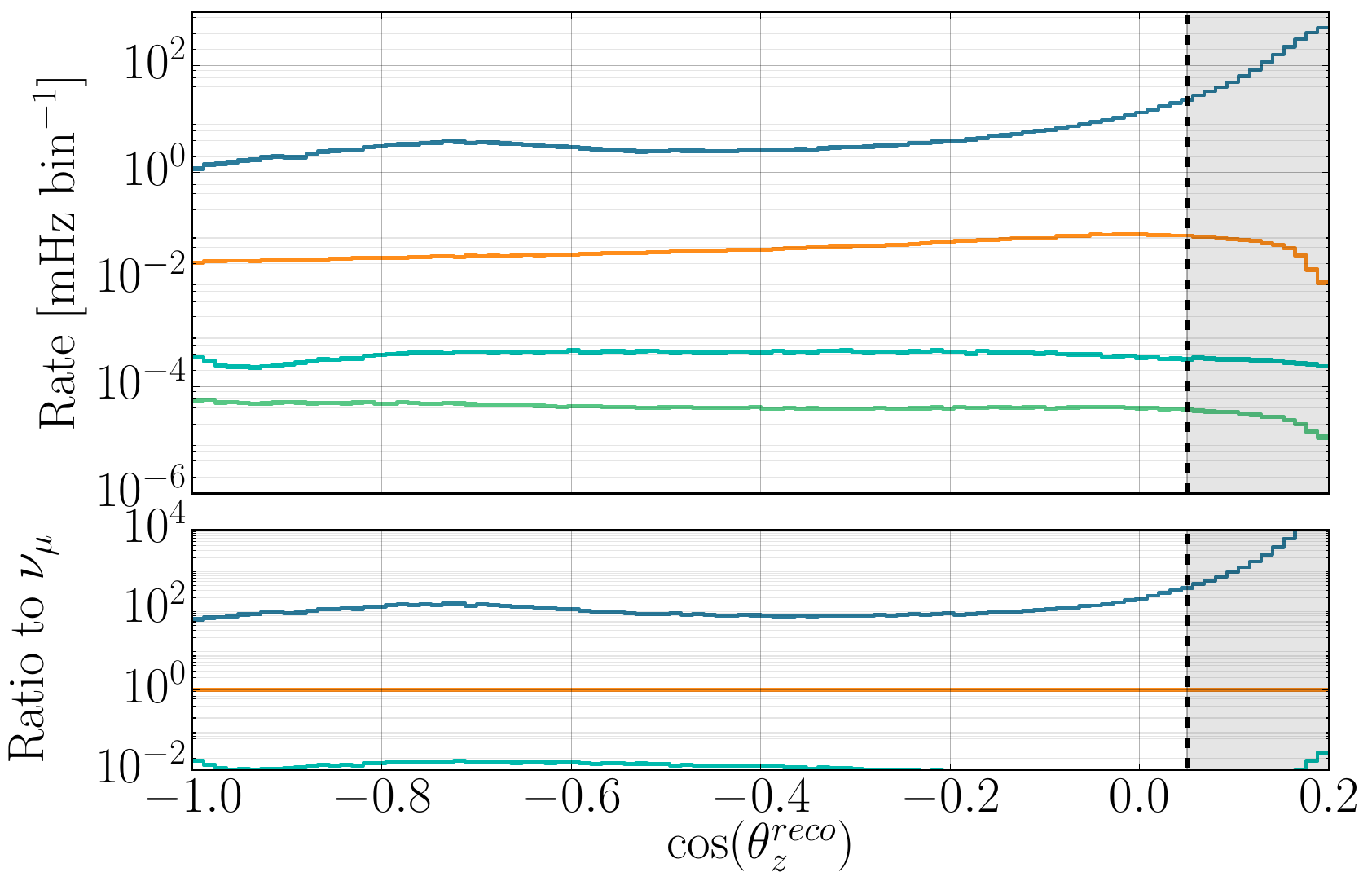}
    \end{minipage}
    \caption{\textbf{\textit{Distributions of variables used in the Diamond filter.}}
    Four different variables are shown, from top to bottom: event charge excluding DeepCore (Qtot NoDC), number of triggered DOMs (NChan NoDC), number of DOMs that see direct light (DirNDoms), and the cosine of the zenith angle ($\cos\theta_z$). 
    The signal (conventional $\nu_\mu$) is shown in orange, while the backgrounds are shown in blue, teal, and green (cosmic ray muons, electron neutrinos, and tau neutrinos respectively). 
    The vertical-dashed line in each plot shows the location of the cut, and the shaded region is rejected.}
	\label{fig::precut}
\end{figure*}

\item Reject all events with a reconstructed zenith angle with $\cosz\geq 0.2$.
The vast majority of these are muons produced in atmospheric showers.
\item  Require at least 15 triggered DOMs per event, and $\geq$ 6 DOMs triggered on direct light. Direct light refers to the Cherenkov photons which arrive at the DOMs without significant scattering, identified via event timing. 
\item The reconstructed track length using direct light (DirL) in the detector must be greater than $\SI{200}\meter$ (${\rm DirL} \geq \SI{200}\meter$), and the absolute value of the smoothness factor (DirS) must be smaller than 0.6 ($|{\rm DirS}|\leq 0.6$). For well-reconstructed events, direct hits should be
smoothly distributed along the track. The DirS variable is a measure of this~\cite{jero2017search}.
The smoothness factor is a measure that defines how uniform the distribution of triggered DOMs is around the reconstructed track.
\end{enumerate}

For every event that passes the precuts, we apply the following reconstruction methodology:

\begin{enumerate}
\item The event passes through an event splitter to separate coincident events into multiple independent sub-events.
A coincident event is defined as an event in which a uncorrelated cosmic-ray muon entered the detector during the readout. We allow a maximum deviation of 800\,ns from the speed of light travel time in which a pair of hits is to be considered correlated.
Approximately 10\% of neutrino events have an accompanying coincident muon in the time window.  
\item Reconstruct the trajectory of each sub-event iteratively, using several timing-based reconstruction algorithms.
The first algorithm uses a simple least-squares linear regression to fit the timing distribution of the first PE observed on each DOM~\cite{pinat2017icecube}. 
Then, algorithms incorporating the single photoelectron and multi-photoelectron information are used to refine the fit. These use likelihood constructions to account for the Cherenkov emission profile as well as the ice scattering and absorption, initially using the first detected photon and  then all detected photons, respectively. 
We require that each fit succeeds in order to keep the event in the sample.
\item Reject events using a likelihood ratio comparison between the unconstrained track reconstruction and one that has a prior on the reconstructed direction.
The prior, defined in Ref.~\cite{weaver2015evidence}, utilizes the fact that the majority of muon tracks are from downward-going cosmic-ray events and are expected to come from the Southern Hemisphere.
\item Calculate a variable to quantify the uncertainty in the reconstructed trajectory~\cite{neunhoffer2006estimating}.
This ``paraboloid sigma'' value encodes the uncertainty on the trajectory reconstruction based on the likelihood profile around the best-fit reconstructed track hypothesis, with a small value indicating better precision in the reconstructed trajectory.
A second variable, called the ``reduced Log likelihood'' (RLogL), uses the best-fit likelihood value as a global measure of the success of the fit.
\item Reconstruct the event energy.
Unlike the trajectory reconstructions, energy reconstruction relies heavily on the intensity of the light incident on each DOM.
Given the trajectory reconstruction, an analytical approximation for the observed light distributions is used, which accounts for the geometry between the emitter and receiver, the ice absorption and scattering, and detector noise.
Stochastic losses from high-energy interactions imply that there will be points along the track with bursts of comparably more intense light.
This is averaged out by broadening the PDF that describes the energy loss expectation.
Further information can be found in Refs.~\cite{icecube_energy, weaver2015evidence}.
\end{enumerate}

The total rate of both signal and background after the precuts is approximately $\SI{1280}\mHz$ (110592 events per day).
This is composed almost entirely of cosmic-ray muons. 

\subsection{The Golden Filter}\label{sec::golden}

The Golden filter was originally designed as the event selection for diffuse astrophysical neutrino searches~\cite{aartsen2015evidence}.
It was optimized to accept high-energy $\nu_\mu$ and was subsequently used in the one-year IceCube high-energy sterile neutrino search~\cite{aartsen2016searches}.
A detailed description of the cuts can be found in Refs.~\cite{weaver2015evidence}. In brief, following simple charge multiplicity cuts, downward-going track-like events are selected and reconstructed using  algorithms of increasing complexity, with successive track quality cuts applied at each stage to reject cosmic muon backgrounds (see Supplementary Material of Ref~\cite{weaver2015evidence}).  The event selection for the one-year diffuse neutrino analysis was determined to have a greater than 99.9\% $\nu_\mu$ purity based on simulated neutrino and cosmic-ray events.
Further scrutiny of approximately 1000 events during the preparation of the present analysis revealed evidence for an approximate 1\% contamination due to coincident cosmic-ray muons.
All these events are reported to have an AQWD greater than $\SI{100}\mppe$.
In many of these events, a coincident poorly reconstructed cosmic-ray muon visibly passed through the detector simultaneously with a neutrino event.
Supplementary cuts were added to the event selection of the 1-year sterile neutrino analysis~\cite{aartsen2016searches} to remove these events.  It was verified that the impact of this contamination on the final analysis result was negligible.

All events passing the Golden Filter are included in our sample, in addition to extra ones recovered using a new, higher-efficiency filter described in Sec.~\ref{sec::diamond}.
The event counts predicted to pass for each sample shown in Table.~\ref{table::evtsel_expectation1}, and for the union in Table.~\ref{table::evtsel_expectation2}.

\begin{figure*}[t!]
    \centering
    \begin{minipage}{0.45\textwidth}
        \centering
        \includegraphics[width=0.9\linewidth]{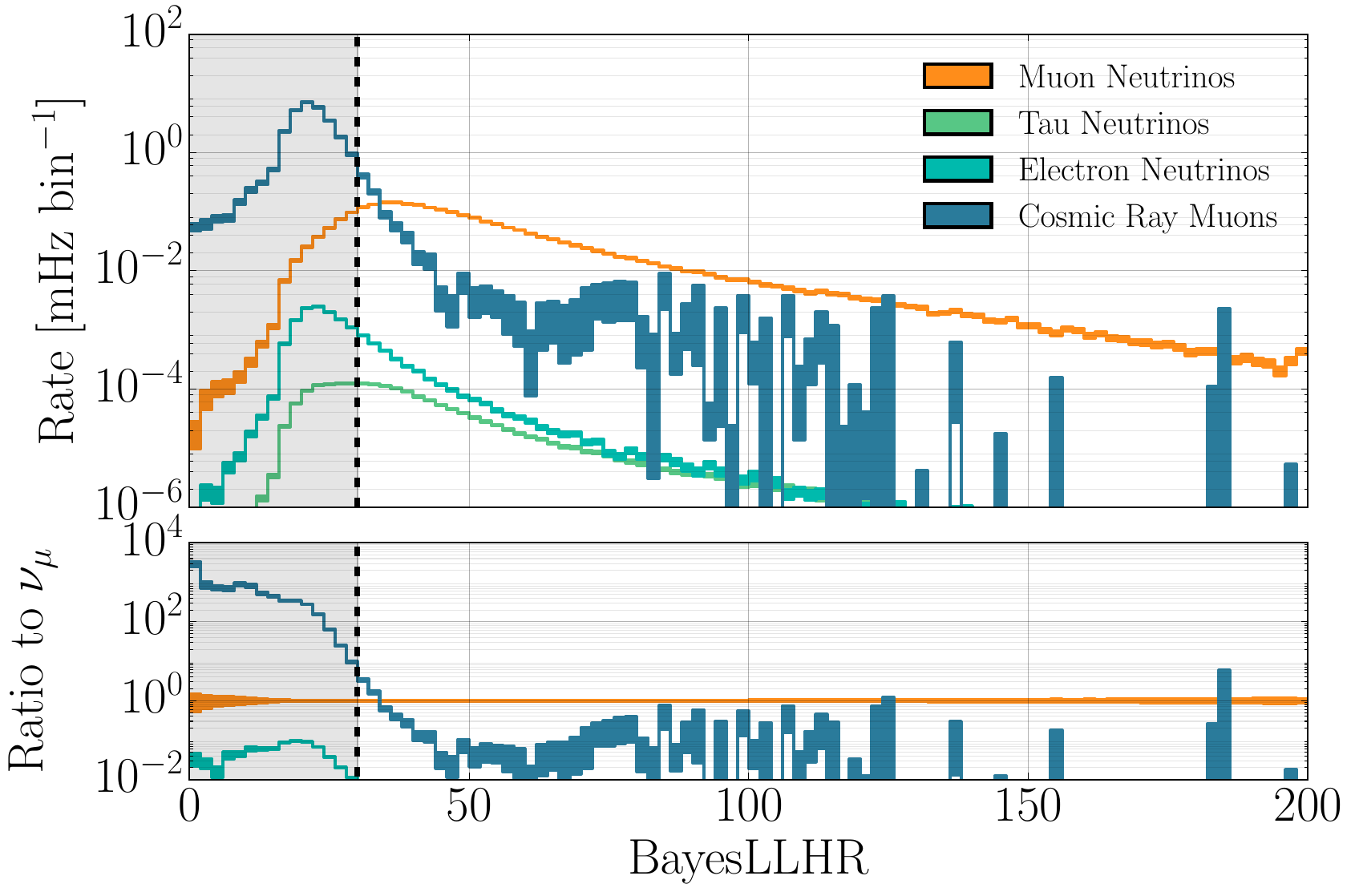}
    \end{minipage}%
    \begin{minipage}{0.45\textwidth}
        \centering
        \includegraphics[width=0.9\linewidth]{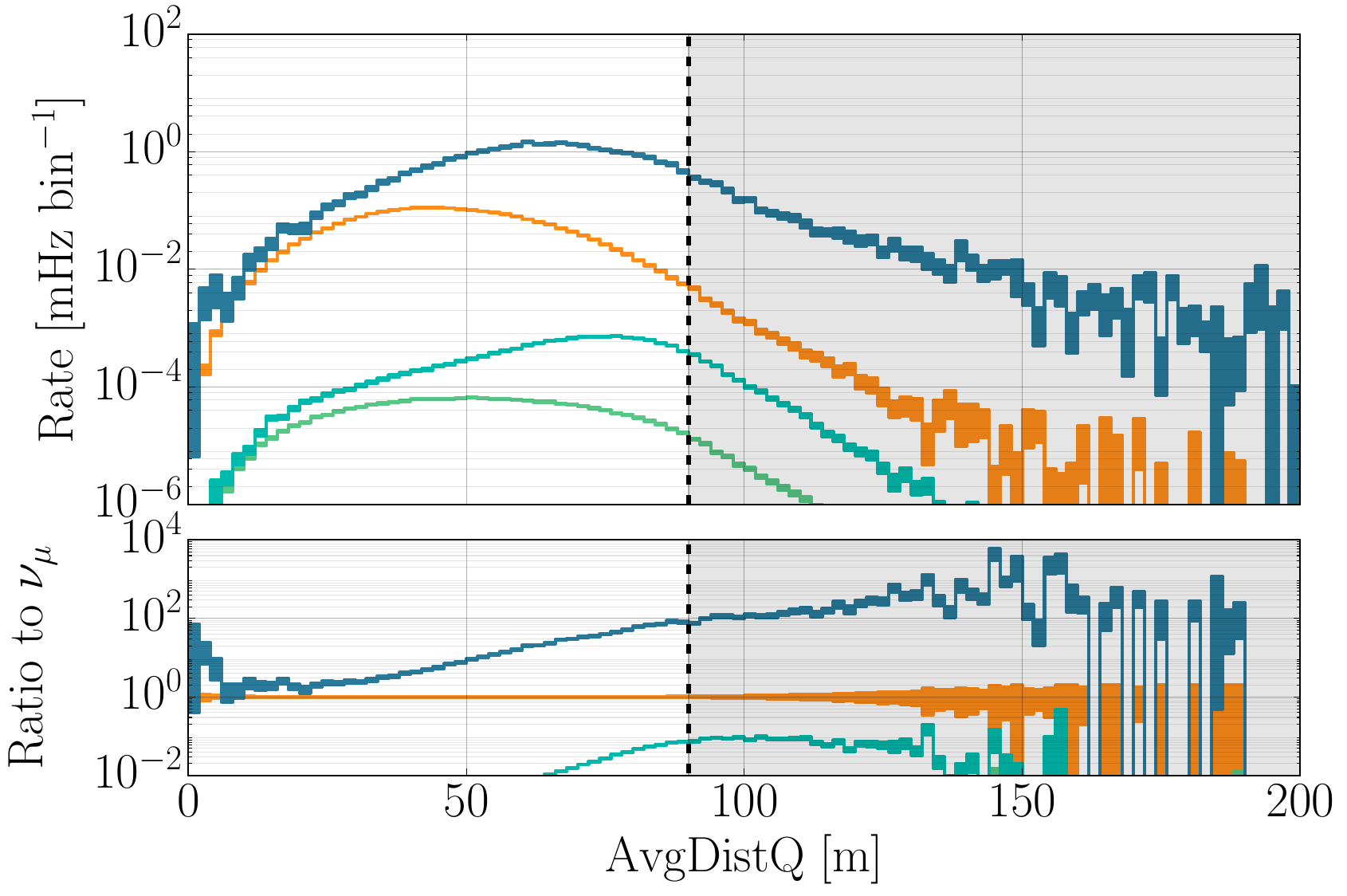}
    \end{minipage}
    \begin{minipage}{0.45\textwidth}
        \centering
        \includegraphics[width=0.9\linewidth]{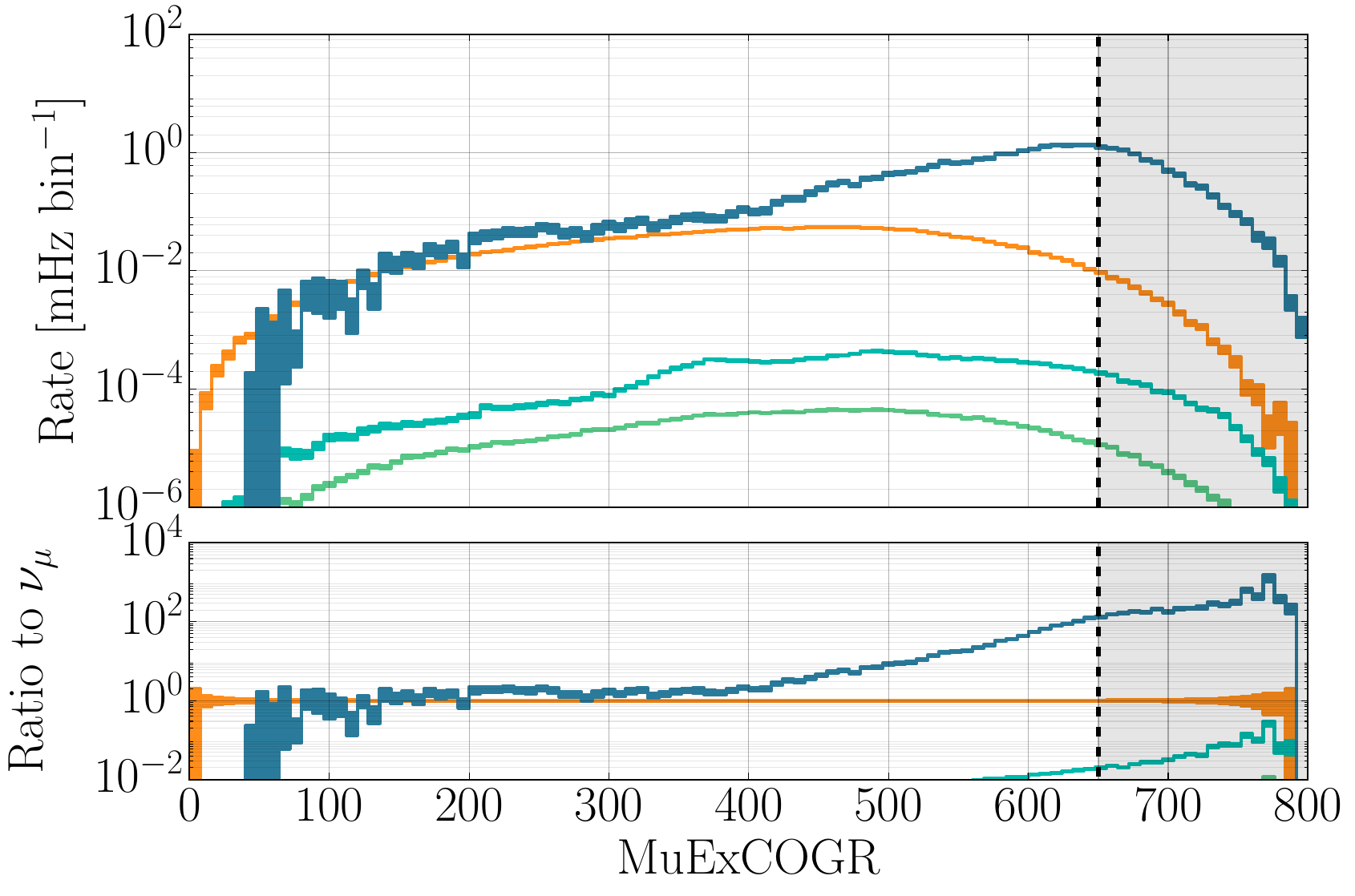}
    \end{minipage}%
    \begin{minipage}{0.45\textwidth}
        \centering
        \includegraphics[width=0.9\linewidth]{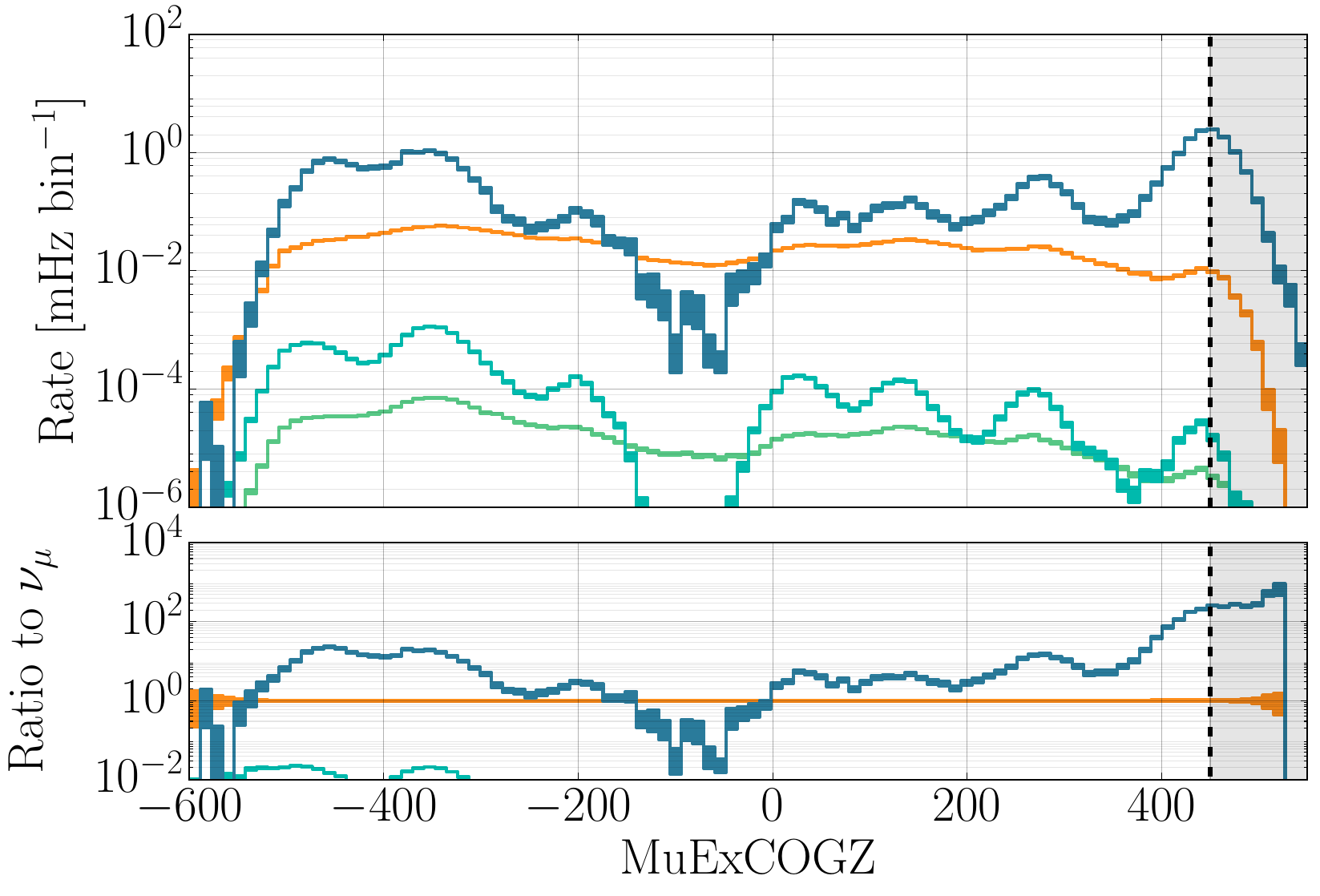}
    \end{minipage}
    \caption{\textbf{\textit{Cuts on variables used to reduce the atmospheric shower background.}}
    The $\nu_\mu$ signal is shown in orange, while the backgrounds are shown in blue, teal, and green representing expectations from cosmic-ray muons, electron neutrinos, and tau neutrinos respectively.
    The vertical-dashed line in each plot shows the location of the cut, and the shaded region is rejected. The notable depth dependent structure is primarily due to the ice optical properties.}
	\label{fig::cr_reduction1}
\end{figure*}

\subsection{The Diamond Filter}\label{sec::diamond}

The Diamond filter represents a new event selection introduced in this work, targeted at improving detection efficiency while maintaining high sample purity.
The Diamond filter begins with a second data reduction step beyond the precuts defined above:

\begin{enumerate}
\item The total charge of the event must be greater than $\SI{20}\pe$ outside of DeepCore (${\rm Qtot} > \SI{20}\pe$).
\item Require the event to have more than 15 triggered DOMs, excluding DeepCore (${\rm NChan}  > 15$).
\item At least 12 DOMs must have seen direct light (${\rm DirNDOMs} \geq 12$).
\item The reconstructed trajectory cannot extend significantly above the horizon ($\cosz < 0.05$).
\end{enumerate}

These cuts reduced the total rate to approximately $\SI{20}\mHz$ (1728 day$^{-1}$) and are each illustrated in Fig.~\ref{fig::precut}.

\begin{figure}[t]  
 \begin{minipage}{\columnwidth}
   \includegraphics[width=\textwidth]{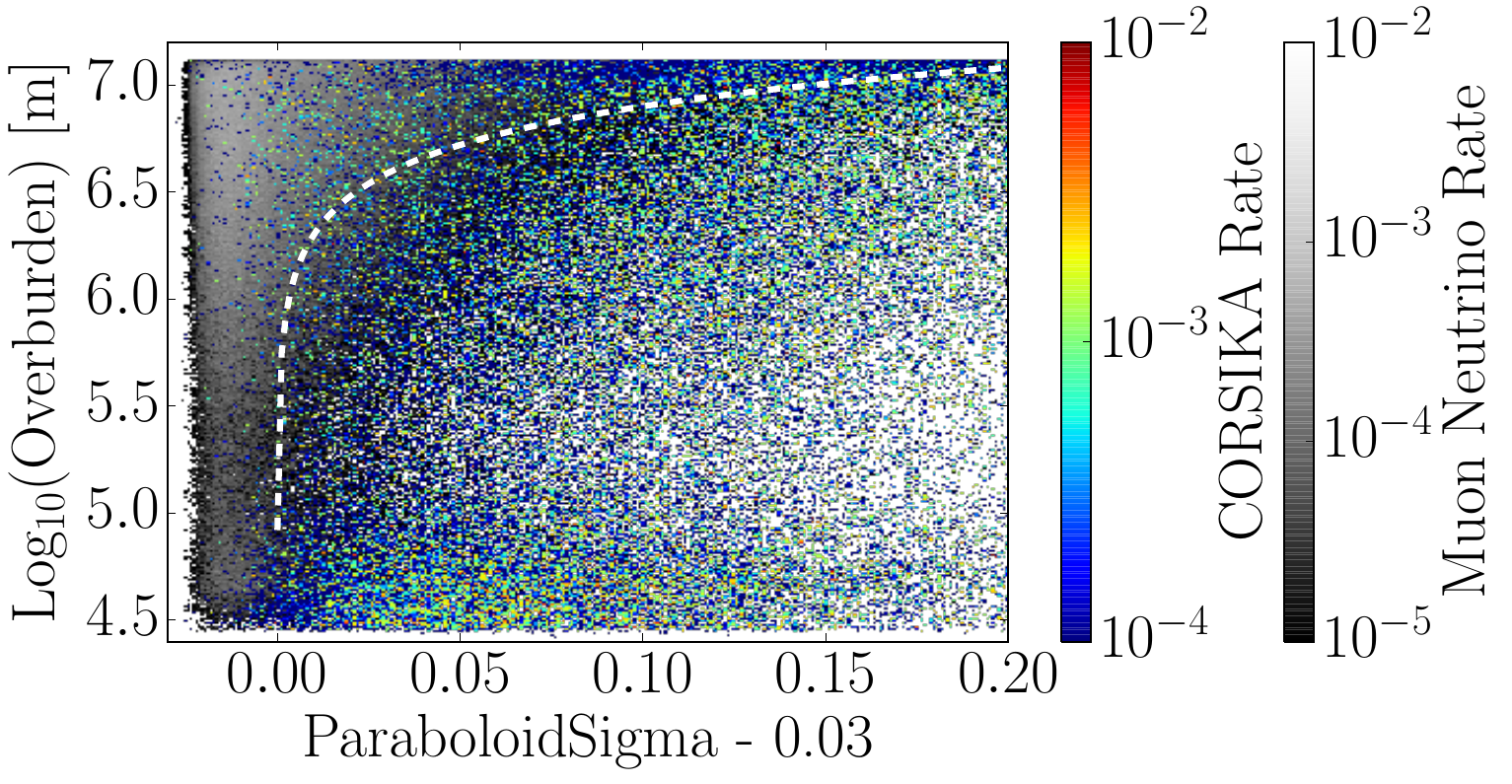}
  \end{minipage}
  \hfill
  \begin{minipage}{\columnwidth}
   \includegraphics[width=\textwidth]{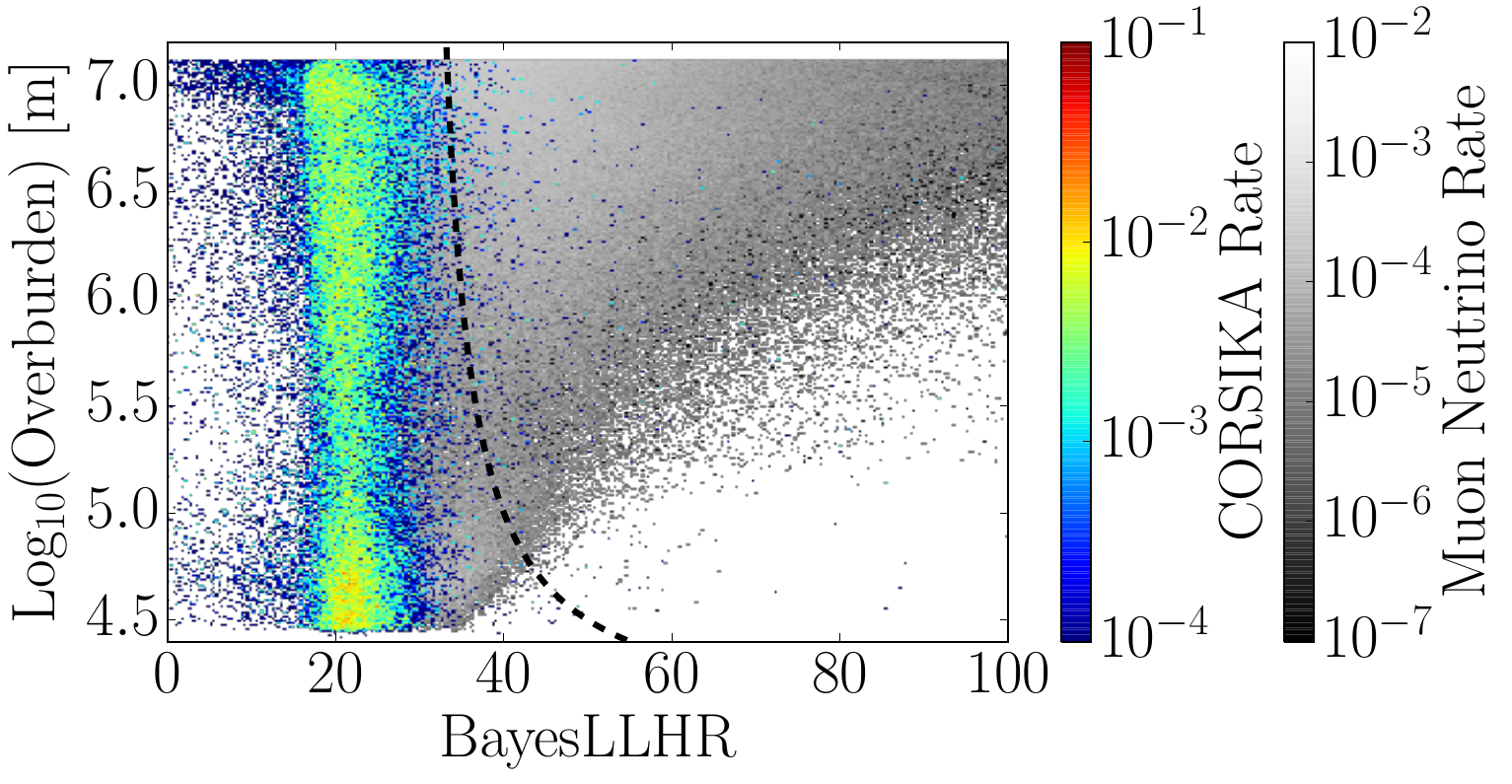}
  \end{minipage}
 \caption{
 \textbf{\textit{Two-dimensional cuts on overburden and reconstruction quality variables.}}
 The top figure uses ParaboloidSigma, while the bottom uses BayesLLHR.
 In these figures the atmospheric muon background is shown with the rainbow color scale labeled CORSIKA, while the signal is shown with the light gray color. For more details, see Ref.~\cite{SpencersThesis}.
 }
 \label{fig::cr_reduction2}
\end{figure}

The ice and rock overburden provides the greatest natural handle on the atmospheric muon contamination.
Horizontal trajectories, for example, have approximately 157 kilometre water equivalent  shielding between the atmosphere and the detector.
Any atmospheric muons reconstructed with a trajectory originating below the horizon will likely have a poor reconstruction, quantified by a large value of paraboloid sigma.
A two-dimensional cosmic-ray muon cut leverages this principle, shown in Fig.~\ref{fig::cr_reduction2}.
At small overburdens, for events near the horizon, we require a smaller uncertainty in the track reconstruction, namely smaller values of paraboloid sigma.
A Bayesian likelihood ratio (BayesLLHR), formed by comparing the reconstruction likelihood with a prior favoring downward-going arrival directions compared to the reconstruction likelihood without this prior, was introduced in Ref.~\cite{Weaver:2015} specifically to reduce the cosmic-ray muon backgrounds.
We include a cut on the Bayesian likelihood ratio as a function of overburden.

A series of straight cuts were then introduced on the center of gravity of the charge in both the vertical direction (COGZ) and the radial direction (COGR).
These cuts reduce the contamination by events near the edge of the detector, known as ``corner-clipping" events, which have a higher probability of being misreconstructed cosmic-ray muons.
We also introduce the same updated AQWD cut found in the Golden filter. Figures~\ref{fig::cr_reduction1} and~\ref{fig::cr_reduction2} show these cuts, and are listed as:

\begin{enumerate}
\item The value paraboloid sigma is greater than 0.03, cut event if $\log_{10}({\rm Overburden}) \leq 0.6 \times \log_{10}({\rm ParaboloidSigma}-0.03)+7.5$.
\item The Bayesian likelihood ratio is less than 33 units (${\rm BayesLLHR} < 33$).
\item The average charge weighted distance is greater than $\SI{90}\mppe$ (${\rm AQWD} > \SI{90}\mppe$).
\item The center of gravity of the charge in the vertical direction is above $\SI{450}\meter$ from the center of IceCube (${\rm COGZ}> \SI{450}\meter$).
\item The center of gravity in the radial direction is greater than $\SI{650}\meter$ (${\rm COGR} > \SI{650}\meter$).
\item Keep the event if $$\log_{10}({\rm Overburden}) \leq 10/({\rm BayesLLHR}-30)+4.$$
\end{enumerate}

Finally, we attempt to remove residual background events with some simple safety cuts.
These are shown in and Fig.~\ref{fig::cleanup2} and  Fig.~\ref{fig::cleanup}.
The two-dimensional RLogL and DirNDoms cuts below are used in the Golden Filter and found to be useful without affecting neutrino data. 
These are given by:

\begin{figure}[ht]  
 \begin{minipage}{\columnwidth}
   \includegraphics[width=\textwidth]{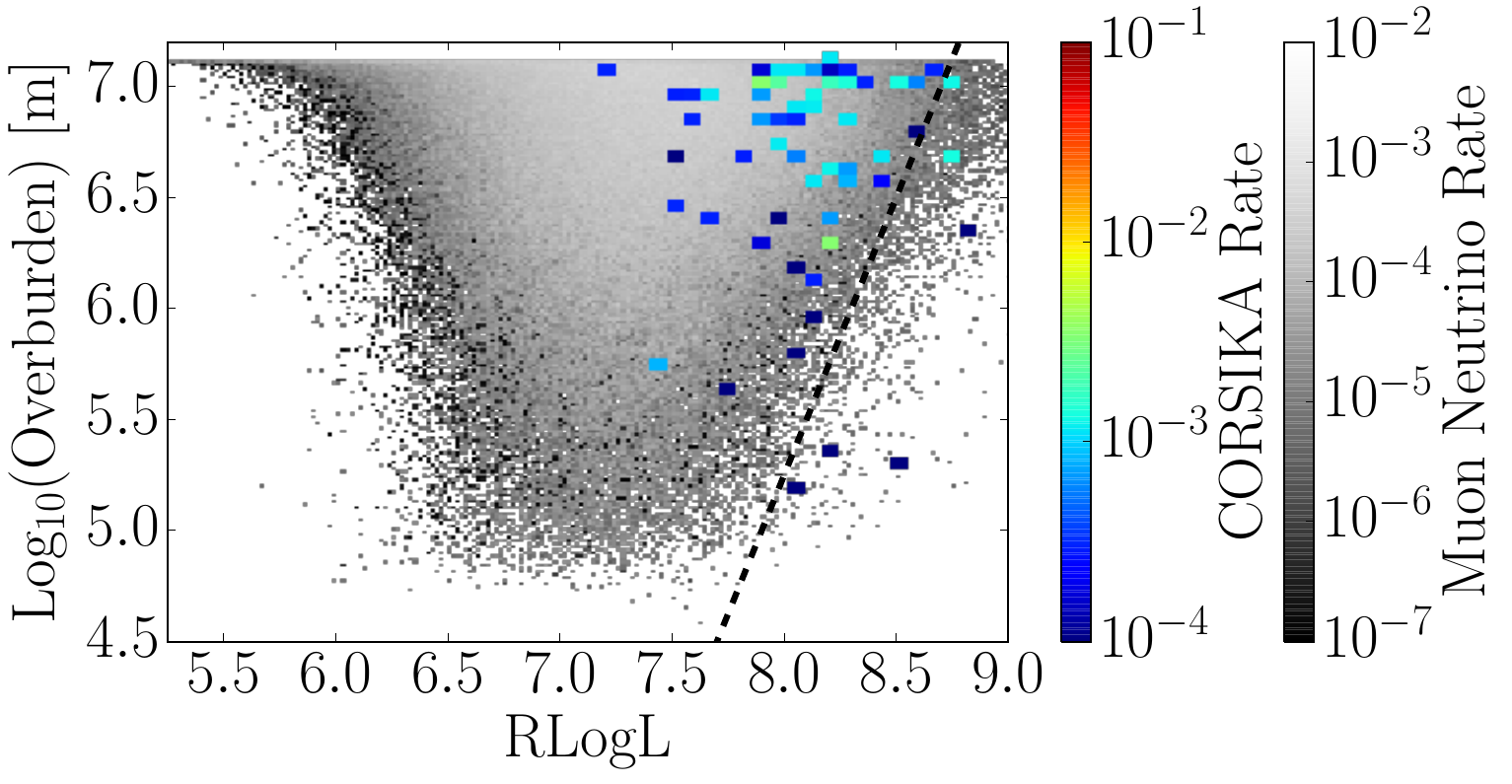}
  \end{minipage}
  \hfill
  \begin{minipage}{\columnwidth}
   \includegraphics[width=\textwidth]{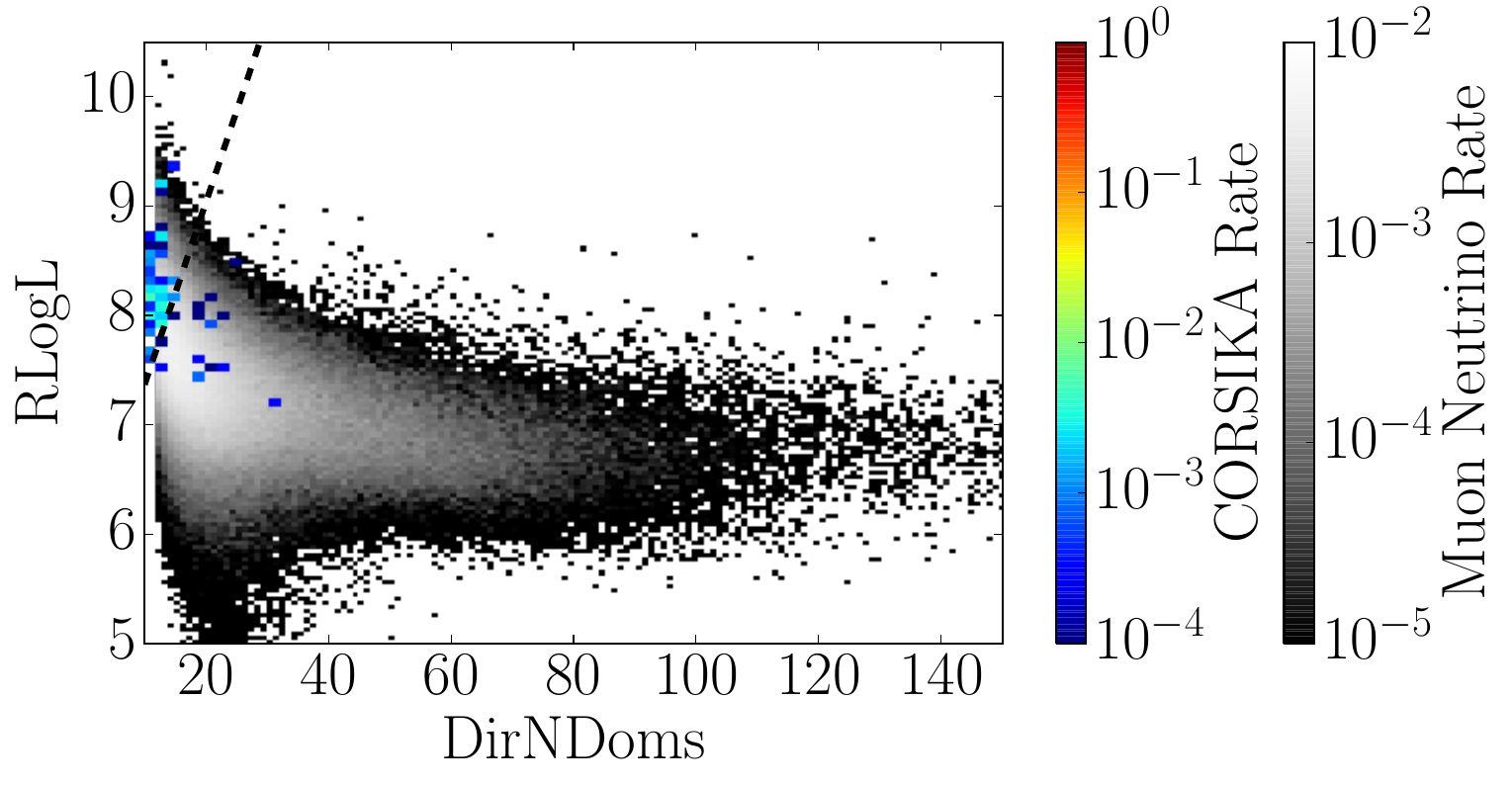}
  \end{minipage}
 \caption{
 \textbf{\textit{Two-dimensional cuts on overburden for RLogL and DirN DOMs.}}
 } These cuts are used to remove residual background contamination in the final sample.
 \label{fig::cleanup2}
\end{figure}

\begin{figure*}[t]
    \centering
    \begin{minipage}{0.45\textwidth}
        \centering
        \includegraphics[width=0.9\linewidth]{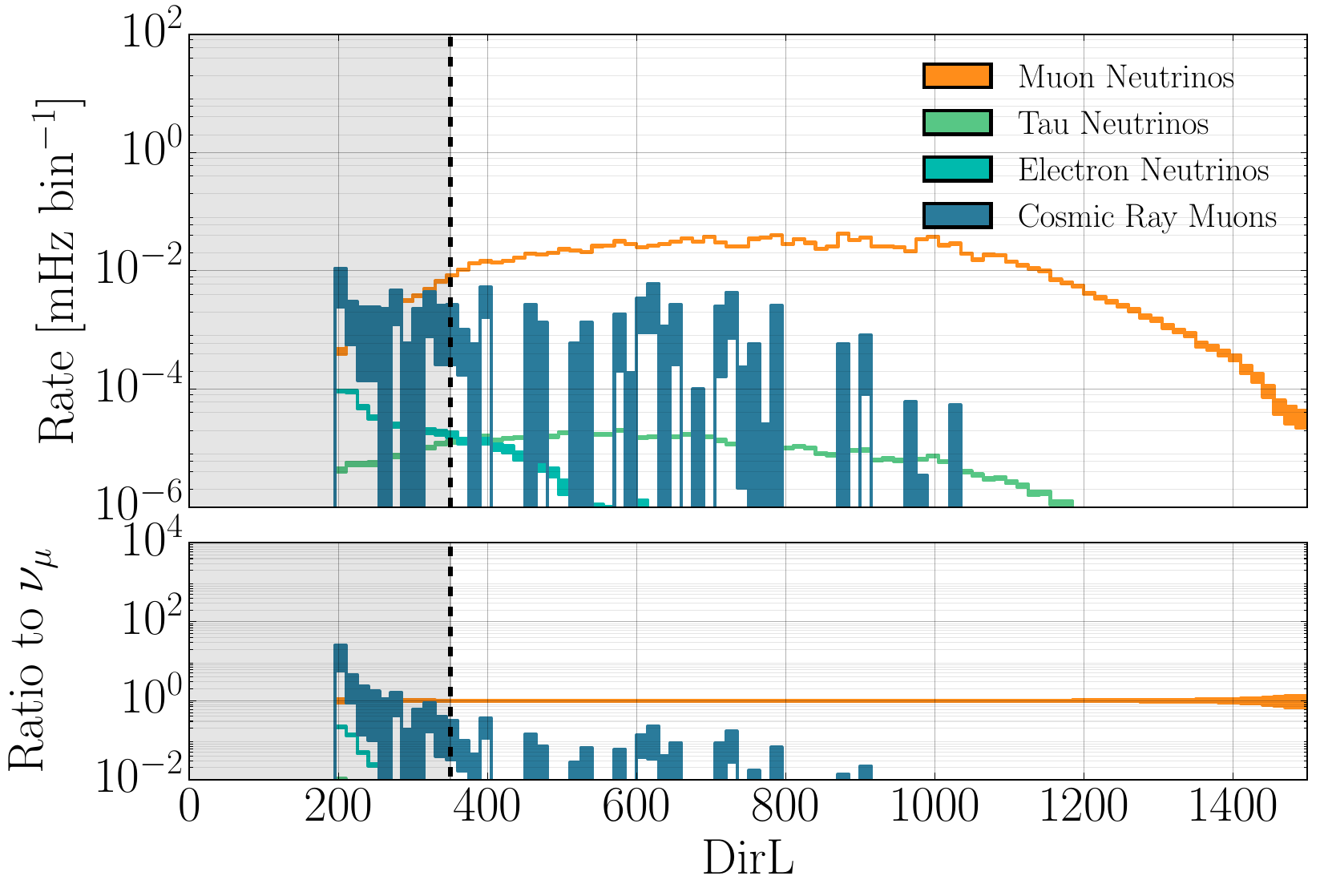}
    \end{minipage}%
    \begin{minipage}{0.45\textwidth}
        \centering
        \includegraphics[width=0.9\linewidth]{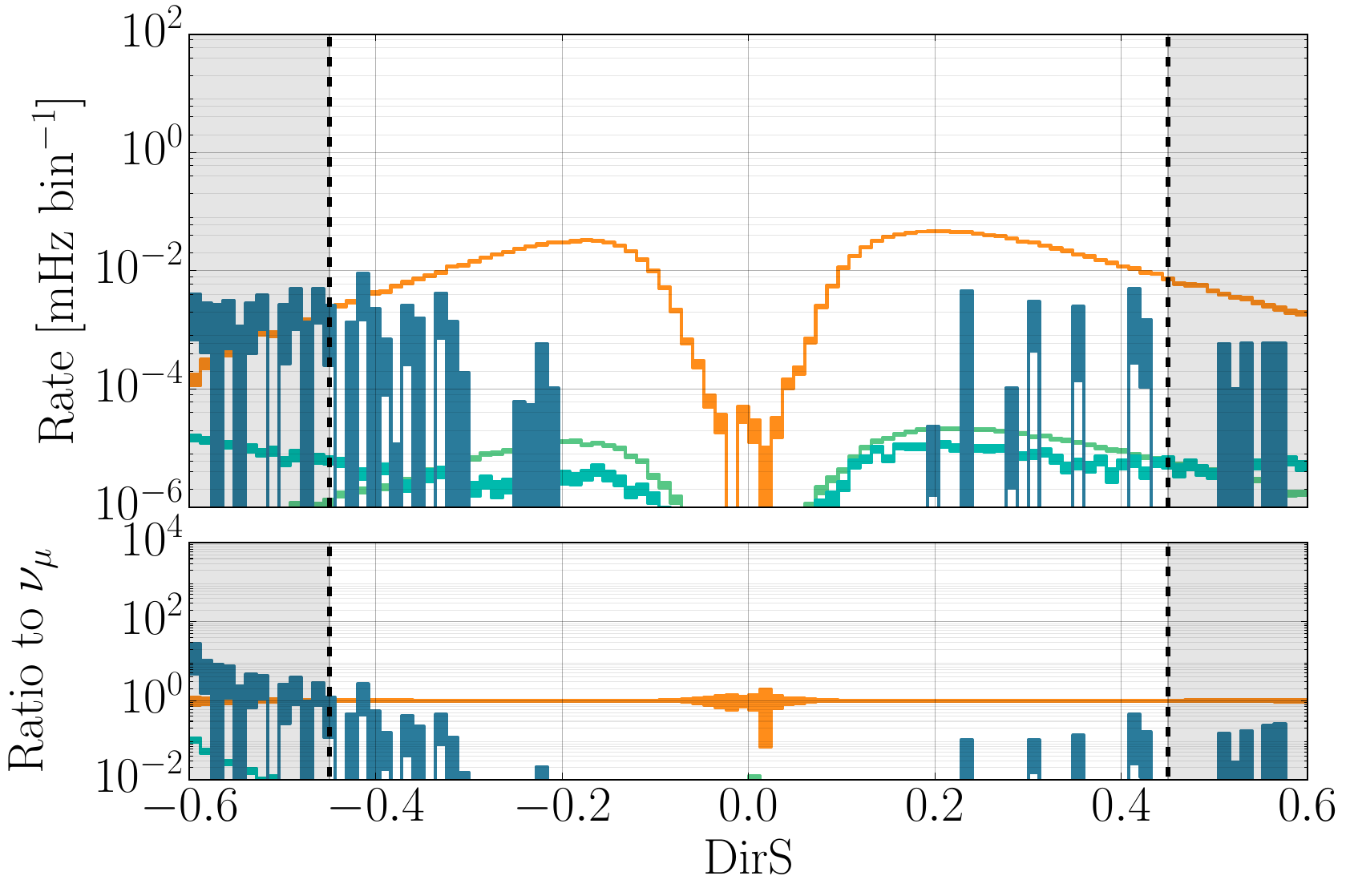}
    \end{minipage}
    \caption{\textbf{\textit{Cuts used in clean-up step.}}
    Top two plots show the 1D cuts used in the clean-up step.
    In these plots the color coding is the same as in \reffig{fig::cr_reduction1} where, again, the vertical-dashed line in each plot shows the location of the cut, and the shaded region is rejected.}
	\label{fig::cleanup}
\end{figure*}

\begin{enumerate}
\item ${\rm BayesLLHR}< 33$.
\item ${\rm AQWD} > \SI{90}\meter$.
\item ${\rm COGZ} > \SI{450}\meter$.
\item ${\rm COGR} > \SI{650}\meter$.
\item Remove event if ${\rm RLogL} > (3/18) \times ({\rm DirNDOMs}) + 5.7$ for all events where $\log_{10}({\rm Overburden}) < 3/1.2 \times ({\rm RLogL}-7.1) +3$.
\end{enumerate}
The resulting event count predictions after these cuts are shown in Table.~\ref{table::evtsel_expectation2}.

\subsection{IceCube data selection}\label{sec::data_selection}

IceCube data are typically broken up into 8-hour runs, which are vetted by the collaboration as having ``good'' data (details can be found in Ref.~\cite{Aartsen:2016nxy}).
For every good run, we additionally require that all 86 strings are active, as well as at least 5,000 active in-ice DOMs.
This data quality condition results in less than 0.4\% loss of livetime.
An average of $5048 \pm 4$ DOMs are active in the detector throughout all seasons used in this analysis.
We observe no significant deviation in the average event rate throughout the years.
The seasonal data rates are shown in Table.~\ref{table::Events}.

\begin{table}[ht]
\footnotesize
\begin{center}
\resizebox{\columnwidth}{!}{
\begin{tabular}{| l c c  c c |}
\hline
\textbf{IceCube Season} & Start date & Number of Events & Livetime [s] & Rate [mHz]\\ [0.5ex]
\hline
\hline
IC86.2011 & 2011/05/13 & 36,293 &  28,753,787 & 1.262 $\pm$ 0.007 \\ [0.5ex]    \hline    
IC86.2012 & 2012/05/15 & 35,728 &  27,935,093 & 1.279 $\pm$ 0.007\\ [0.5ex]    \hline 
IC86.2013 & 2013/05/02 & 37,823 & 29,864,837  & 1.266 $\pm$ 0.007\\ [0.5ex]    \hline  
IC86.2014 & 2014/05/06 & 38,926 & 30,874,233  & 1.261 $\pm$ 0.006\\ [0.5ex]    \hline  
IC86.2015 & 2015/05/18 & 39,930 &  31,325,569 & 1.275 $\pm$ 0.006\\ [0.5ex]    \hline  
IC86.2016 & 2016/05/25 & 38,765 &  30,549,512 & 1.269 $\pm$ 0.006\\ [0.5ex]    \hline 
IC86.2017 &  2017/05/25    &44,403 & 34,712,607  & 1.279 $\pm$ 0.006 \\ [0.5ex]    \hline 
IC86.2018 &  2018/06/19   &33,867 & 26,732,203  & 1.267 $\pm$ 0.007 \\ [0.5ex]    \hline 
\hline
\textbf{Total}     &    & 305,735 & 240,747,841  & 1.270 $\pm$ 0.002\\ [0.5ex]    \hline  
\end{tabular}}
\caption{\textbf{\textit{Number of events per season.}}
The total number of \numu events, livetimes, and rates from each IceCube season considered in this work.}
\label{table::Events}
\end{center}
\end{table}

The PMT gain is known to vary with time. At the beginning of every season the DOM high voltage is adjusted accordingly to maintain a gain of $10^7$.
To verify the stability of the extracted charge as a function of time, an analysis into the time variation of the single photoelectron charge distribution was performed.
The components used to describe the single photoelectron charge distribution, the sum of two exponentials and a Gaussian, are found to have no systematic variation as a function of time greater than that observed by random scrambling of the years. This is reported in Ref.~\cite{spe_paper} and in agreement with the stability checks performed in Ref.~\cite{icecube_instrumentation}.

\section{Systematic uncertainties \label{ch::systematics}}

Compared to the one-year high-energy sterile neutrino search by IceCube~\cite{aartsen2016searches}, the number of $\nu_\mu$ events used in this analysis is nearly a factor of 14 larger.
This corresponds to a significant decrease in the bin-wise statistical uncertainty in a large portion of the reconstructed energy-\cosz plane.
A considerable effort has been devoted to properly modeling and understanding the systematic uncertainties at the requisite 1\%-per-bin level.
Each uncertainty reported in this section will be described in terms of the shape it generates on the reconstructed energy-\cosz plane as it is perturbed within $1\sigma$ of its Gaussian prior.  Maps of these effects as used in the analysis are provided in the Supplementary Material.
The full list of nuisance parameters and their priors and boundaries are shown in Table.~\ref{table::Priors}.

\begin{table}[h]
\begin{center}
\resizebox{\columnwidth}{!}{
 \begin{tabular}{ l c c c }
 \hline
 \hline
\textbf{Parameter} & Central & Prior (Constraint) & Boundary  \\ [0.5ex]
\hline
\hline
\multicolumn{4}{c}{\textbf{Physics Mixing Parameters}}\\
\hline
$\Delta m^2_{41}$  & none &  flat log prior & [0.01, 100]~eV$^2$  \\
\hline
sin$^2(\theta_{24})$   & none &  flat log prior & [10$^{-2.6}$, 1.0]    \\
\hline
sin$^2(\theta_{34}$) & none & flat log prior  & [10$^{-3.1}$, 1.0]   \\
\hline
\multicolumn{4}{c}{\textbf{Detector parameters}}\\
\hline
DOM efficiency  & 0.97 & 0.97 $\pm$ 0.10  & [0.94, 1.03]  \\
\hline
Bulk Ice Gradient 0 & 0.0 &  0  $\pm$ 1.0* & NA  \\
\hline
Bulk Ice Gradient 1 & 0.0  & 0  $\pm$ 1.0* & NA \\
\hline
Forward Hole Ice (p$_2$)   & -1.0  & -1.0  $\pm$ 10.0  & [-5, 3]    \\
\hline
\multicolumn{4}{c}{\textbf{Conventional Flux parameters}}\\
\hline
Normalization ($\Phi_{\mathrm{conv.}}$)  &  1.0 & 1.0  $\pm$ 0.4 & NA  \\
\hline
Spectral shift ($\Delta\gamma_{\mathrm{conv.}}$) & 0.00 & 0.00  $\pm$ 0.03  & NA \\
\hline
Atm. Density   &  0.0 & 0.0  $\pm$ 1.0& NA \\
\hline
Barr WM       &  0.0 & 0.0  $\pm$ 0.40 & [-0.5, 0.5] \\
\hline
Barr WP       &  0.0 & 0.0  $\pm$0.40   & [-0.5, 0.5] \\
\hline
Barr YM       &  0.0 & 0.0  $\pm$ 0.30  & [-0.5, 0.5] \\
\hline
Barr YP       &  0.0 & 0.0  $\pm$0.30 & [-0.5, 0.5]  \\
\hline
Barr ZM       &  0.0 & 0.0  $\pm$ 0.12 & [-0.25, 0.5]   \\
\hline
Barr ZP       &  0.0 & 0.0  $\pm$ 0.12  & [-0.2, 0.5]  \\
\hline
\multicolumn{4}{c}{\textbf{Astrophysical Flux parameters}}\\
\hline
Normalization ($\Phi_{\mathrm{astro.}}$)    &  0.787 & 0.0  $\pm$ 0.36*  & NA \\
\hline
Spectral shift ($\Delta\gamma_{\mathrm{astro.}}$)      &  0 & 0.0  $\pm$ 0.36*  & NA \\
\hline
\multicolumn{4}{c}{\textbf{Cross sections}}\\
\hline
Cross section $\sigma_{\nu_\mu}$    &  1.00 & 1.00  $\pm$ 0.03  & [0.5, 1.5]  \\
\hline
Cross section $\sigma_{\overline{\nu}_\mu}$      &  1.000 & 1.000  $\pm$ 0.075 & [0.5, 1.5]   \\
\hline
Kaon energy loss $\sigma_{KA}$     &  0.0 & 0.0  $\pm$ 1.0  &  NA \\
\hline
\hline
\end{tabular}
}
\caption{\textbf{\textit{Summary of physics and nuisance parameters used in the analysis.}}
Each row specifies the constraint (prior) used in the frequentist (Bayesian) analysis for each physics or nuisance parameter.
All constraints (priors) used in the analysis are one dimensional Gaussian functions, except in the case of the bulk ice and astrophysics flux parameters (marked with an asterisk) where a correlated prior is employed. }
\label{table::Priors}
\end{center}
\end{table}

\subsection{DOM efficiency\label{sec::dom_eff}}

The term \textit{DOM efficiency} is used to describe the effective photon detection efficiency of the full detection unit.
In addition to DOM-specific effects like photocathode efficiency, collection efficiency, wavelength acceptance, \textit{etc.}, it also encompasses any physical property that changes the percentage of photons that deposit a measurable charge in the detector globally.
This includes properties external to the DOM such as the cable shadow, hole ice properties, and some aspects of bulk ice properties.

The secondary particles in simulated neutrino-nucleon interactions are propagated through the ice with an overabundance of photons produced along their track.
During the detector level simulation, photons are down-sampled, \textit{i.e.} a percentage of the propagated photons are randomly destroyed to achieve the desired DOM efficiency in simulation.
Events are generated at a relative DOM efficiency of 1.10, then down-sampled to the central value of 0.97, as determined by calibration measurements.
Five systematically different data sets were simulated relative to the central value at +6.3\%, +4.7\%, +2.4\%, -1.6\%, and -3.1\%, which allow us to probe DOM efficiency values between approximately 0.93 and 1.03. 

The five discrete DOM efficiency datasets are converted into a continuous parameterization using penalized splines~\cite{whitehorn2013penalized} fitted to the reconstructed energy and \cosz distributions.
This procedure is also used for various systematic data sets, namely the hole ice and Earth opacity effects, and allows us to re-weight each event to any systematic value within the uncertainty range.

An example of the shape-only (mean normalization removed), effect of perturbing the DOM efficiency by $\pm1$\% relative to the central MC set is shown in  the Supplementary Material, panels i.a and i.b.
As expected from a change in the average observed charge, the shape is manifest primarily as a shift in reconstructed energy scale, with lower DOM efficiencies pulling mean reconstructed energy to lower values.
The prior is chosen to be $\pm 10\%$, which was determined independently using minimum ionizing atmospheric muons~\cite{icecube_energy}.
In practice, while the DOM efficiency prior is wide, the constraint imposed by the observed energy scale in data implies that DOM efficiency has a tight posterior in the analysis, and is not an especially limiting source of uncertainty.

\subsection{Bulk ice}\label{sec::systematics_bulk_ice}
\begin{figure}[b!]
\centering
\includegraphics[width=\columnwidth]{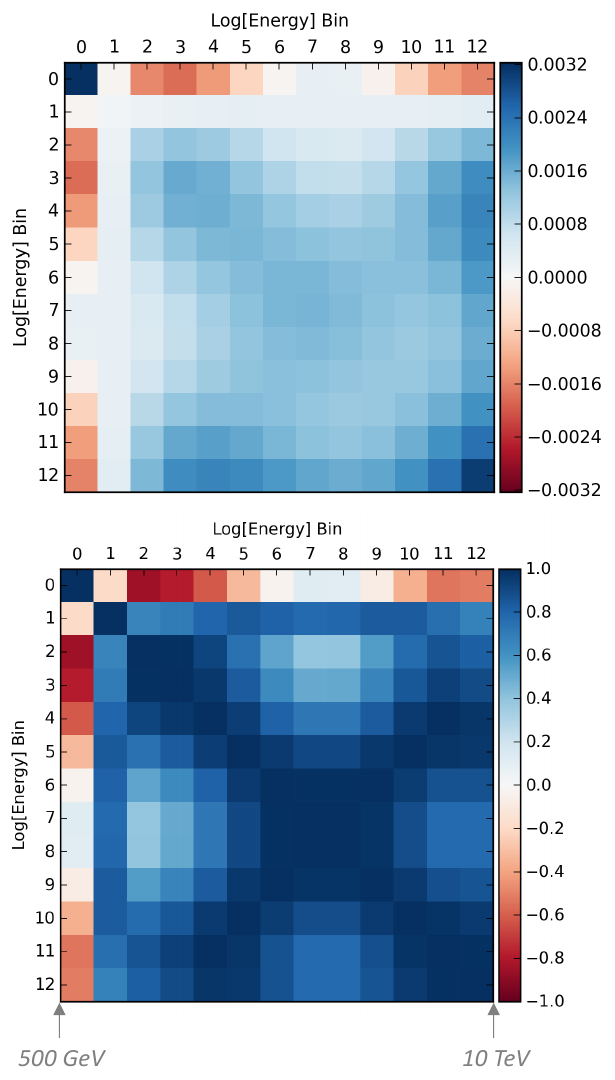}
\caption{\textbf{\textit{Energy distribution ice uncertainty covariance (top) and correlation matrix (bottom).}}
The color scale shows the covariance / correlation between energy bins.}
\label{fig:A_Cov}
\end{figure}
Bulk ice refers to the undisturbed ice between the IceCube strings, through which photons must propagate between emission and detection.
The bulk ice is highly transparent, but residual  impurities, commonly referred to as ``dust,'' introduce both scattering~\cite{Askebjer:1997ep} and absorption phenomena~\cite{Price:97}.
The dust concentration within IceCube accumulated from snowfall over the 100,000 year history of the ice~\cite{GRL:GRL13620,Icecube:2017vgw}, with a concentration that correlates with the climate history of the Earth.
To model the depth dependence of optical scattering and absorption, IceCube uses a layered ice model~\cite{ackermann2006optical,aartsen2013measurement}, wherein absorption and scattering are parameterized for every $\SI{10}\meter$ layer.
The layers are non-planar, to account for the buckling of the glacier as it has flowed, as measured with ``dust-logger'' devices~\cite{GRL:GRL20535} deployed into some of the holes before DOM deployment.
The ice also demonstrates anisotropic light propagation~\cite{AnisotropyPaper}, governed by the direction of glacial flow.
These effects are all incorporated into an ice model calibrated to LED flasher data~\cite{aartsen2013measurement}.
The ice model used in this analysis is several generations newer than the one used for the one-year sterile neutrino search, and includes anisotropy, tilt, and the present best-fit layered ice coefficients. 

Assessing the uncertainty on this model is very challenging, because it depends on a large number of parameters, all constrained by common calibration data.
A new method of treating IceCube's bulk ice uncertainties, called the ``SnowStorm'', was developed for this analysis, and has been published in Ref.~\cite{Aartsen:2019jcj}.
In a SnowStorm Monte Carlo ensemble, every event is generated with a distinct set of nuisance parameters, drawn randomly from a multivariate Gaussian distribution.
Manipulation of the ensemble by either cutting or weighting can be used to study the impact of each one of many potentially correlated uncertainties on the analysis, and construct a covariance matrix.
Full mathematical details are provided in Ref.~\cite{Aartsen:2019jcj}.  

To maintain a manageable number of nuisance parameters for the SnowStorm method, instead of treating each ice  layer coefficient as a free parameter, we select the most important ice  uncertainty contributions by working in Fourier space.
Perturbations to the ice model that distort the scattering or absorption parameters over detector-size scales are expected to impact analyses, whereas very localized effects are not, after averaging over incoming neutrino directions and energies.
Thus the uncertainty on the lowest Fourier modes of the continuous ice model encode the majority of the uncertainty on the layered ice.
Previous analyses had only used the zeroth mode, or overall absorption and scattering scale, as an uncertainty. Here, however, we find substantial impact on the analysis space from modes up to the fourth.
The allowed mode variations and their covariances are constrained using flasher calibration data to yield a nuisance parameter covariance matrix.
This matrix, along with nuisance parameter gradients derived using the SnowStorm method, are used to construct an energy-dependent covariance matrix in analysis space.
The effect of the ice uncertainty on zenith distribution is found to be far sub-leading.  

The ice covariance matrix and associated correlation matrix is shown in Fig.~\ref{fig:A_Cov}.
In order to incorporate this covariance matrix into our nuisance parameter formalism, it must be decomposed in terms of a set of nuisance parameters that can vary within prescribed priors. To avoid incorporation of the full set of nine new parameters into the fit for each phase and amplitude each mode, plus a constant, we instead define two effective gradients that vary within a correlated prior to yield the same covariance matrix.  Such a decomposition encodes the ice uncertainty budget from the first four modes in only two effective nuisance parameters. 
Variation of these parameters within a suitably correlated Gaussian prior reproduces the full covariance matrix to high precision,  encoding several distinct sources of systematic uncertainty into two effective parameters. The effects of perturbing these parameters at the 1$\sigma$ level are shown in the Supplementary Material, panels i.c and ii.a.

\subsection{Hole ice}\label{sec::systematics_hole_ice}

\begin{figure}[b!]
    \centering
        \includegraphics[width=0.99\linewidth]{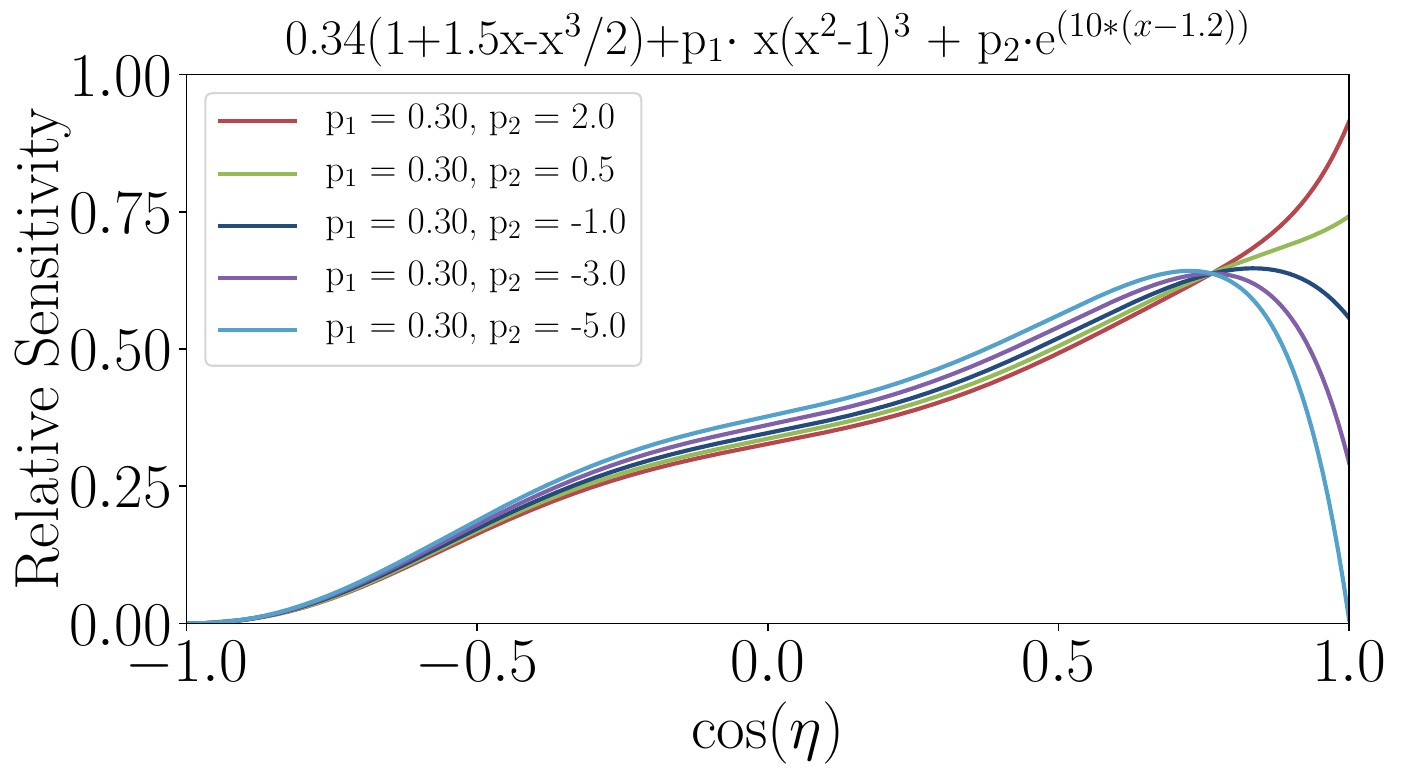}
    \caption{\textbf{\textit{DOM angular efficiency variations.}}
    Different allowed angular acceptance curves are shown as a function of the incident photon angle.}
    \label{fig::domice}
\end{figure}

Each photon detected by a DOM must also propagate through the refrozen ice in the boreholes, known as ``hole ice''~\cite{fiedlschuster2019effect}, which were drilled to deploy the strings.
Recorded images of the refreezing process~\cite{rongen2016measuring} suggest that the hole ice has a transparent component extending from the edge of the hole inwards, and a central column of bubbles or impurities, roughly 8 to $\SI{10}\cm$ in diameter.
The primary effect of hole ice is to introduce additional optical scattering near the DOM, which effectively perturbs the angular acceptance curve relative to that measured in laboratory conditions. 

An empirical parameterization has been derived from microscopic simulations of light interacting with the hole ice, depending on two free parameters $p_1$ and $p_2$:
\begin{equation}
A(\eta) = 0.34(1+1.5\eta^3/2)+p_1\eta(\eta^2-1)^3+p_2 e^{(10(\eta-1.2))},
\end{equation}
where $\eta$ is the angle of the incoming photon, as indicated in the left side of Fig.~\ref{fig::domice}.
The $p_2$ parameter primarily varies the upward-going photon acceptance ($\cos(\eta)=1$), the subset most strongly scattered by the bubble column.
This parameter is referred to as the ``forward hole ice'' and will be included as a systematic uncertainty in this analysis.
The $p_1$ parameter, on the other hand, is found to have a minimal impact and fixed at its default value.

Five identical sets of Monte Carlo were generated with the only difference being the description of the angular acceptance forward hole ice parameter.
These sets were produced with hole ice parameters $p_2 = -5, -3, -1, 1, 3$ and $p_1=0.3$ and are shown in Fig.~\ref{fig::domice}.
Each of these curves is commonly normalized to maintain a constant overall efficiency factor.
Penalized splines were generated to re-weight each event to any continuous value for $p_2$ between -5 and 3.
The central MC set was chosen to be $p_2 = -1.0$ and $p_1 = 0.3$ and we assign a wide prior to the forward hole ice parameter, namely $p_2 = \pm 10$.
The shape generated by perturbing the forward hole ice to -3 and +1 relative to the central set is shown in the Supplementary Material, panels ii.b and ii.c.

\subsection{Atmospheric neutrino flux\label{sec::atm_uncer}}

Unlike the one year high-energy sterile neutrino search, we have transitioned from using discrete variants of the cosmic ray and hadronic interaction models to continuously parameterized fluxes controlled by nuisance parameters.
The envelope of parameterized models is consistent with the spread of discrete models considered in the earlier analysis, with the benefit of enabling effective interpolation between them, guided by physics-motivated tunable parameters.

The uncertainty in the conventional neutrino spectrum is factorized into the uncertainty in the meson production in the atmosphere, the overall normalization, the cosmic ray spectral index, the atmospheric density, and the rate of meson energy loss in air.
The following subsections give an overview of how each of these are implemented. 

\begin{figure*}[t]
    \centering
    \includegraphics[width=0.85\textwidth]{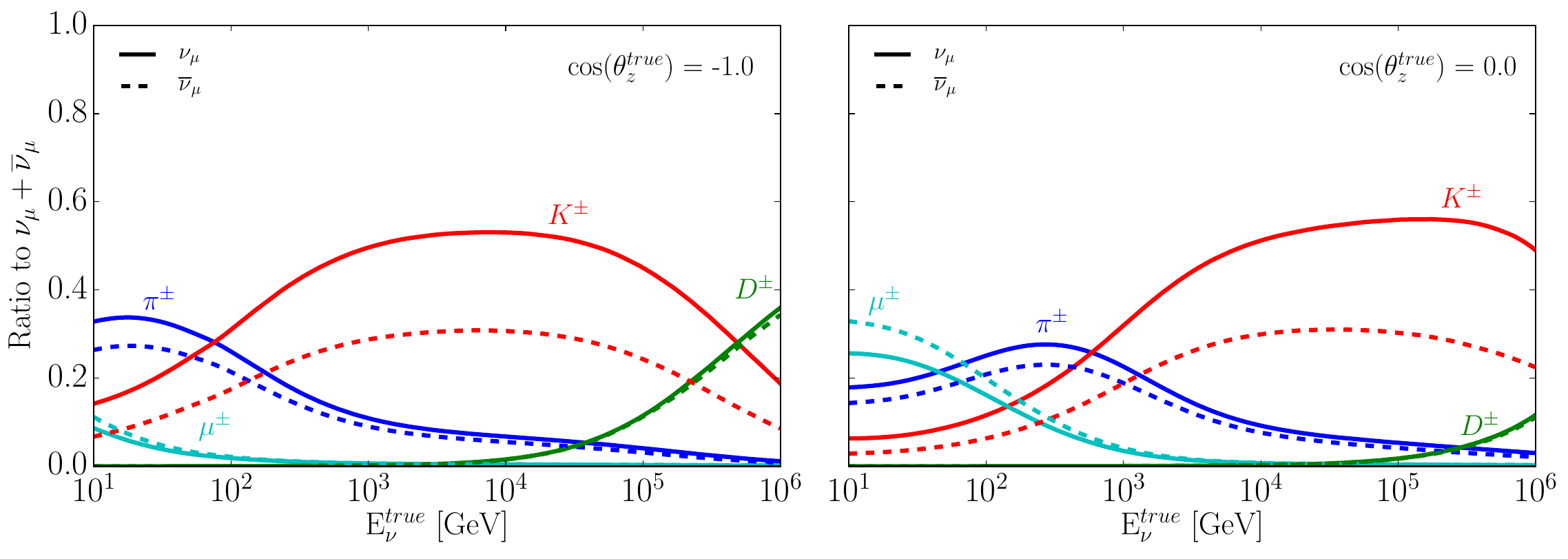}
    \caption{
    \textbf{\textit{Contribution of different parent particles to the atmospheric neutrino flux.}}
    The neutrino flux from a given parent particles -- pion in blue, kaon in red, muon in teal, and D-meson in green -- is compared to the total $\nu_\mu$ and antineutrino flux at IceCube. The right panel is for horizontal neutrinos while the left column is for vertically upward-going neutrinos.
    }
    \label{fig::mceqflux}
\end{figure*}

Fig.~\ref{fig::mceqflux} shows how different components are manifest in the total $\nu_\mu$ flux (HillasGaisser2012) as a function of energy and zenith angle. The plot on the left shows the upward-going flux and the plot on the right shows the flux from the horizon. 
In this figure, the color combination represents the neutrino flux from a given progenitor labeled at the top of the figure.

\subsubsection{Hadronic uncertainties using the Barr parameterization}\label{sec::barr}

The Barr parameterization~\cite{barr2006uncertainties} describes the uncertainty associated with the production of pions and kaons in hadronic interactions based on accelerator data.
The uncertainties are estimated as a function of the incident particle energy, $E_i$ and $x_{\mathrm{lab}}= E_i/E_s$, where $E_s$ is the energy of the secondary total energy.  They are independently calculated for positive and negative mesons.
In the energy range of interest to this analysis,  $\SI{100}\GeV$ to $\SI{10}\TeV$, the neutrino flux is dominated by neutrinos produced from kaon decay.
Using the notation of Ref.~\cite{barr2006uncertainties}, the Barr parameters responsible for describing the uncertainties associated with pion production (A$^\pm$ to I$^\pm$) are found to be negligible at analysis level. We thus restrict our consideration to only those that impact the kaon production above $\SI{30}\GeV$: W$^\pm$, Y$^\pm$, and Z$^\pm$.
The relevant phase space for each parameter in terms of $x_{\mathrm{lab}}$ and primary energy are shown in the second and third column of Table.~\ref{table::Barrparams}.  For full details of the Barr scheme, we refer the reader to Ref.~\cite{barr2006uncertainties}.

\begin{table}[h]
\begin{center}
\resizebox{\columnwidth}{!}{

 \begin{tabular}{|c c c c c  |}
 \hline
\textbf{Parameter} & x$_{\mathrm{lab}}$ & Energy [GeV] & Meson & Uncertainty  \\ [0.5ex]
\hline
\hline
W$^\pm$ & 0.0-0.1 & 30 - 1$\times$10$^{11}$ &  K$^{\pm}$  & 40\%  \\
\hline
Y$^\pm$  & 0.1-1.0 & 30 - 1$\times$10$^{11}$  &  K$^{\pm}$ & 30\% \\
\hline
Z$^\pm$  & 0.1-1.0 & 500 - 1$\times$10$^{11}$ &   K$^{\pm}$  & 12.2\% log$_{10}$(E/500GeV) \\
\hline
\end{tabular}
}
\caption{\textbf{\textit{Summary of Barr parameters definitions and allowed regions.}}
The uncertainties associated with the three relevant Barr parameters, along with the description of the phase space in which they are valid.}
\label{table::Barrparams}
\end{center}
\end{table}

The flux gradients are constructed by computing the flux difference close to the nominal values.
Then these gradients are multiplied by the variation of the parameter corresponding parameter to obtain the effective shapes in the reconstructed energy - \cosz plane.
We use the nomenclature ``P'' and ``M'' on the Barr parameters to denote whether they are used for the positively or negatively charged mesons.
The effects of varying the relevant Barr gradients are shown in the Supplementary Material, rows iii and iv.

\subsubsection{Conventional neutrino flux normalization}

At large values of $\Delta m^2_{41}$ there are regions in the physics parameter space with small signal shape and large normalization shifts, caused by fast energy dependent oscillation which is unresolved within detector resolution.
To control against spurious fits to sterile neutrino hypotheses in these regions it is vital to include an appropriate uncertainty on the conventional neutrino flux normalization. 
This is was primarily derived from the theoretical uncertainty reported in  Ref.~\cite{fedynitch2012influence} and an extrapolation from the uncertainties quoted in the HKKM calculation given in Ref.~\cite{honda2007calculation}. 

The theoretical uncertainty reported in Ref.~\cite{fedynitch2012influence} accounts for both the cosmic ray and hadronic interaction model in the energy range of interest for this analysis.
Up to approximately $\SI{1}\TeV$, the hadronic interaction model represents the majority of the uncertainty since the cosmic-ray models in this regime are relatively well established.
Above this energy, the uncertainty arising from features around the cosmic-ray knee dominates the total uncertainty.
The sub-TeV uncertainty is in agreement with the calculated total uncertainty found in the HKKM calculation~\cite{honda2007calculation}.
At $\SI{1}\TeV$, the uncertainty is reported as 25\% and consists of the uncertainties associated with the pion and kaon production, hadronic interaction cross section, and atmospheric density profile.

Based on the findings described above, we include a 40\% uncertainty on the conventional atmospheric neutrino normalization.
The shape and normalization exhibited when perturbing the conventional atmospheric normalization by $\pm 1 \sigma$ is shown in the Supplementary Material, panels v.a and v.b.
As well as changing the normalization, this uncertainty introduces a small relative shape effect from the changing ratio of contributions from conventional, astrophysical and prompt neutrino fluxes.

\subsubsection{Cosmic-ray spectral slope}

In the energy range of interest for this analysis, the cosmic ray spectrum responsible for producing the atmospheric neutrinos follows approximately an $E^{-2.65}$ energy dependence.
We attribute a spectral shift, $\Delta \gamma$, to the energy dependence as:
\begin{equation}
\phi (E; \Delta \gamma) = \phi (E) \Big(\frac{E}{E_0}\Big)^{-\Delta \gamma},
\end{equation}
where $E_0$ has been chosen to be $\SI{2.2}\TeV$ in order to approximately preserve the total normalization.
The measured cosmic-ray spectral index from the recent measurements is shown in Table.~\ref{table::cosmic_ray_slope}.
Based on these measurements, we assign a prior width on the cosmic ray spectral shift of $\Delta \gamma = 0.03$.
The shape of the cosmic-ray spectral shift at $\pm 1 \sigma$ is shown Supplementary Material panel v.c and vi.d.

\begin{table}[h]
\footnotesize
\begin{center}
 \begin{tabular}{|c c c c  |}
 \hline
\textbf{Experiment} & Year & Energy Range &  C.R. Slope   \\ [0.5ex]
\hline
\hline
CREAM-III~\cite{yoon2017proton} & 2017 & 1TeV - 200TeV & -2.65 $\pm$ 0.03  \\
\hline
HAWC~\cite{alfaro2017all}  & 2017 & 10TeV - 500TeV & -2.63 $\pm$ 0.01  \\
\hline
Argo-YBJ ~\cite{iacovacci2013cosmic} & 2016 & 3TeV - 300TeV & -2.64 $\pm$ 0.01 \\
\hline
PAMELA~\cite{karelin2011proton} & 2011 & 50TeV - 15TeV & -2.70 $\pm$ 0.05  \\
\hline
\end{tabular}
\caption{\textbf{\textit{The measured cosmic-ray spectral slope and uncertainty for several experiments.}}}
\label{table::cosmic_ray_slope}
\end{center}
\end{table}

\subsubsection{Atmospheric density}\label{sec::atm_denstiy_unce}

The pions and kaons produced in the hadronic showers induced by cosmic rays can either interact or decay, with the latter producing the conventional neutrino flux.
The competition between the two processes depends on the local atmospheric density.
IceCube has previously shown that the atmospheric conditions presented to the cosmic-ray flux can affect the atmospheric neutrino spectrum~\cite{gaisser2013seasonal}. 

We ascribe an uncertainty to the atmospheric density by perturbing the Earth's atmospheric temperature within a prior range given by the NASA Atmospheric InfraRed Sounder ({\tt AIRS}) satellite~\cite{AIRS} temperature data.
The satellite provides open source atmospheric data for weather forecasting and climate science and reports the temperature profile as a function of atmospheric depth and location.
Using monthly averaged temperature data arranged on a 180~$\times$~360 grid (each element representing a 1$^\circ$~$\times$~1$^\circ$ area on the surface of the Earth), we calculate the density at 24 discrete altitudes assuming the ideal gas law, from which we linearly interpolate to describe the atmospheric density profile.
A random z-score is chosen and all data points are shifted according to reported systematic error on {\tt AIRS} measurement.  This protocol is used to ascribe uncertainty on the neutrino flux due to atmospheric density  uncertainties.  The 1976 United States Standard\cite{atmosphere1976national} atmosphere model was used as a cross-check, and falls within this envelope.
The resulting atmospheric profile is injected into {\tt MCEq} to generate a neutrino flux.
This is performed independently for a variety of cosmic ray and hadronic interaction models:

\begin{enumerate}
\item The hadronic atmospheric shower model.
\begin{itemize}
\item {\tt QGSJET-II-04}~\cite{lattes1980hadronic}
\item {\tt SIBYLL 2.3 RC1}~\cite{fletcher1994s}
\end{itemize}
\end{enumerate}
\begin{enumerate}
\item The cosmic-ray flux model.
\begin{itemize}
\item Zatsepin-Sokolskaya/PAMELA ~\cite{zatsepin2006three,adriani2011pamela}
\item Hillas-Gaisser/Gaisser-Honda~\cite{hillas2006cosmic,fletcher1994s,barr1989flux}
\item Poly-gonato~\cite{hoerandel2003knee,ter2000eas}
\end{itemize}
\end{enumerate}

\begin{figure}[t]
    \begin{center}
    \includegraphics[width=.99\columnwidth]{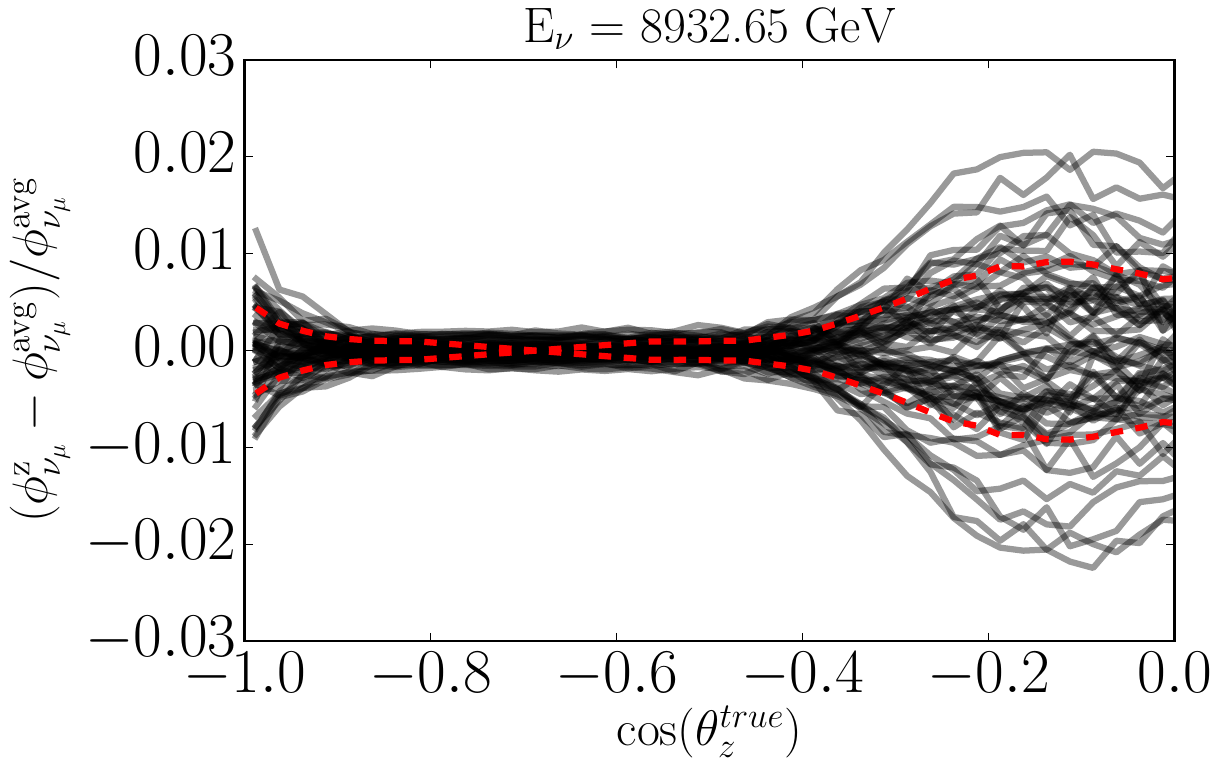}
    \caption{\textbf{\textit{Effects of atmosphere temperature variations in the atmospheric neutrino angular distribution.}}
    The change in neutrino flux relative to the average flux at $\SI{8.932}\TeV$ given temperature variations from the {\tt AIRS} satellite data (black).
    The standard deviation of the distribution of temperature fluctuations is shown as a dashed red line.}
    \label{fig::AtmFluxTempEffect}
    \end{center}
\end{figure}

For a given model and neutrino energy, we average over all months and longitudinal variations to determine the change in the zenith distribution associated with the temperature profile perturbation.
Fig.~\ref{fig::AtmFluxTempEffect} shows an example for true $\nu_\mu$ energy of $\SI{8.9}\TeV$.
The standard deviation, shown as the dotted red line is computed for each zenith angle and is assigned as the atmospheric density uncertainty.
Note that we force the lines to cross near $\cos(\theta_z) = -0.7$ in order to account for the $180^\circ$ temperature offset between the northern and southern hemispheres.
The shape generated when perturbing the atmospheric density to $\pm 1\sigma$ is shown in the Supplementary Material panels vi.b and vi.c.
It appears primarily as a zenith dependent effect.  In total, 4450 different combinations of temperature shifts (z-score perturbations), hadronic interaction models, cosmic ray models, monthly variations, and sampling longitudes are used to assess the spread attributed to the temperature uncertainty.

\subsubsection{Kaon-nuclei total cross section \label{sec::kaonenergyloss}}
We must account for the uncertainty in the charged-meson energy losses during the air shower development.
Of the mesons responsible for the $\nu_\mu$ flux, we are particularly interested in the uncertainty associated with the kaon re-interaction with oxygen (O) and nitrogen (N) nuclei within the atmosphere.
Uncertainties on the KO(N) total interaction cross section in principle influence the energy spectrum of emitted neutrinos, and have been investigated and parameterized.

The total cross section for K$^{\pm}$-nucleon has not been measured above $\SI{310}\GeV$~\cite{pdg_hadronic}, the lower end of our energy spectrum.
From proton-proton ($pp$) cross section measurements, one can theoretically derive the kaon-nucleus cross section through a Glauber~\cite{glauber1955cross,glauber1970high} and Gribov-Regge~\cite{gribov1968reggeon} multiple scattering formalism.
This approach has been experimentally verified across a wide range of energies and projectile-target nuclear composition: $\sqrt{s} = \SI{5.02}\TeV$ for proton-lead ($p$Pb) collisions~\cite{khachatryan2016measurement}, $\sqrt{s} = \SI{2.76}\TeV$ for PbPb collisions~\cite{toia2011bulk}, and $\sqrt{s} = \SI{57}\TeV$ for pAir~\cite{abreu2012measurement}.
However, verification that this approach also holds for $p$O (and thus KO(N)) interactions has yet to be realized and is currently the subject of a planned LHC run in 2021-2023~\cite{dembinski2019future}.

At high-energies, above $\sqrt{s} \gtrsim \SI{50}\GeV$, the total hadron-hadron cross section as a function of center of mass energy, $\sqrt{s}$, is given by

\begin{equation}
\sigma_{\mathrm{tot}} \approx Z_{ab} + B_{ab} \log^2\left(\frac{s}{s_0^{ab}}\right) ,
\end{equation}

where $B_{ab}$ describes the shape and is universal for all hadron-hadron interactions, namely $B_{pp} = B_{\pi p} = B_{Kp} = B_{pn} \equiv B$) at high energies; $Z_{ab}$ is a normalization factor dependent on the projectile; and $s_0^{ab}$ is a scale factor for the collision.
High energy $\pi p$ (up to $\sqrt{s}= \SI{600}\GeV$) and $pp$ (up to $\sqrt{s} =\SI{50}\TeV$) data exist and is available to constrain the universality constant $B$, as well as the scaling of $Z_{ab}$ between projectiles.
Ref.~\cite{halzen2012total} finds $B_{Kp} = 0.293 \pm 0.026_{\mathrm{sys}} \pm 0.04 _{\mathrm{stat}}~\si\mb$ and $Z_{Kp} =17.76\pm 0.43~\si\mb$.
At energies above $\sqrt{s}=\SI{40}\GeV$, the total uncertainty becomes dominated by the uncertainty in the $B$ parameter.
By perturbing the total cross section within the uncertainties of $B$ and $Z_{ab}$, we determine that the uncertainty over the range of interest for this analysis ($\sqrt{s} \approx \SI{20}\GeV$ to $\SI{500}\GeV$) is at the few-percent level with a modest dependence on energy.

Recent measurements indicate that the high-energy $pp$ total cross section uncertainty is known to $\sim 3.7\%$~\cite{allbrooke2016measurement} and $p$Pb to within $\sim 3.4\%$~\cite{khachatryan2016measurement}, in agreement with the Glauber and Gribov-Regge predictions.
We include a conservative estimate on the total kaon-nuclei total cross section of $\pm 7.5\%$. The shape generated when perturbing the  kaon-nuclei total cross section terms by $\pm 1\sigma$ is provided in the Supplementary Material panels vii.a and vii.b.

\subsection{Astrophysical neutrino flux}\label{sec::astrophyisical}

The astrophysical neutrino flux is modeled as having an unbroken ``single power law'' energy spectrum, equal $\nu_\mu$ to $\bar{\nu}_\mu$ contributions, and isotropic angular distribution.
The initial energy distribution of the astrophysical neutrino flux should not be affected by the presence of a sterile neutrino, but the normalization of each flavor component is affected; see Ref.~\cite{Brdar:2016thq,Arguelles:2019tum} for a detailed discussion.
Thus, in this analysis, the energy spectrum of astrophysical neutrinos is defined by the added $\nu_\mu$ and antineutrino components normalizations, $\Phi_{\rm astro}$, at $\SI{100}\TeV$ and the change in the astrophysical spectral index, $\Delta\gamma_{\mathrm{astro}}$, relative to a central value of $\gamma_{\rm astro}=-2.5$, namely

\begin{figure}[b!]
    \centering
    \includegraphics[width=\columnwidth]{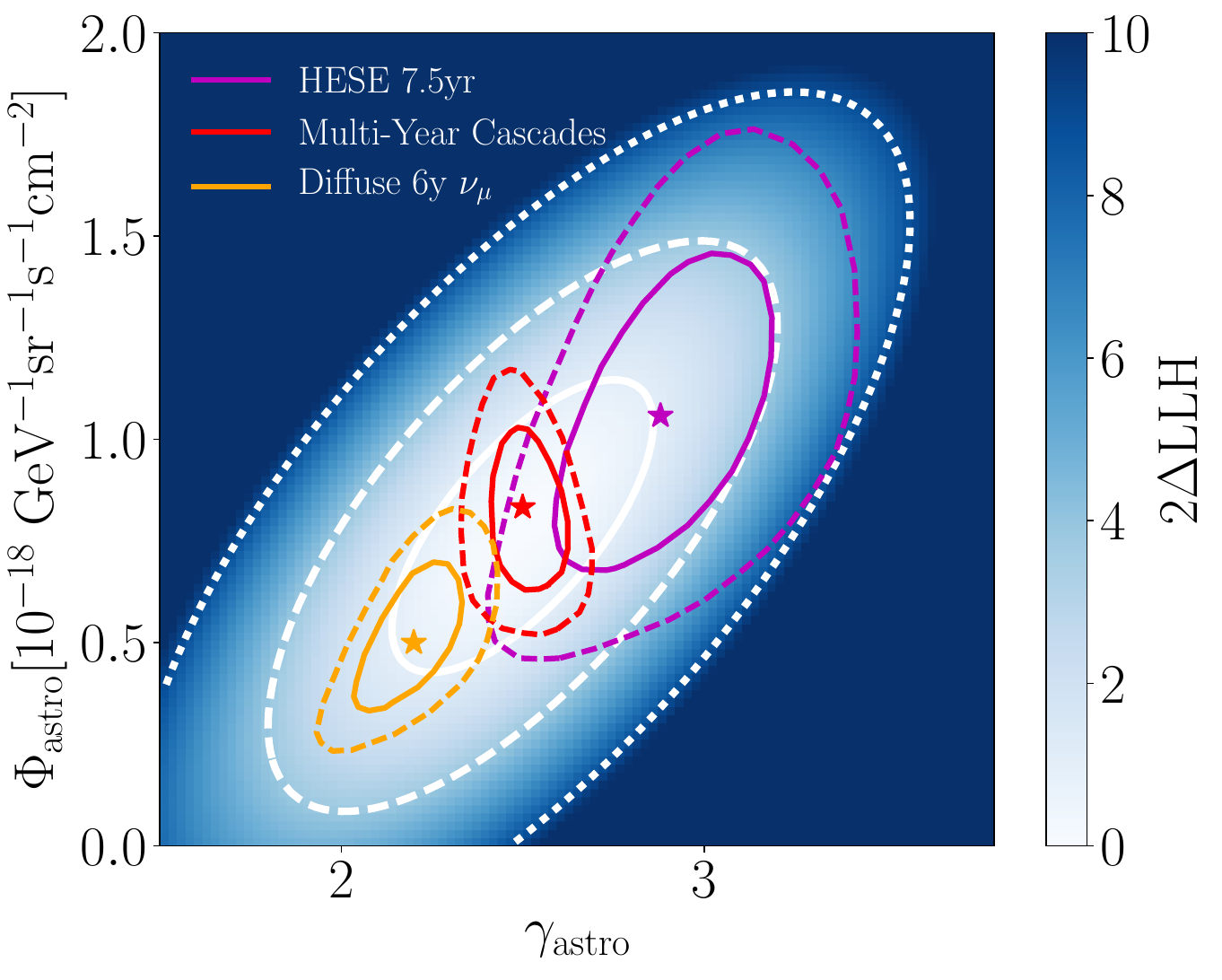}
    \caption{\textbf{\textit{Results from three measurements of the astrophysical neutrino flux performed by IceCube~\cite{schneider2019characterization}.}}
    The vertical axis shows the overall six-neutrino normalization at $\SI{100}\TeV$ assuming an unbroken power-law and democratic neutrino flavor composition at Earth.
    The horizontal axis shows single-power law spectral index.
    The stars correspond to the location of the best-fit point of each measurement, while the solid (dashed) lines correspond to the 68.3\% (95.4\%) confidence regions.
    The color scale shows the shape of the correlated two-dimensional Gaussian constraint (prior) at the 68.3\% (white solid), 95.4\% (white dashed), and 99.7\% (white dotted) levels used in the frequentist (Bayesian) analysis.}
  \label{fig::astro_prior}
\end{figure}

\begin{equation}
\frac{dN_\nu}{dE} = \Phi_{\mathrm{astro}} \Big( \frac{E_\nu}{\SI{100}\TeV}\Big)^{-2.5 +\Delta\gamma_{astro}}.
\end{equation}
The central astrophysical neutrino flux has a normalization at $\SI{100}\TeV$ of
\begin{equation}
\Phi_{\mathrm{astro}} = 0.787 \times 10^{-18} {\rm GeV}^{-1}{\rm sr}^{-1}{\rm s}^{-1}{\rm cm}^{-2},
\end{equation}
and 
\begin{equation}
\Delta\gamma_{\mathrm{astro}}=0.
\end{equation}
Both parameters are included as nuisance parameters in this analysis constrained by a correlated uncertainty constructed to span IceCube's various astrophysical neutrino measurements, shown in Fig.~\ref{fig::astro_prior}.
This figure also shows three previous single power-law fits to the astrophysical neutrino flux performed by IceCube~\cite{schneider2019characterization,aartsen2016observation,chianese2017interpreting,stachurska2019icecube}.

As with the atmospheric neutrino flux, the astrophysical neutrino flux is propagated through the Earth using \texttt{nuSQuIDS} accounting for the in-Earth sterile neutrino oscillation physics, as well as high-energy neutrino attenuation within the Earth. The effects of varying the astrophysical normalization and index are shown in Supplementary Material panels vii.c and viii.a.
\subsection{Neutrino-nucleon interaction}\label{sec::crosssectionSys}

In our energy range, the interaction between the neutrino and matter is dominated by neutrino nucleon deep inelastic scattering (DIS).
The neutrino-nucleon cross section enters the analysis in two different parts of the simulation: during the neutrino propagation through the Earth and at the interaction in the proximity of the detector.
The latter was previously investigated thoroughly in Ref.~\cite{jones2015sterile, delgado2015new} and found to have a minimal impact on the final event distribution.
The effect of the propagation through the Earth required further investigation for this analysis and has now been included as a nuisance parameter. 

The neutrino-nucleon DIS cross section increases with neutrino energy, which makes the Earth opaque to high-energy neutrinos.
We use the cross sections described in Ref.~\cite{cooper2011high} for both the neutrino-nucleon interaction during propagation and the interaction near the IceCube detector.
Uncertainties are provided for both NC and CC interaction channels from $\SI{50}\GeV$ to $5\times 10^{20}\si\GeV$.
From approximately $\SI{10}\TeV$ upwards, the neutrino and antineutrino $\nu_\mu$ charged-current cross section are predicted to within 2\% and 5\%, respectively.
Below this energy the Earth opacity is negligible.
We include separate systematic uncertainties for the neutrinos and antineutrinos, with prior width 3.0\% for neutrinos and 7.5\% for antineutrinos in order to account for additional potential corrections from nuclear parton distribution function~\cite{Bertone:2018dse}.
The uncertainties are implemented via penalized splines constructed over 30 support points in the cross section scaling parameter, ranging from 50\% to 150\% of the nominal value.
The shape in reconstructed energy and \cosz when perturbing the cross sections for neutrinos or antineutrinos by +10\% is shown in the Supplementary Material panel viii.b.
As expected the shape changes are primarily localized at high-energy and Earth crossing trajectories.
Due to limited statistics in this region the impact of this systematic is small compared to the flux uncertainty, see Fig.~\ref{fig::NM1_sensitiviity}.

\subsection{Effect on expected sensitivity}

\begin{figure}[t]  
    \begin{minipage}{\columnwidth}
        \includegraphics[width=0.9\textwidth]{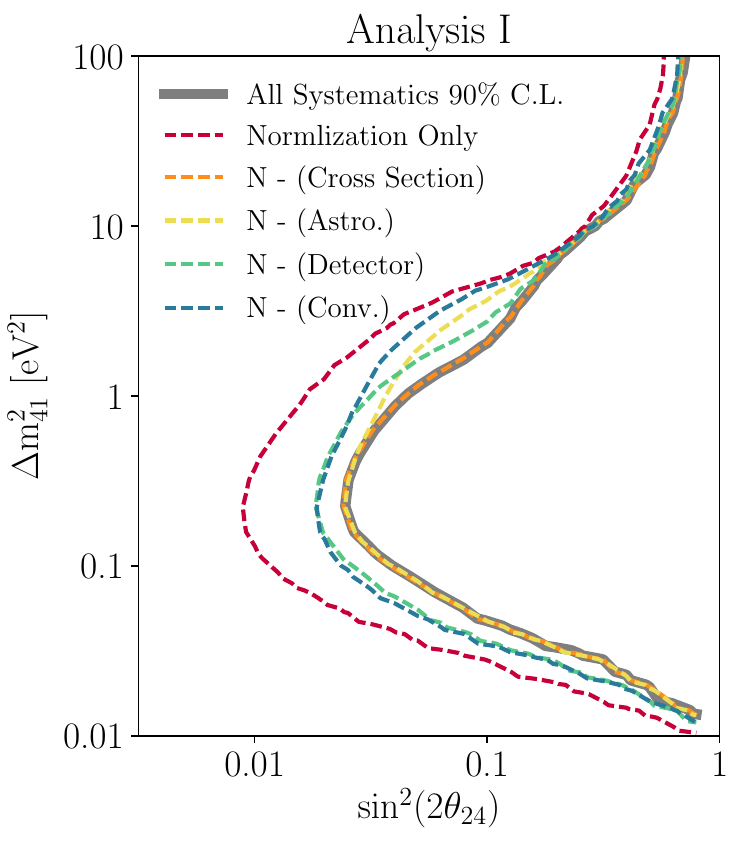}
        \includegraphics[width=0.9\textwidth]{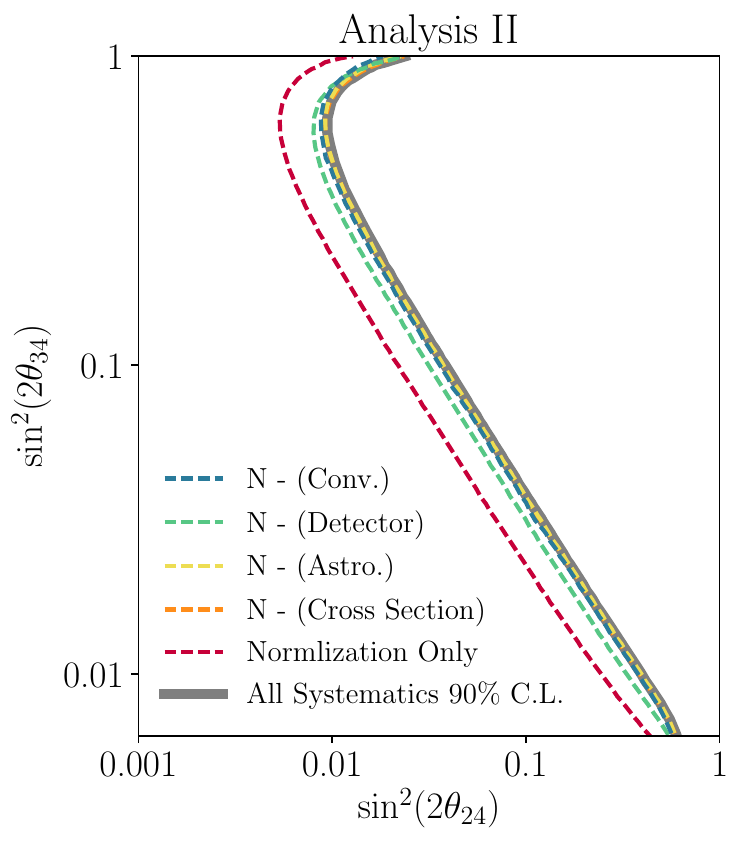}
    \end{minipage}
    \caption{\textbf{\textit{Effects of different systematic groups on the sensitivity.}}
    The analyses sensitivity at 90\% C.L., estimated using an Asimov set, are shown as a solid gray line.
    The color lines show the estimated sensitivity when a given systematic uncertainty category is removed.
    }
  \label{fig::NM1_sensitiviity}
\end{figure}

In the fit all nuisance parameters are allowed to vary within their constraints at each physics hypothesis point.
The freedom allowed by these parameters introduces uncertainty that weakens the sensitivity of the analysis to sterile neutrinos.
Fig.~\ref{fig::NM1_sensitiviity} quantifies the impact of each source of uncertainty on the expected Asimov sensitivity of each analysis.
For each curve, one class of uncertainties (normalization, cross sections, detector response, astrophysical, or conventional flux) is held fixed at its central values while leaving the others to vary within their constraints.
The largest impact in both analyses comes from the normalization freedom, followed by conventional flux, and detector response.
The effect of cross section and astrophysical flux uncertainties is comparably very small.  

\subsection{Posteriors and pulls}

\begin{figure*}[t]  
    \centering
    \subfloat{\includegraphics[width=\columnwidth]{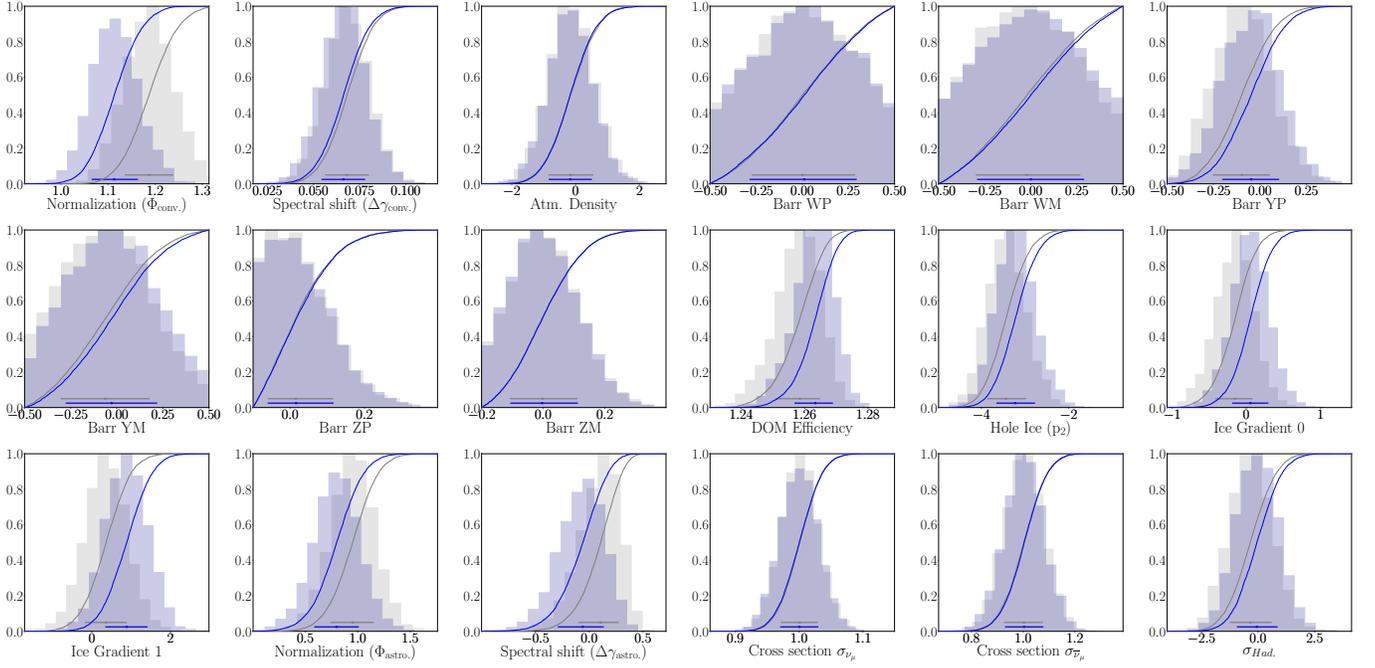}}
\caption{\textbf{\textit{Nuisance parameters posterior distributions at best-fit sterile parameter point.}}
    Each subplot shows the posterior distribution for Analysis I (grey) and Analysis II (blue) as histograms.
    In each of them we also show the cumulative distribution as solid lines, as well as the standard deviation as a horizontal error bar.}
    \label{fig::posterior}
\end{figure*}

The 18 nuisance parameters relating to the conventional neutrino flux, astrophysical neutrino flux, cross sections, and detector uncertainties are fit to data at each point in the parameter space.
Table~\ref{table::final_fits} shows the minimized values at the best-fit sterile neutrino parameter points for both analyses.
Each nuisance parameter includes a Gaussian constraint and central value defined in Table.~\ref{table::Priors}.
Figure~\ref{fig::posterior} shows the posterior distribution of each nuisance parameter for our Bayesian analyses at the best-fit points of Analysis I and Analysis II as grey and blue histograms respectively. The posteriors in both analyses are rather similar, reflecting the lack of strong dependence of the nuisance parameter allowed regions on the best fit point.

\begin{table}[b!]
\resizebox{\textwidth}{!}{%
\begin{tabular}{| l c c|}
\hline
\textbf{Parameter} & Analysis I & Analysis II \\
\hline
\multicolumn{3}{c}{\textbf{Physics Mixing Parameters}}\\
\hline
$\Delta m^2_{41}$ & $\SI{4.47}\eV^2$ & $>\SI{10}\eV^2$\\
\hline
$\sin^2(2 \theta_{24})$ & 0.10 & 0.006\\
\hline
$\sin^2(2 \theta_{34})$ & 0.0 & 0.40\\
\hline
\multicolumn{3}{c}{\textbf{Detector parameters}}\\
\hline
DOM Efficiency &  0.961 $\pm$ 0.005 &  0.965 $\pm$ 0.005\\
\hline
Ice Gradient 0 &  -0.15 $\pm$ 0.25 & 0.05 $\pm$ 0.24\\
\hline
Ice Gradient 1 &  0.36 $\pm$ 0.53 & 0.89 $\pm$ 0.54\\
\hline
Hole Ice ($p_2$) &  -3.44 $\pm$ 0.44 &  -3.23 $\pm$ 0.44\\
\hline
\multicolumn{3}{c}{\textbf{Conventional Flux parameters}}\\
\hline
Normalization ($\Phi_{\mathrm{conv.}}$)   &  1.19 $\pm$ 0.05 &  1.11 $\pm$ 0.05\\
\hline
Spectral shift ($\Delta\gamma_{\mathrm{conv.}}$)   &  0.068 $\pm$ 0.012 &  0.066 $\pm$ 0.012 \\
\hline
Atm. Density      &  -0.16 $\pm$ 0.71 &  -0.17 $\pm$ 0.68\\
\hline
Barr WM           &  -0.02 $\pm$ 0.28 & 0.00 $\pm$ 0.29\\
\hline
Barr WP           &  0.00 $\pm$ 0.28 &  0.01 $\pm$ 0.29\\
\hline
Barr YM           &  -0.06 $\pm$ 0.24 & -0.03 $\pm$ 0.25\\
\hline
Barr YP           &  -0.10 $\pm$ 0.15 &  -0.05 $\pm$ 0.15\\
\hline
Barr ZM           &  -0.00 $\pm$ 0.11 & -0.00 $\pm$ 0.11\\
\hline
Barr ZP           &  0.01 $\pm$ 0.09 & 0.016 $\pm$ 0.089\\
\hline
\multicolumn{3}{c}{\textbf{Astrophysical Flux parameters}}\\
\hline
Normalization ($\Phi_{\mathrm{astro.}}$)     &  0.95 $\pm$ 0.21 & 0.80 $\pm$ 0.21\\
\hline
Spectral shift ($\Delta\gamma_{\mathrm{astro.}}$)   &  0.11 $\pm$ 0.19 & -0.06 $\pm$ 0.21\\
\hline
\multicolumn{3}{c}{\textbf{Cross sections}}\\
\hline
Cross section $\sigma_{\nu_\mu}$   &  1.00 $\pm$ 0.03 & 1.000 $\pm$ 0.03\\
\hline
Cross section $\sigma_{\overline{\nu}_\mu}$    &  1.003$\pm$ 0.075 & 1.004 $\pm$ 0.074\\
\hline
Hadronic energy loss $\sigma_{KA}$ &  -0.35 $\pm$ 0.93 & -0.06 $\pm$ 0.90\\
\hline
\end{tabular}%
}  
\caption{\textbf{\textit{The measured model parameters for Analysis I (left) and Analysis II (right) at their respective best-fit points.}}
The reported $\pm$1$\sigma$ uncertainties on each of the 18 nuisance parameters are derived from the calculated standard deviations of the posterior distributions shown in Fig.~\ref{fig::posterior}.
A description of the priors on each nuisance parameter can be found in Table~\ref{table::Priors}.}
\label{table::final_fits}
\end{table}


\begin{figure}[t]
\begin{center}
    \includegraphics[width=0.99\columnwidth]{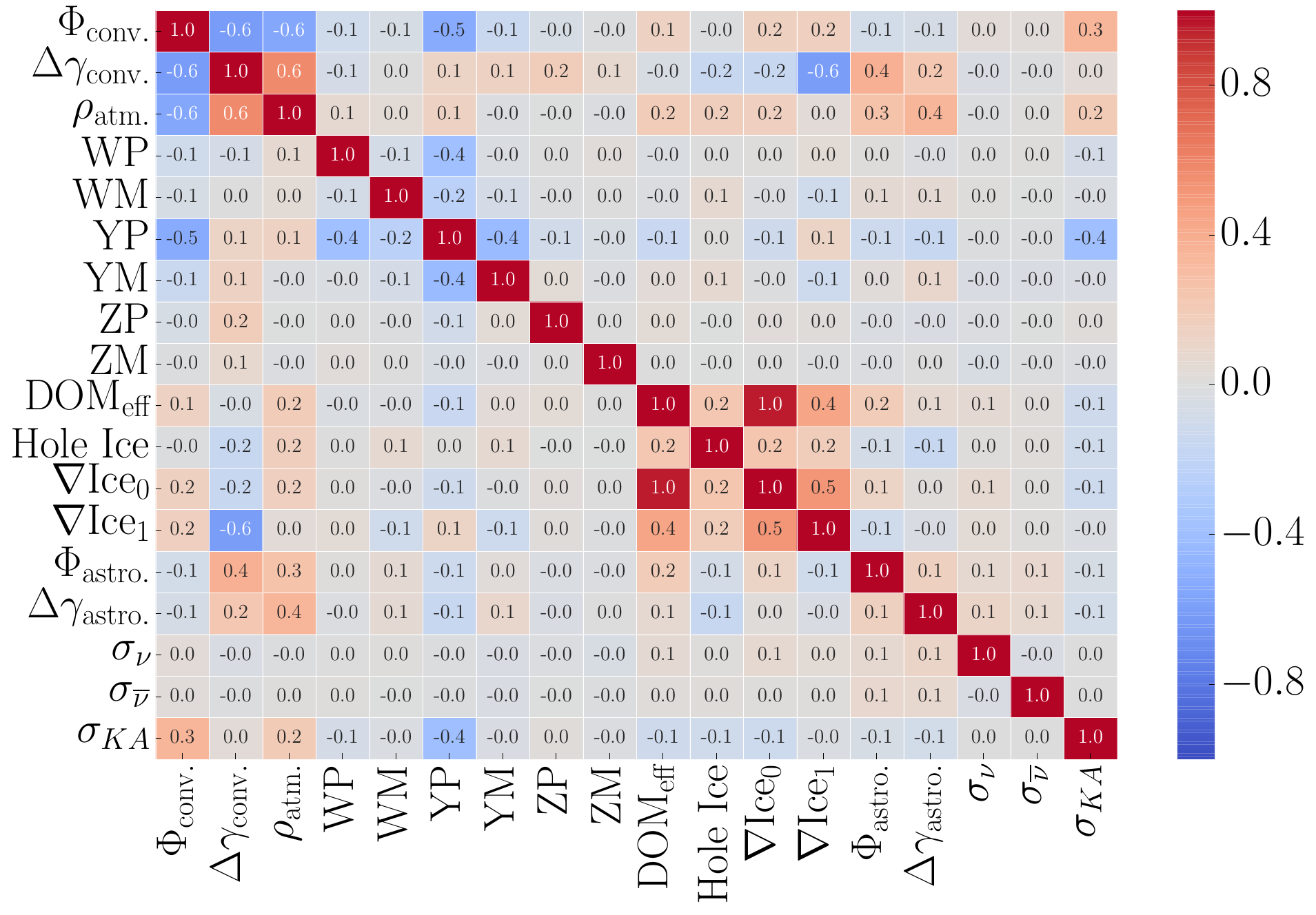}
    \caption{\textbf{\textit{Correlation between nuisance parameters.}}
    The correlation matrix has been calculated for the best-fit point of Analysis I.}
    \label{fig::correlations}
\end{center}
\end{figure}

Fig.~\ref{fig::correlations} shows the correlations between each of the 18 nuisance parameters at the best-fit point of Analysis I; Analysis II is not shown, but was found to be largely the same.
Correlations between subsets of nuisance parameters are observable, as shown in Fig.~\ref{fig::correlations}.
For example, we find the conventional flux normalization to be anti-correlated with the cosmic-ray spectral index as well as the atmospheric density; the DOM efficiency to be highly correlated with the ice properties; and the astrophysical normalization to be correlated with the astrophysical spectral index.

\section{Results\label{sec::results}}

Following a staged unblinding and post-unblinding process to check data consistency, the results of Analyses I and II were obtained for the full IceCube data sample.
During the unblinding process some {\em a-posteriori} changes were implemented to alleviate moderate data / Monte Carlo disagreement at high energies.  This is primarily related to updating the prior size on the astrophysical component to match IceCube's recent astrophysical neutrino measurements, as well as expanding treatments of cross section uncertainties, and the kaon energy losses.  The analysis result was qualitatively unchanged, following these modifications.

\begin{figure}[ht!]  
 \begin{minipage}{\columnwidth}
   \includegraphics[width=\textwidth]{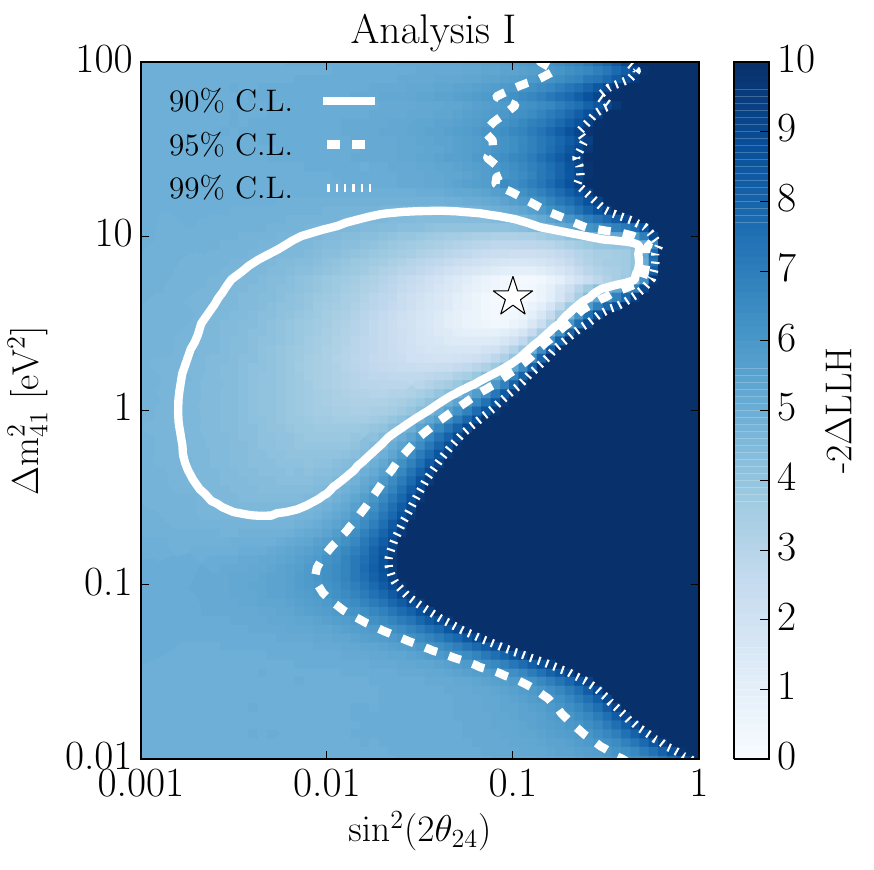}
  \end{minipage}
  \begin{minipage}{\columnwidth}
   \includegraphics[width=\textwidth]{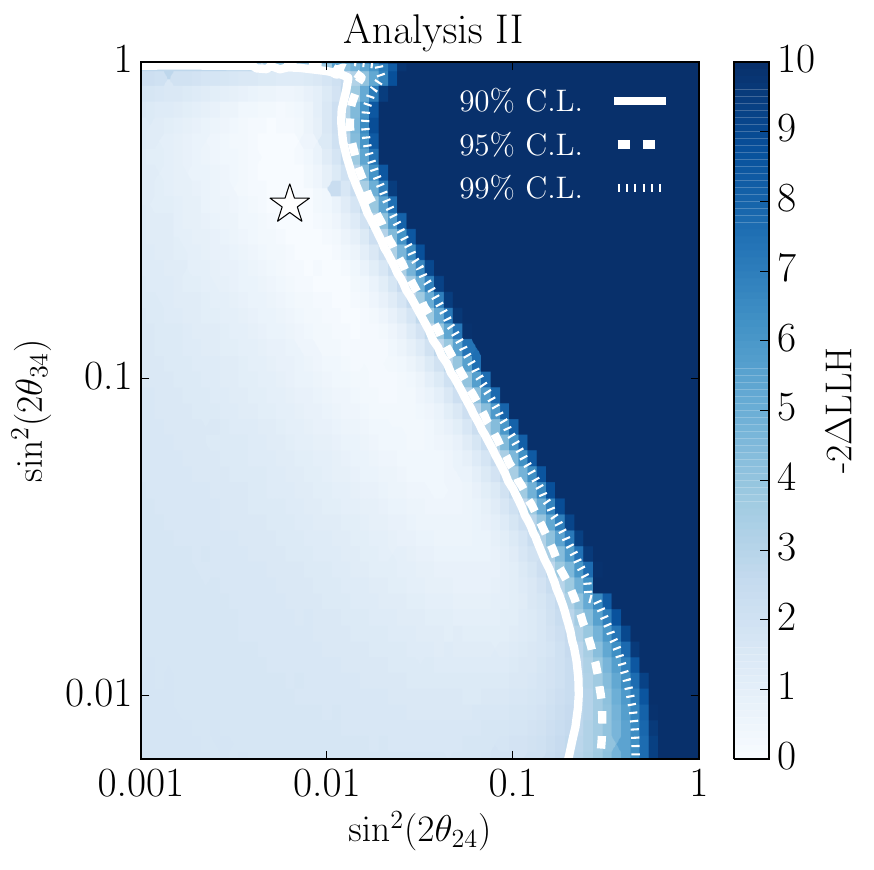}
 \end{minipage}
 \caption{\textbf{\textit{Frequentist result for analyses I and II.}}
 The result of Analysis I (top) and Analysis II (bottom). 
 The best-fit points are marked with white stars, and the 90\% (solid), 95\% (dashed), and 99\% (dotted) C.L. contours are drawn assuming Wilks' theorem with two degrees of freedom.
 The color scale shows the likelihood difference with respect to the best-fit point scaled by two in both analyses.}
 \label{fig::results}
\end{figure}

The analysis results are shown as likelihood maps in Fig.~\ref{fig::results}, with overlaid  90\%, 95\%, and 99\% C.L. contours, calculated assuming Wilks' theorem.
Analysis I was found to have a best-fit point at $\Delta m^2_{41}=\SI{4.47}\eV^2$ and $\sin^2(2 \theta_{24})=0.10$.
The $TS$ compared to the no sterile neutrino hypothesis is $-2\Delta \log \mathcal{L}_{\rm profile} =4.94$, corresponding to a p-value of 8\% when assuming two degrees of freedom.
Analysis II was found to have a best-fit point at $\sin^2(2 \theta_{34})=0.40$, $\sin^2(2 \theta_{24})=0.006$, with a $-2\Delta \log \mathcal{L}_{\rm profile}=1.74$ corresponding to a p-value of 19\% when assuming one degree of freedom.  The validity of Wilks' theorem was tested for both analyses at several points along the contour using the Feldman-Cousins method~\cite{Feldman:1997qc}, with Analysis I likelihoods following a $\chi^2$ distribution with two degrees of freedom, and Analysis II likelihoods following a $\chi^2$ distribution with two, as expected.  A full discussion of these analysis results and interpretation of the closed 90\% CL contour in Analysis I follows in Sec.~\ref{sec:sens}.

\begin{figure}[tbh]  
 \begin{minipage}{\columnwidth}
   \includegraphics[width=\textwidth]{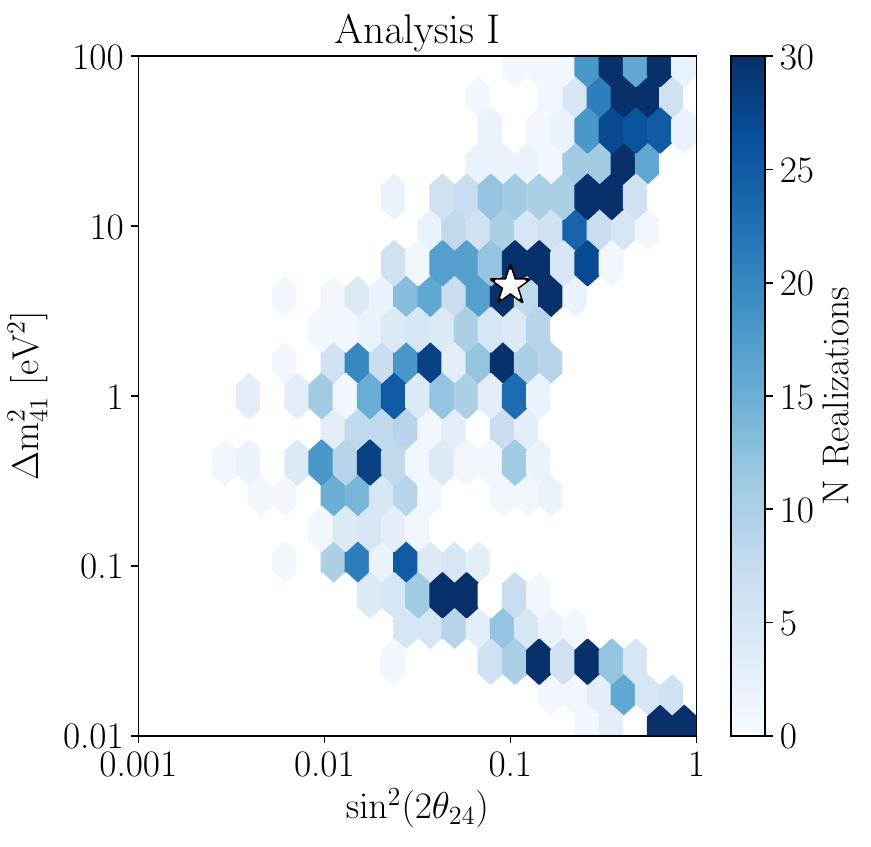}
  \end{minipage}
  \hfill
  \begin{minipage}{\columnwidth}
   \includegraphics[width=\textwidth]{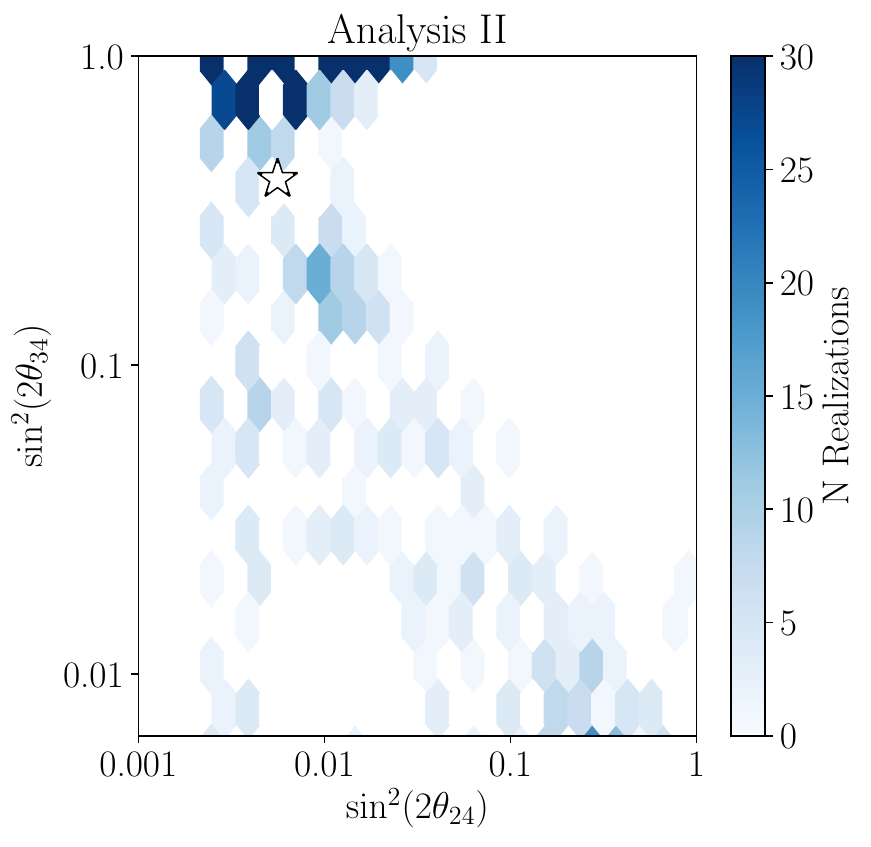}
 \end{minipage}
 \caption{\textbf{\textit{Best-fit points distribution from null hypothesis pseudodata for analyses I and II.}}
 The distribution of the best-fit points given two thousand null realizations (blue) throughout the physics parameter space.
 Also shown is the location of the analyses best-fit points, as white stars. }
 \label{fig::best_fit_points}
\end{figure}

Using two thousand pseudo-experiments with the nuisance parameters set to the central values, the distribution of best-fit points throughout the parameter space for pseudo-experiments with no injected sterile neutrino signal are determined.
These are shown in Fig.~\ref{fig::best_fit_points}. 
The purpose of this test is to establish whether the best-fit point of each analysis has fallen in a location where one might expect it to, given an experiment with no injected signal.
As expected, the statistical fluctuations tend to populate best-fit points around the edge of the  90\% C.L. sensitivity.
The distribution of best-fit points for Analysis I shows a slight clustering at large values of $\Delta m^2_{41}$, above $\sim \SI{10}\eV^2$.
It was found in Ref.~\cite{jones2015sterile} that the fast oscillations in this region average out pulling the normalization downward with very little signal shape; \textit{e.g.} Fig. 3.4.6 of Ref.~\cite{jones2015sterile}.
This implies that statistical fluctuations of experiments with no true signal can find best-fit points in this region of parameter space, given a modest normalization shift.
The observed best-fit points for both analyses are shown as the white stars and are found in regions consistent with statistical fluctuations of the no sterile neutrino model.

\begin{figure}[tbh]  
 \begin{minipage}{\columnwidth}
   \includegraphics[width=\textwidth]{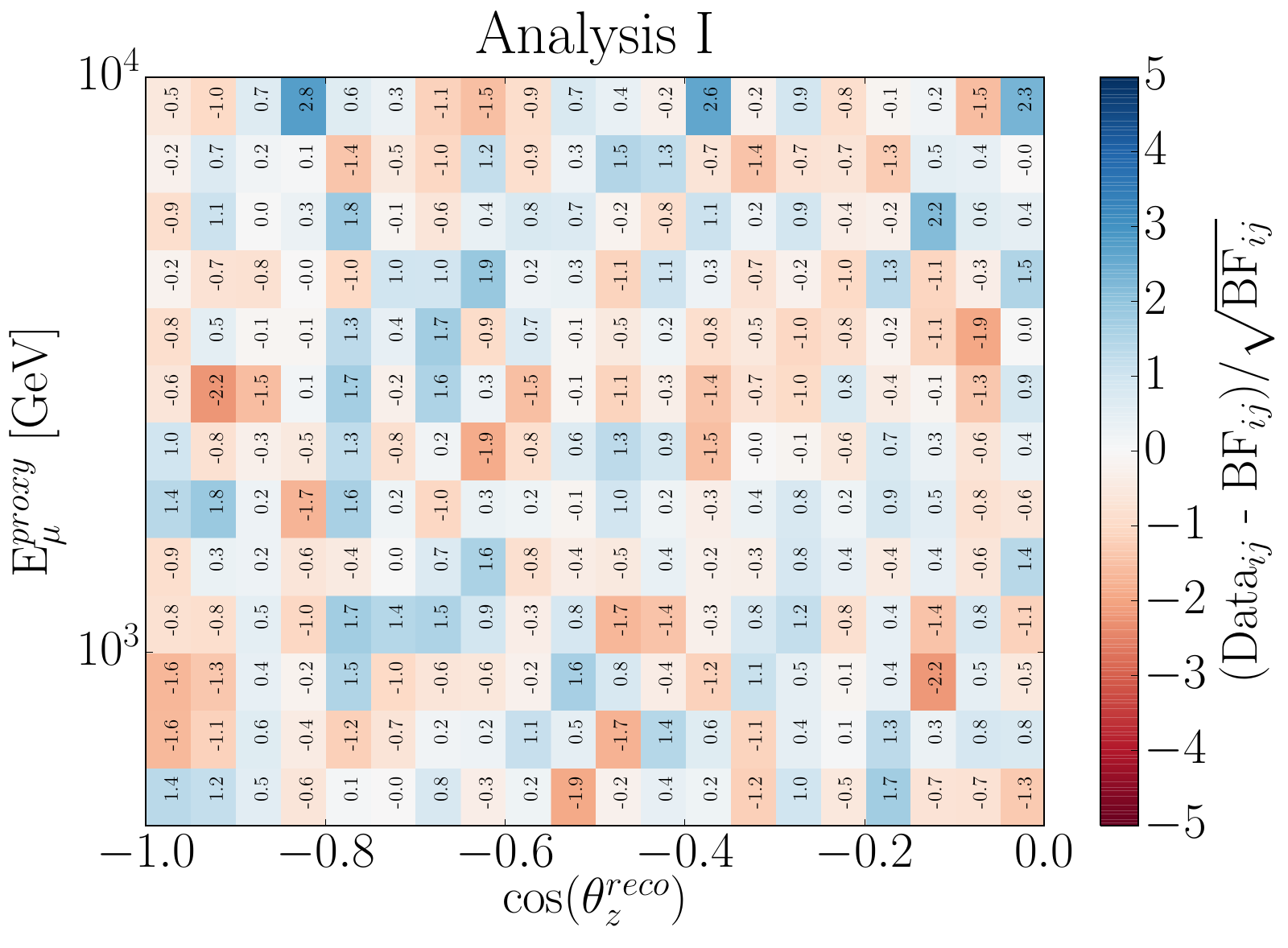}
  \end{minipage}
  \hfill
  \begin{minipage}{\columnwidth}
   \includegraphics[width=\textwidth]{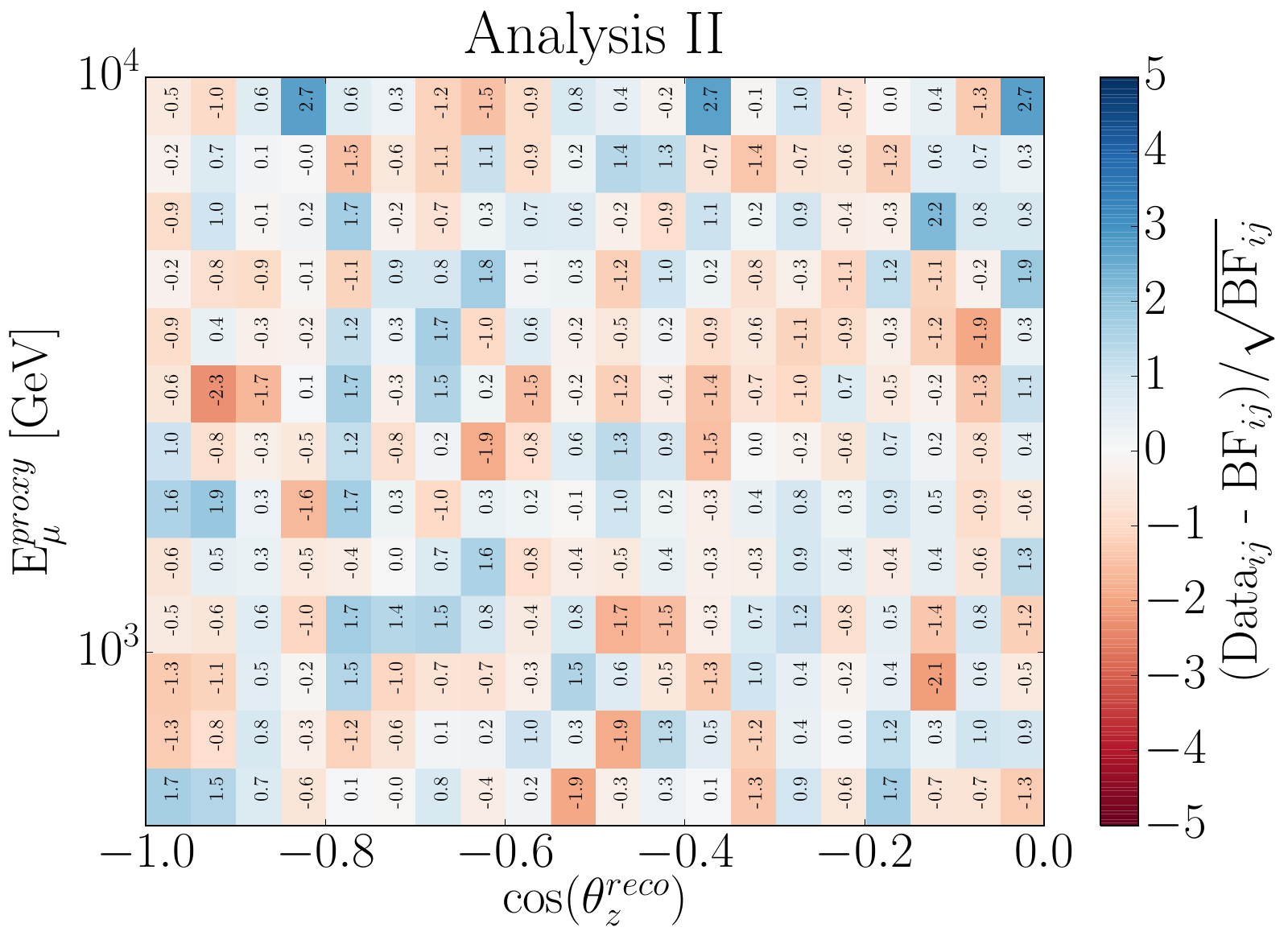}
  \end{minipage}
 \caption{\textbf{\textit{Data pull distribution with at analyses best-fit points.}}
 The observed statistical-only pull distribution in reconstructed energy and \cosz for Analysis I (top) and Analysis II (bottom) at their respective best-fit points.
 }
 \label{fig::2d_and_systematic_pulls1}
\end{figure}

The data pull relative to best fit (BF) in bin $i$ was calculated as $\mathrm{Pull}_i =  (\mathrm{Data}_i - \mathrm{BF}_i)/\sqrt{\mathrm{BF}_i}$.
The 2D gaussian statistical pull distribution of the result compared to the measured best-fit point for each analysis is shown in Fig.~\ref{fig::2d_and_systematic_pulls1}.
We observe the maximum statistical per-bin pull out of 260 bins to be $+2.7\sigma$ and the minimum to be $-2.2\sigma$.
The p-value with which one would expect at least one bin perturbed at least this far in either direction, given null realizations, is calculated to be 60.4\% for Analysis I and 61.1\% for Analysis II based on 10,000 pseudo-experiments, demonstrating that these excursions are not larger than expectations.

\begin{figure}[tbh!]  
 \begin{minipage}{\columnwidth}
   \includegraphics[width=\textwidth]{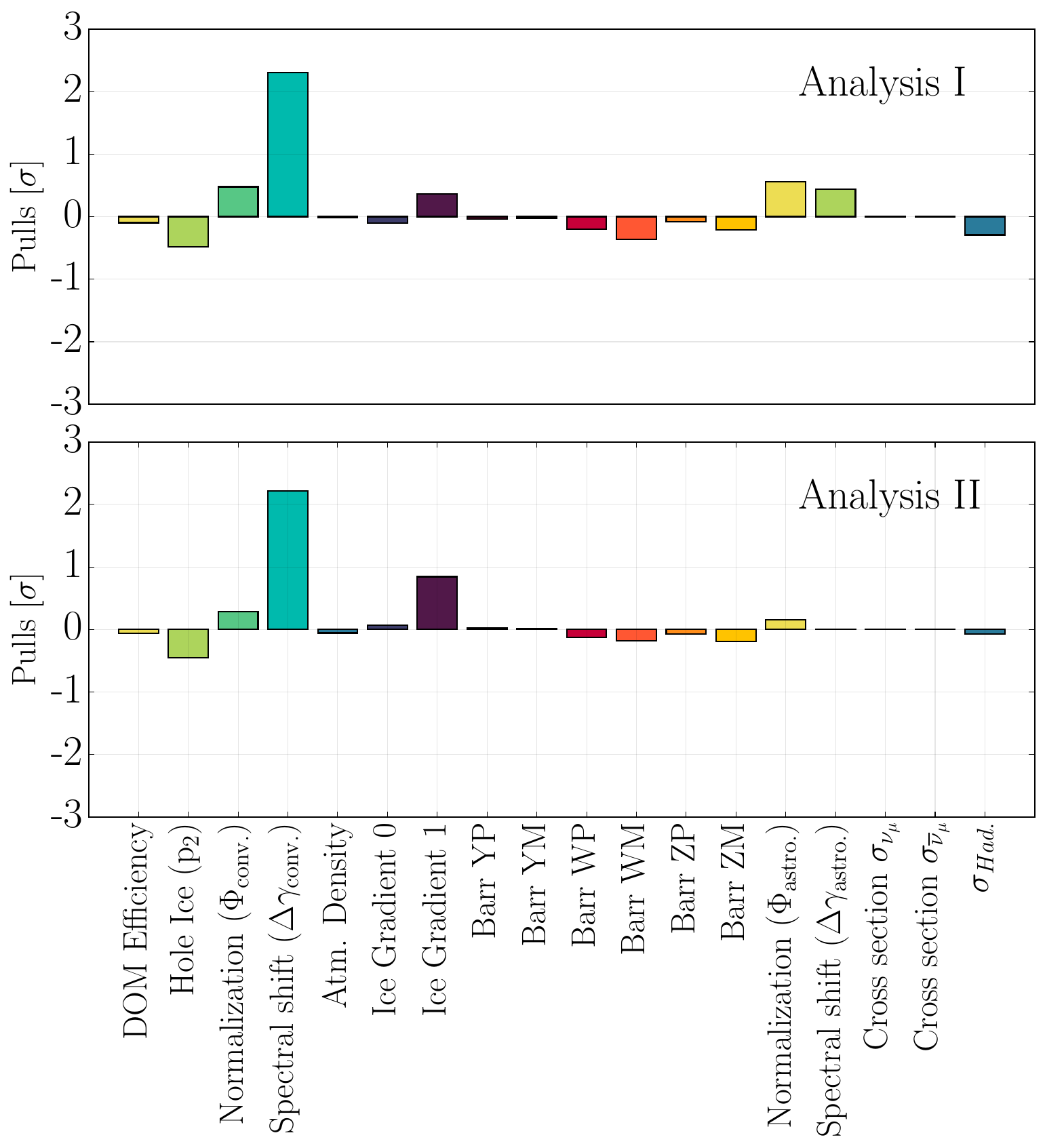}
  \end{minipage}
 \caption{\textbf{\textit{Most important systematic pulls at best-fit point.}}
 The systematic nuisance parameters pulls for both analyses at their respective best-fit points.
 }
 \label{fig::systematic_pulls2}
\end{figure}

\begin{figure}[tbh]  
 \begin{minipage}{\columnwidth}
   \includegraphics[width=\textwidth]{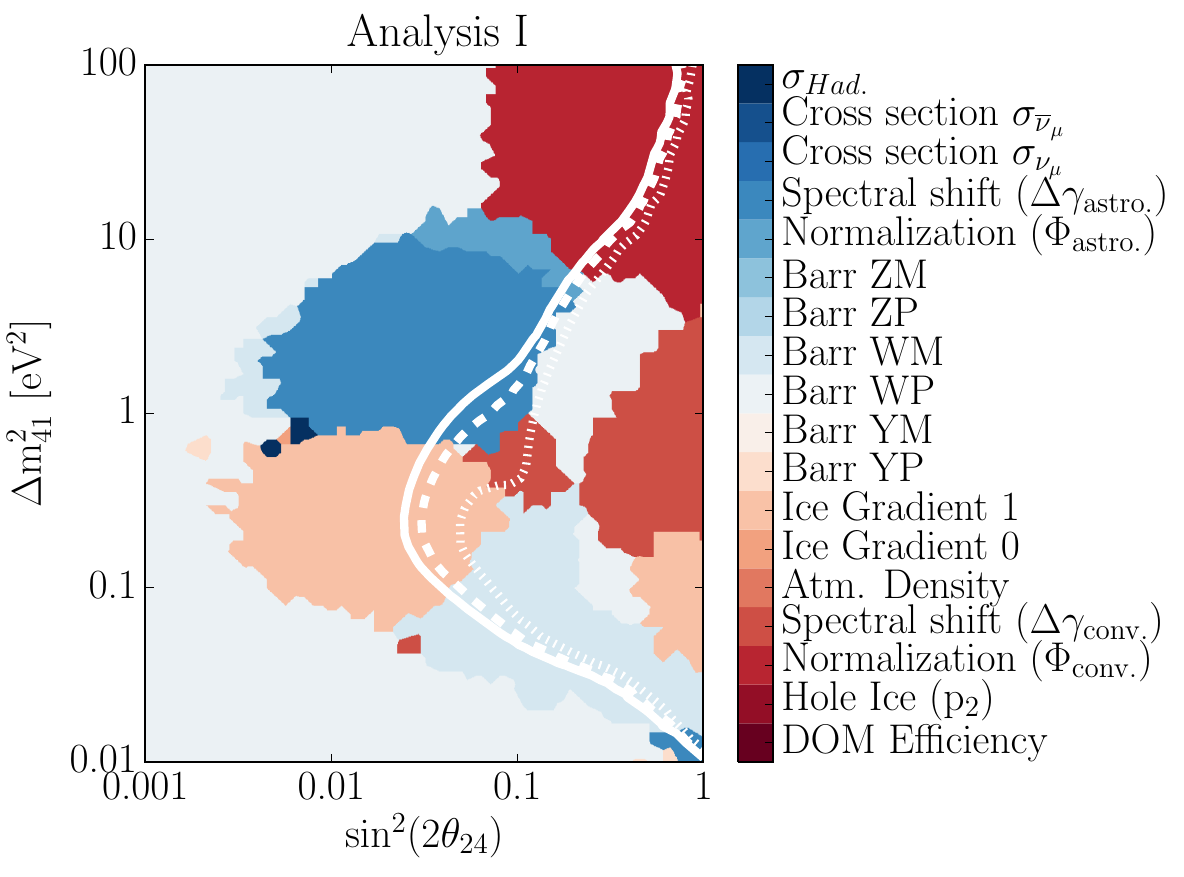}
  \end{minipage}
  \hfill
  \begin{minipage}{\columnwidth}
   \includegraphics[width=\textwidth]{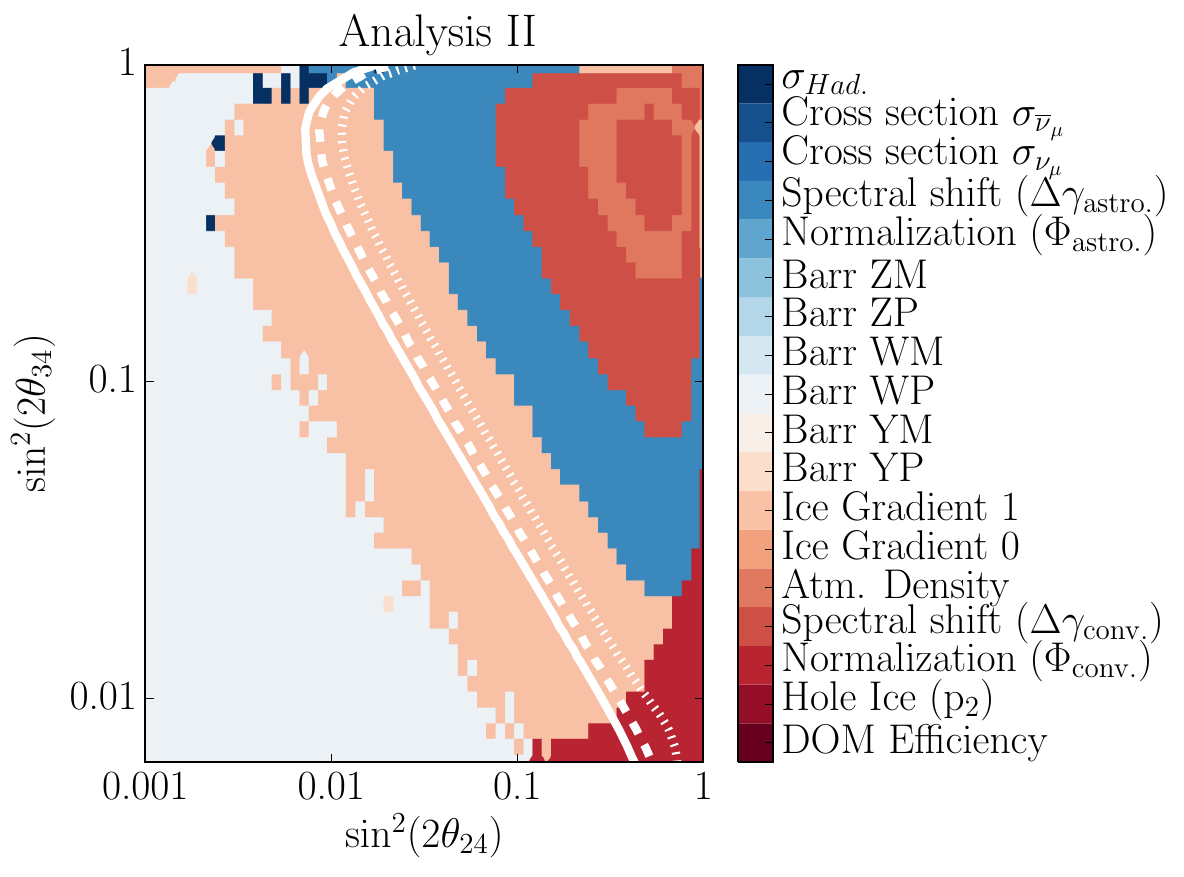}
  \end{minipage}
 \caption{\textbf{\textit{Expected dominant systematics across the parameter space.}}  The expected strongest-pulling nuisance parameter at each point in the physics space for each analysis.
 Please.
 }
 \label{fig::dominant_sys}
\end{figure}   

The nuisance parameter pulls, defined as $\mathrm{Sys\,Pull} =  (\mathrm{Fit\,Value}  - \mathrm{Prior\,Center} )/\mathrm{Prior\,Width}$, at the best-fit points of each analysis are shown in Fig.~\ref{fig::systematic_pulls2}.
None of the nuisance parameters, for either analysis, are in tension with their associated priors at a pull greater than $\pm 2.3\sigma$.

Both analyses appear to prefer similar systematic pulls.
The largest difference observed is between the measured conventional atmospheric neutrino normalization, where they are within 8\% of each other, corresponding to approximately $1.1\sigma$ given the posterior width.
It is also noted that the posterior width of the neutrino-nucleon cross section is identical to the prior width, indicating that we do not have significant sensitivity to this particular source of systematic uncertainty.

\section{Comparison to expected frequentist sensitivity~\label{sec:sens}}

\begin{figure}[thb!]  
 \begin{center}
  \begin{minipage}{\columnwidth}
   \includegraphics[width=0.9\textwidth]{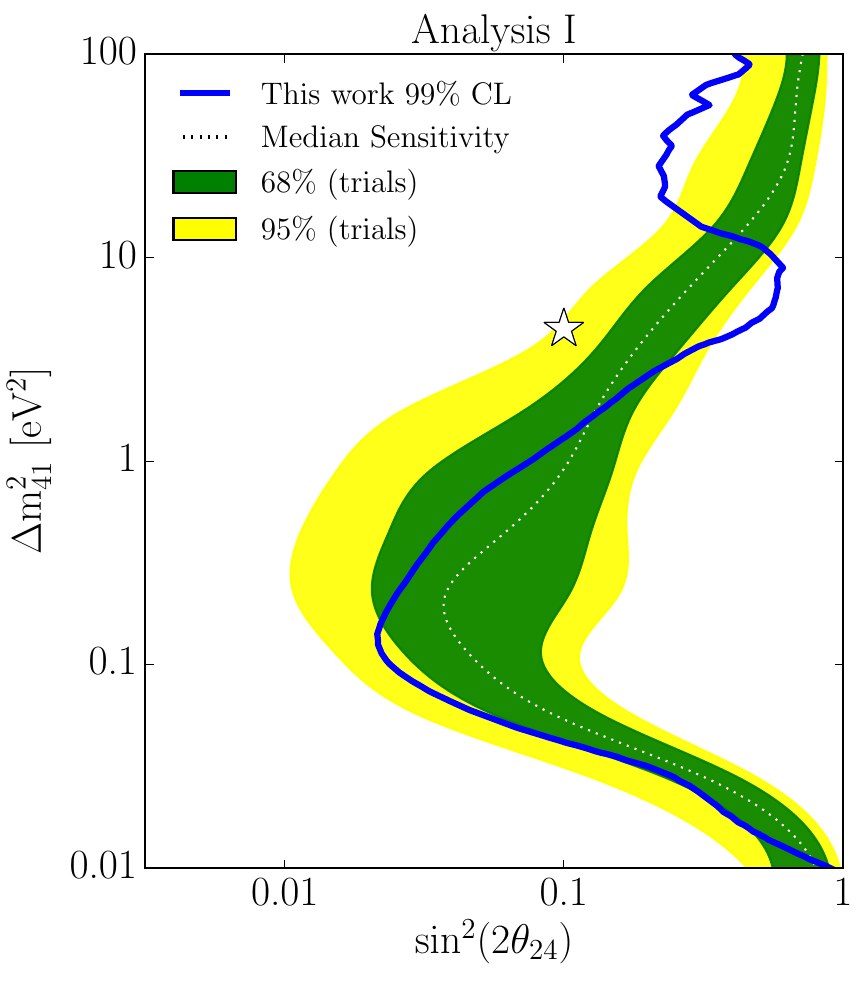} \includegraphics[width=0.9\textwidth]{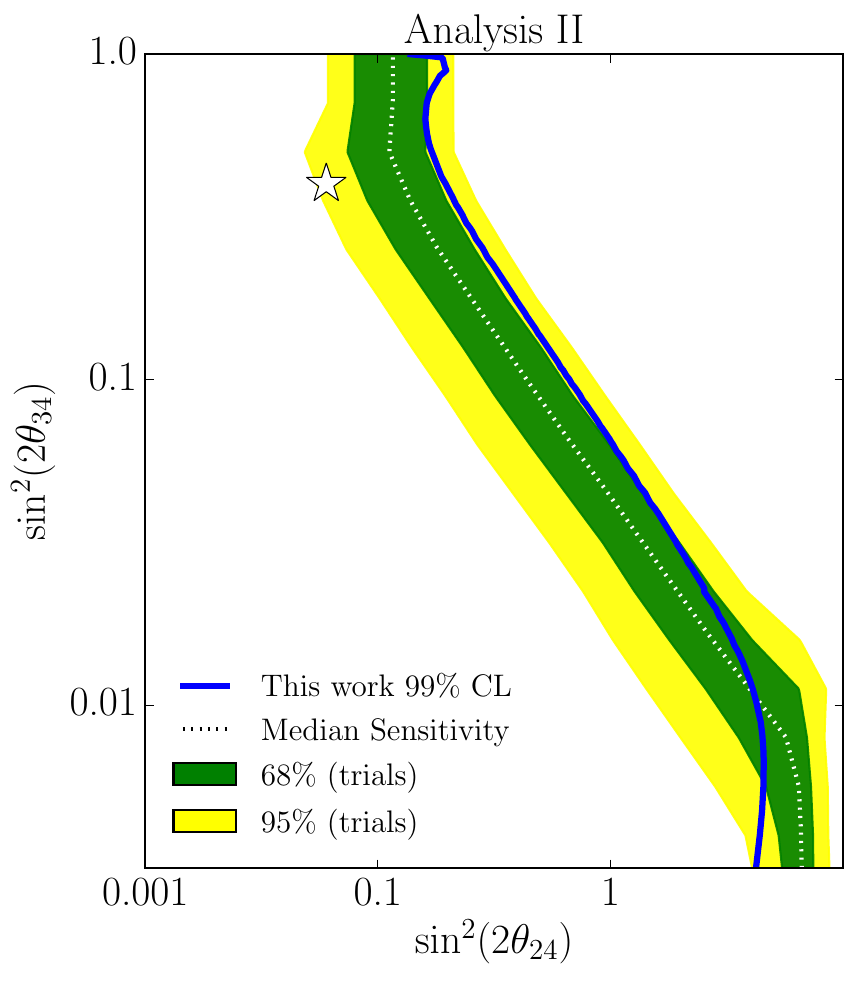}
  \end{minipage}
 \end{center}
 \caption{\textbf{\textit{Analyses frequentist result with expected sensitivities.}}
 The 99\% C.L. Brazil Bands for Analysis I (top) and Analysis II (bottom), overlaid with the analysis result shown as a black line.
 The yellow band corresponds to the 95\% spread, while the green to the 68\%.
 The median sensitivity is shown as a dashed white line, while the bet fit points for each analysis are shown as a white star.
 }
 \label{fig::brazil}
\end{figure}

The median sensitivity is defined by the values of the sterile neutrino mixing parameters that can be excluded in case of the no sterile neutrino model, at various confidence levels, in 50\% of pseudo-experiments. The reported sensitivity of this analysis is calculated through an ensemble of two thousand simulated pseudo-experiments, each of which was generated by drawing from the expected distribution at the no sterile neutrino model with the nuisance parameters at their central value.
For a given realization, we construct a confidence interval at the 90\% C.L. and 99\% C.L..
For every value of  $\Delta m^2_{41}$, or $\sin^{2}(\theta_{34})$ in the case of Analysis II, the coordinate of the contour in $\sin^{2}(\theta_{24})$ is recorded.
The distribution of the crossing values for $\sin^{2}(\theta_{24})$ are then used to define the 68.3\% (1$\sigma$) and 95.4\%~($2\sigma$) confidence intervals.
If the contour crosses more than once, we take the maximum  $\sin^{2}(\theta_{24})$ value of the crossing.
This procedure is performed for each value in of $\Delta m^2_{41}$, for both the 90\% CL and 99\% C.L. contours.
The resulting distributions produce ``Brazil bands'' and the median sensitivity values.
The width of the Brazil band indicates the expected scale of statistical variations of the result over repeated pseudo-experiments, given no injected signal.  Comparison of this width to the scales of effects from adding or removing systematic uncertainties provides a semi-quantitative method to define whether the analyses are statistically or systematically limited. 
These bands for both analyses at the 99\% C.L. are shown in Fig.~\ref{fig::brazil}.

In both analyses, the scale of the effects of systematic uncertainties remain significantly smaller than the scale of statistical fluctuations of the final result embodied in the sensitivity interval, which is an indication that both analyses remain statistics limited.

The 99\% C.L. contour is relatively consistent with its sensitivity envelope, largely enclosed within the 95\% region for both analyses.
The 90\% C.L. contour is closed in Analysis I.
By construction, comparison of the 90\% contour with the Brazil band should be made using its right-most edge, which also appears consistent with the sensitivity from pseudo-experiments.
Also by construction, a closed 90\% C.L. contour is expected in approximately 10\% of pseudo-experiments, and the best-fit point of Analysis I falls in a location consistent with expectations from null realizations. The p-value of the likelihood at this best-fit point is 8\% relative to the no sterile neutrino model.
We therefore conclude that this particular data realization is unexceptional relative to results of pseudo-experiments generated under the no sterile neutrino hypothesis.

\section{Tests of result robustness}

\begin{figure}[t]  
 \begin{minipage}{\columnwidth}
   \includegraphics[width=\textwidth]{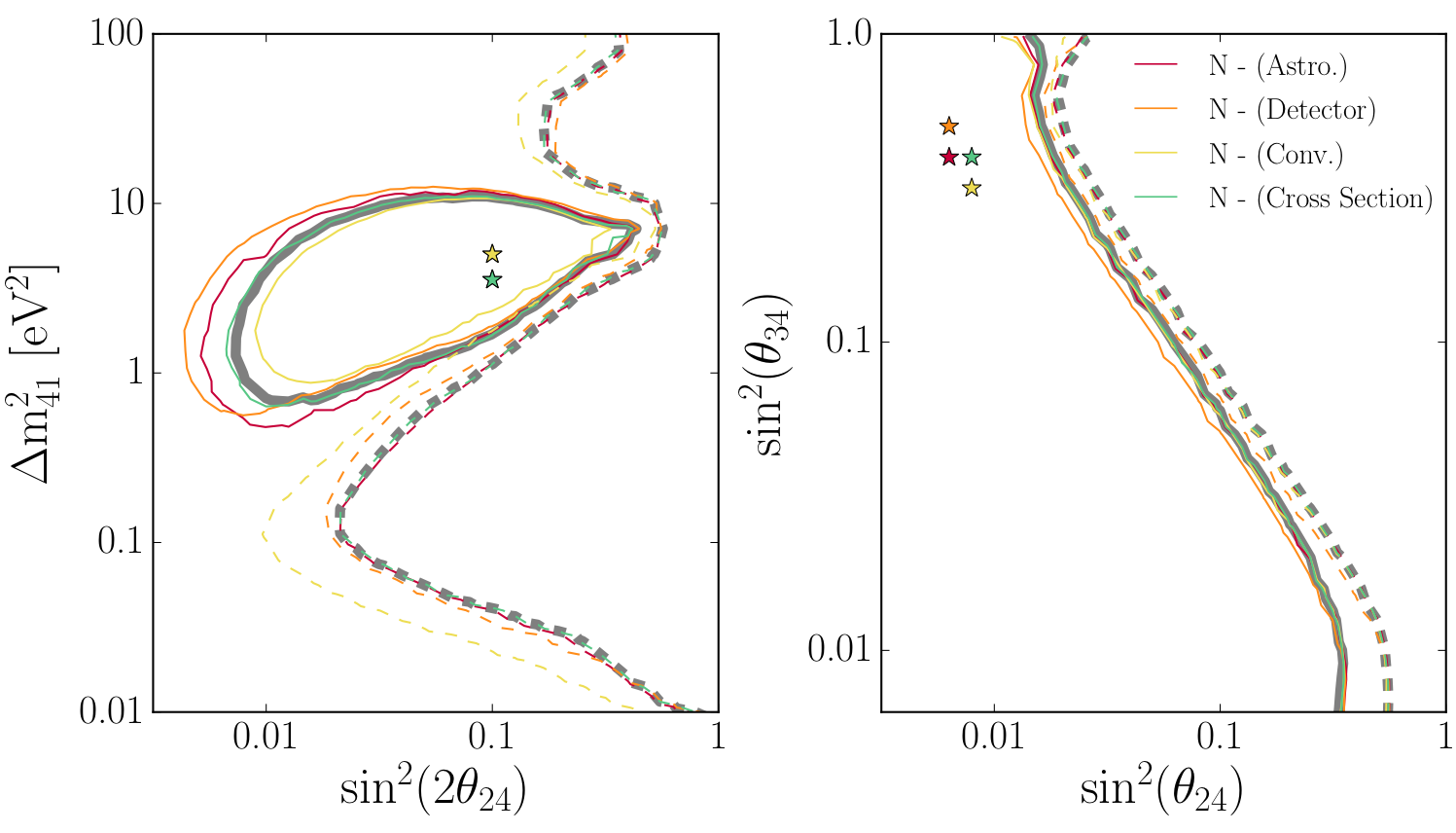}
 \end{minipage}
 \caption{\textbf{\textit{Effect of removing a systemic category on the frequentist results.}}
 Each color line corresponds to the analysis performed without a single systematic group and the star of the same color is the corresponding best-fit point. Left: Analysis I; right: Analysis II. The solid (dashed) lines show the 90\% C.L. (99\% C.L.) and the stars represent the best-fit point.}
 \label{fig::compare_nm1}
\end{figure}

Fig.~\ref{fig::compare_nm1} shows the impact on the result after removing various groups of systematic uncertainty categories from the analyses.
This tests whether the result is especially sensitive to any specific group of systematic effects.
The solid (dashed) lines in these figures show the 90\% C.L. (99\% C.L.) and the stars represent the best-fit parameters location. 
We find the main analysis results are robust in all cases.  

\begin{figure}[t]  
 \begin{minipage}{\columnwidth}
   \includegraphics[width=\textwidth]{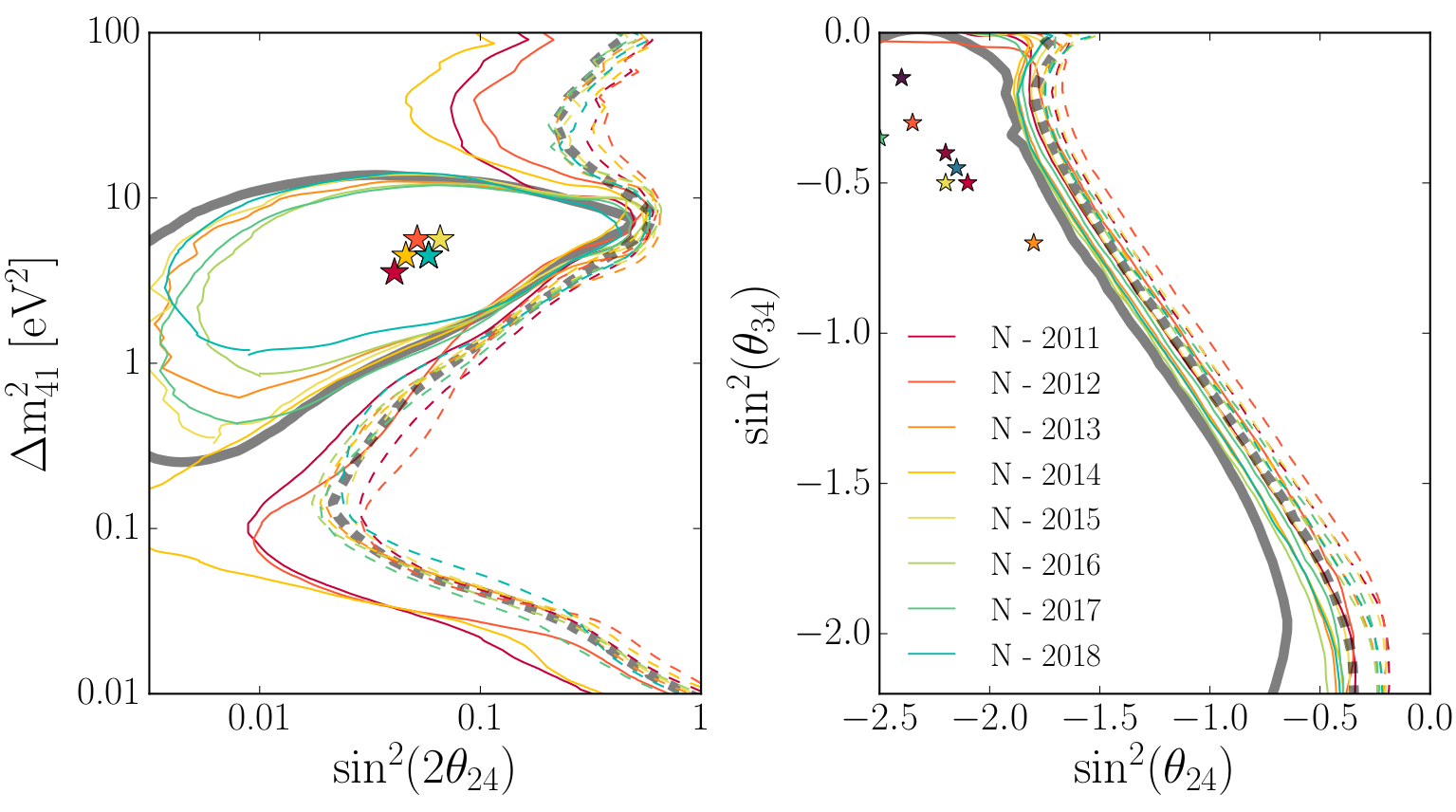}
 \end{minipage}
 \caption{\textbf{\textit{Effect of removing a given year on the frequentist results.}}
 Each color line corresponds to the analysis performed without a given year.
 Left: Analysis I; right: Analysis II. The solid (dashed) lines show the 90\% C.L. (99\% C.L.) and the stars represent the best-fit point.}
 \label{fig::compare_ym1}
\end{figure}

Fig.~\ref{fig::compare_ym1} shows the impact of removing any one year of data, to test for the effects of time-localized excursions on the result.
The contour moves due to statistical fluctuations for each entry, but its shape is broadly unchanged when removing any one year of data, demonstrating stability against time-localized statistical or systematic fluctuations.

Other studies are also made to test for result robustness, including loosening priors on the systematic uncertainties with the largest pulls and testing consistency of each year of data one-by-one with the total accumulated data set.
In all cases, strong consistency is observed.  

\section{Discussion}

\begin{figure}[t]  
 \begin{minipage}{\columnwidth}
   \includegraphics[width=\textwidth]{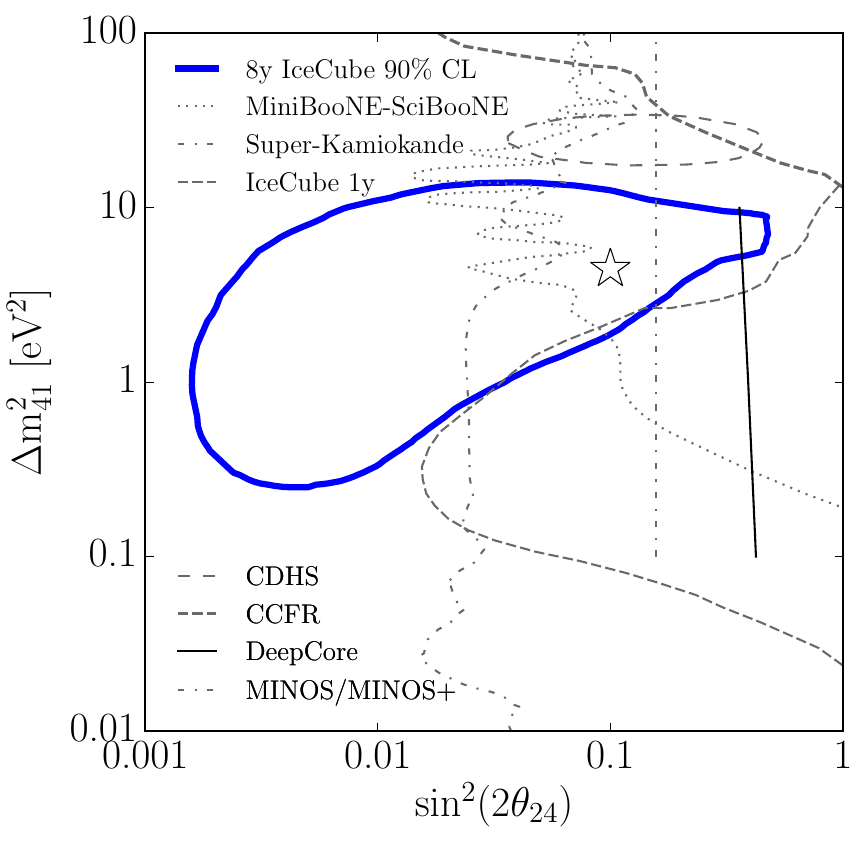}
  \end{minipage}
  \hfill
  \begin{minipage}{\columnwidth}
   \includegraphics[width=\textwidth]{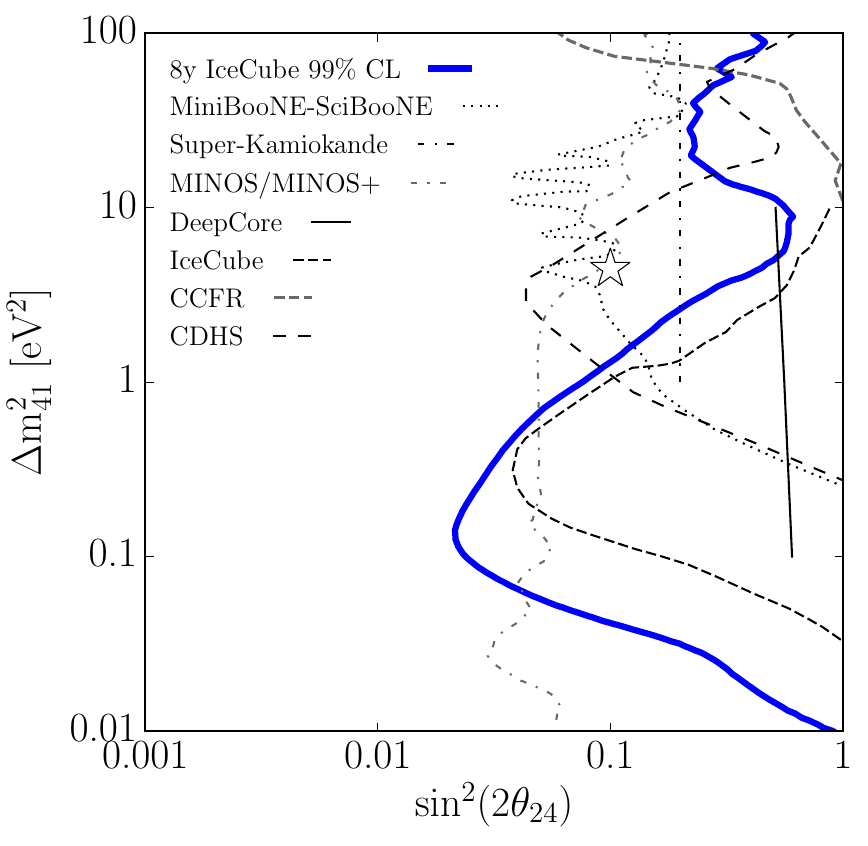}
 \end{minipage}
 \caption{\textbf{\textit{Comparison to other $\nu_\mu$ disappearance results.}}
    The solid blue line in the top (bottom) panel shows the Analysis I frequentist result at 90\% C.L. (99\% C.L.) compared to other experiments' results shown as thin black lines~\cite{adamson2011active,adamson2012improved, adamson2011search,adamson2016search,adamson2017search,abe2015limits,mahn2012dual,cheng2012dual,stockdale1984limits}. Where results were not available at 99\% C.L., methods of Ref.\cite{Diaz:2019fwt} were applied using public data releases.
    }
 \label{fig::compare_world}
\end{figure}

There are three presently published results from IceCube on the 3+1 sterile neutrino model in a comparable parameter space to Analysis I. The three year DeepCore result~\cite{aartsen2017search} placed a limit on the sterile neutrino parameters at  $\Delta m^2_{41} \approx \SI{1}\eV^2$ and $\sin^2(2 \theta_{24})=0.39$ at the 90\% C.L.. The result was re-derived at several other $\Delta m^2_{41}$ values between 0.1 eV$^2$ and 10 eV$^2$ with only a very small dependence on $\Delta m^2_{41}$ observed. The one-year high-energy sterile neutrino search excluded the region from approximately $\SI{0.1}\eV^2 \leq \Delta m^2_{41} \leq \SI{2.0}\eV^2$ above $\sin^2(2 \theta_{24})=0.1$, extending to approximately $\sin^2(2 \theta_{24})=0.016$ and $\Delta m^2_{41} = \SI{0.27}\eV^2$. Alongside the publication of the full-detector (IC86) result, an independent measurement using a partial IceCube configuration with 59 active strings (IC59) was reported at 99\% C.L..  These results are collected and compared with the result of this analysis and other world data in Fig.~\ref{fig::compare_world}.

The result of Analysis I shown in blue is in good  agreement with the previous IceCube limits at 90\% C.L.. The 99\% C.L. exclusion region over sterile neutrino mixing parameters  is expanded relative to previous analyses.
In the region below $\Delta m^2_{41}=\SI{0.1}\eV^2$, the confidence reaches down to a factor seven smaller mixing amplitudes, largely due to the improved statistics at low energies.   

\begin{figure}[t]  
 \begin{minipage}{\columnwidth}
   \includegraphics[width=\textwidth]{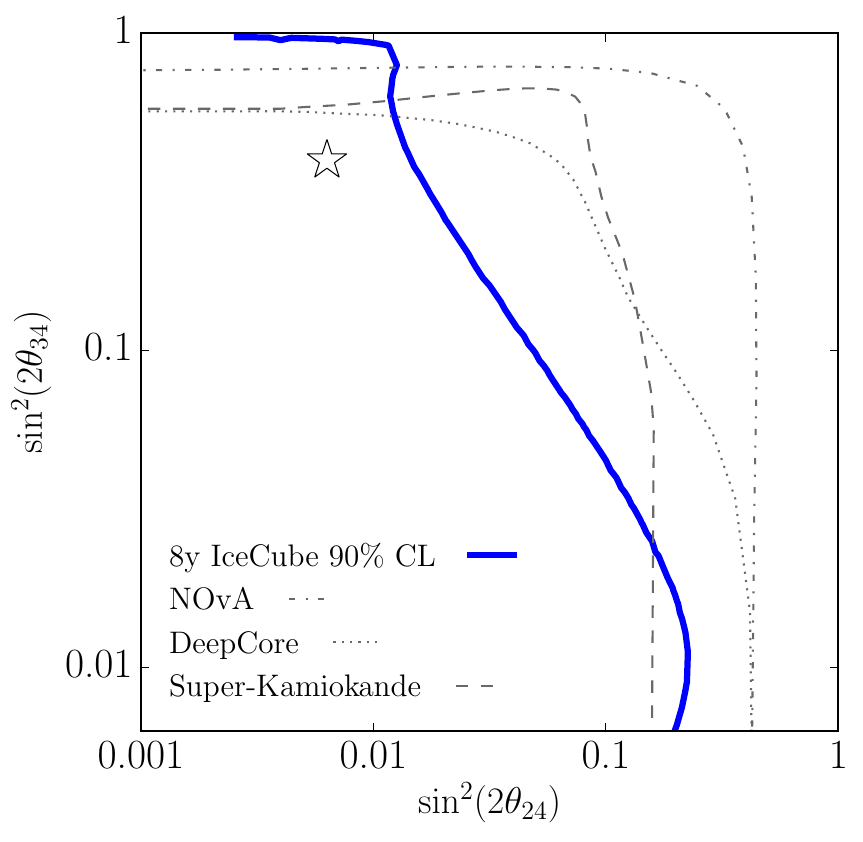}
  \end{minipage}
  \hfill
  \begin{minipage}{\columnwidth}
   \includegraphics[width=\textwidth]{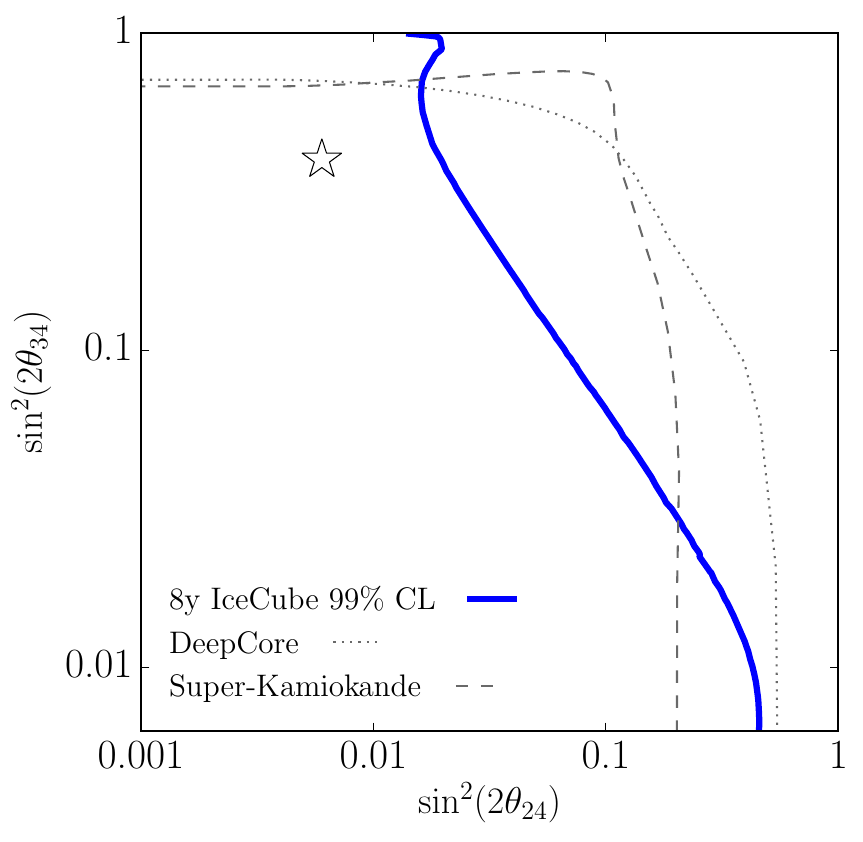}
 \end{minipage}
 \caption{\textbf{\textit{Analysis II frequentist result compared with other experiments' results.}}
 The solid blue line in the top (bottom) panel shows the Analysis II frequentist result at 90\% C.L. (99\% C.L.) compared to other experiments' results shown as thin black lines.
 }
 \label{fig::compare_world2}
\end{figure}

The three-year DeepCore sterile neutrino analysis has also placed limits in the comparable space to Analysis II.  We find that the result of Analysis II improves the limit on the sterile neutrino mixing parameters below approximately $\sin^2(2 \theta_{34})=0.4$.
Here, the confidence interval is shown to increase by a factor ranging from two to approximately five. This comparison is shown alongside other world data in Fig.~\ref{fig::compare_world2}.

Prior measurements of $\nu_\mu$ disappearance have been made by MINOS~\cite{adamson2011active,adamson2012improved, adamson2011search,adamson2016search}, MINOS+~\cite{adamson2019search}, NO$\nu$A~\cite{adamson2017search}, DeepCore, Super-Kamiokande~\cite{abe2015limits},   MiniBooNE-ScibooNE~\cite{mahn2012dual,cheng2012dual}, and CDHS~\cite{stockdale1984limits}.
One can also compare these results to the results of global fits.  The 99\% C.L. limit excludes part of the allowed region from Ref.~\cite{alex_global}, and the lower island from Ref.~\cite{collin2016sterile}.
The best-fit point from Ref.~\cite{collin2016sterile} is centrally within the allowed region at 90\%~C.L..
Despite the existence of a non-trivial allowed region in Analysis I, comparison with the preferred region from appearance experiments where $\nu_\mu\rightarrow\nu_e$ anomalies are observed shows a strong tension with the IceCube result, as it does with all other $\nu_\mu$ or $\overline{\nu}_{\mu}$ disappearance searches.
The increased extent of the 99\% contour in the relevant parameter space suggests that, despite finding a closed 90\% contour, this result is in increased tension with the allowed region from appearance experiments, relative to the previous IceCube result.

The equivalent comparison to world data for Analysis II is shown in Fig.~\ref{fig::compare_world2}.
Here, data is compared at the 90\% C.L. (top) and 99\% CL (bottom) to other results in this parameter space from Super-Kamiokande~\cite{abe2015limits} and DeepCore~\cite{aartsen2017search}.
This analysis provides world leading limits in the region $\Delta m^2_{41} \geq \SI{10}\eV^2$ from approximately $0.024 \leq \sin^2(2 \theta_{34}) \leq 0.54$ and $0.012 \leq \sin^2(2 \theta_{24}) \leq 0.16$. 

\section{Conclusion}

We have presented a detailed description of an eight-year search for sterile neutrinos in two parameter spaces.
The result uses a new high-purity and high-efficiency event selection for upward-going track-like events, and incorporates detailed treatments of systematic uncertainties stemming from ice properties, detector response, atmospheric, and astrophysical neutrino fluxes. 

The results obtained by analyzing 305,735 atmospheric and astrophysical $\nu_\mu$ and $\bar{\nu}_\mu$ events are used to generate confidence intervals in the space of $\Delta m^2_{41}$ vs. $\sin^2(2\theta_{24})$ assuming $\theta_{34}$ and all $CP$-phases to be zero (Analysis I) and in the space of $\sin^2(2\theta_{24})$ vs. $\sin^2(2 \theta_{34})$ for $\Delta m^2_{41} \geq \SI{20}\eV^2$ and again assuming all $CP$-phases to be zero (Analysis II).
In both parameter spaces, strong exclusions are obtained at 99\% C.L., increasing tensions with the global preferred regions from appearance experiments.
A closed contour is observed at 90\% C.L. in Analysis I, which includes parts of the allowed regions from global fits to world data.
The best-fit likelihood is found to be consistent with fluctuations of the no sterile neutrino model with a p-value of 8\%, and the best-fit point is unexceptional relative to observed closed-contour results obtained from pseudo-experiments.  
However, a consistent result obtained with each year of data is suggestive of a small systematic effect rather than a fluctuation of purely statistical origin.
Therefore, while this result is not considered as strong evidence for sterile neutrinos, it is likely to be impactful on the landscape of 3+1 global fits due to its high statistical power in the relevant parameter space.

\begin{acknowledgments}

The IceCube collaboration acknowledges the significant contributions to this manuscript from the Massachusetts Institute of Technology and University of Texas at Arlington groups.

We acknowledge the support from the following agencies:  USA {\textendash} U.S. National Science Foundation-Office of Polar Programs,
U.S. National Science Foundation-Physics Division,
Wisconsin Alumni Research Foundation,
Center for High Throughput Computing (CHTC) at the University of Wisconsin-Madison,
Open Science Grid (OSG),
Extreme Science and Engineering Discovery Environment (XSEDE),
U.S. Department of Energy-National Energy Research Scientific Computing Center,
Particle astrophysics research computing center at the University of Maryland,
Institute for Cyber-Enabled Research at Michigan State University,
and Astroparticle physics computational facility at Marquette University;
Belgium {\textendash} Funds for Scientific Research (FRS-FNRS and FWO),
FWO Odysseus and Big Science programmes,
and Belgian Federal Science Policy Office (Belspo);
Germany {\textendash} Bundesministerium f{\"u}r Bildung und Forschung (BMBF),
Deutsche Forschungsgemeinschaft (DFG),
Helmholtz Alliance for Astroparticle Physics (HAP),
Initiative and Networking Fund of the Helmholtz Association,
Deutsches Elektronen Synchrotron (DESY),
and High Performance Computing cluster of the RWTH Aachen;
Sweden {\textendash} Swedish Research Council,
Swedish Polar Research Secretariat,
Swedish National Infrastructure for Computing (SNIC),
and Knut and Alice Wallenberg Foundation;
Australia {\textendash} Australian Research Council;
Canada {\textendash} Natural Sciences and Engineering Research Council of Canada,
Calcul Qu{\'e}bec, Compute Ontario, Canada Foundation for Innovation, WestGrid, and Compute Canada;
Denmark {\textendash} Villum Fonden, Danish National Research Foundation (DNRF), Carlsberg Foundation;
New Zealand {\textendash} Marsden Fund;
Japan {\textendash} Japan Society for Promotion of Science (JSPS)
and Institute for Global Prominent Research (IGPR) of Chiba University;
Korea {\textendash} National Research Foundation of Korea (NRF);
Switzerland {\textendash} Swiss National Science Foundation (SNSF);
United Kingdom {\textendash} Department of Physics, University of Oxford.
\end{acknowledgments}

\bibliography{Master}
\bibliographystyle{apsrev4-1}
\clearpage


\onecolumngrid
\appendix

\ifx \standalonesupplemental\undefined
\setcounter{figure}{0}
\setcounter{table}{0}
\setcounter{equation}{0}
\fi

\renewcommand{\figurename}{SUPPL. FIG.}
\renewcommand{\tablename}{SUPPL. TABLE}
\renewcommand{\theequation}{A\arabic{equation}}

\newpage

\begin{sidewaystable}
        \centering
        Supplementary Material: Effect of Each Systematic Uncertainty (1)
        \begin{tabular}{cM{70mm}M{70mm}M{70mm}}
           \toprule
            Key. & a & b & c  \\
            \midrule
            i & 
            \includegraphics[width=69.5mm]{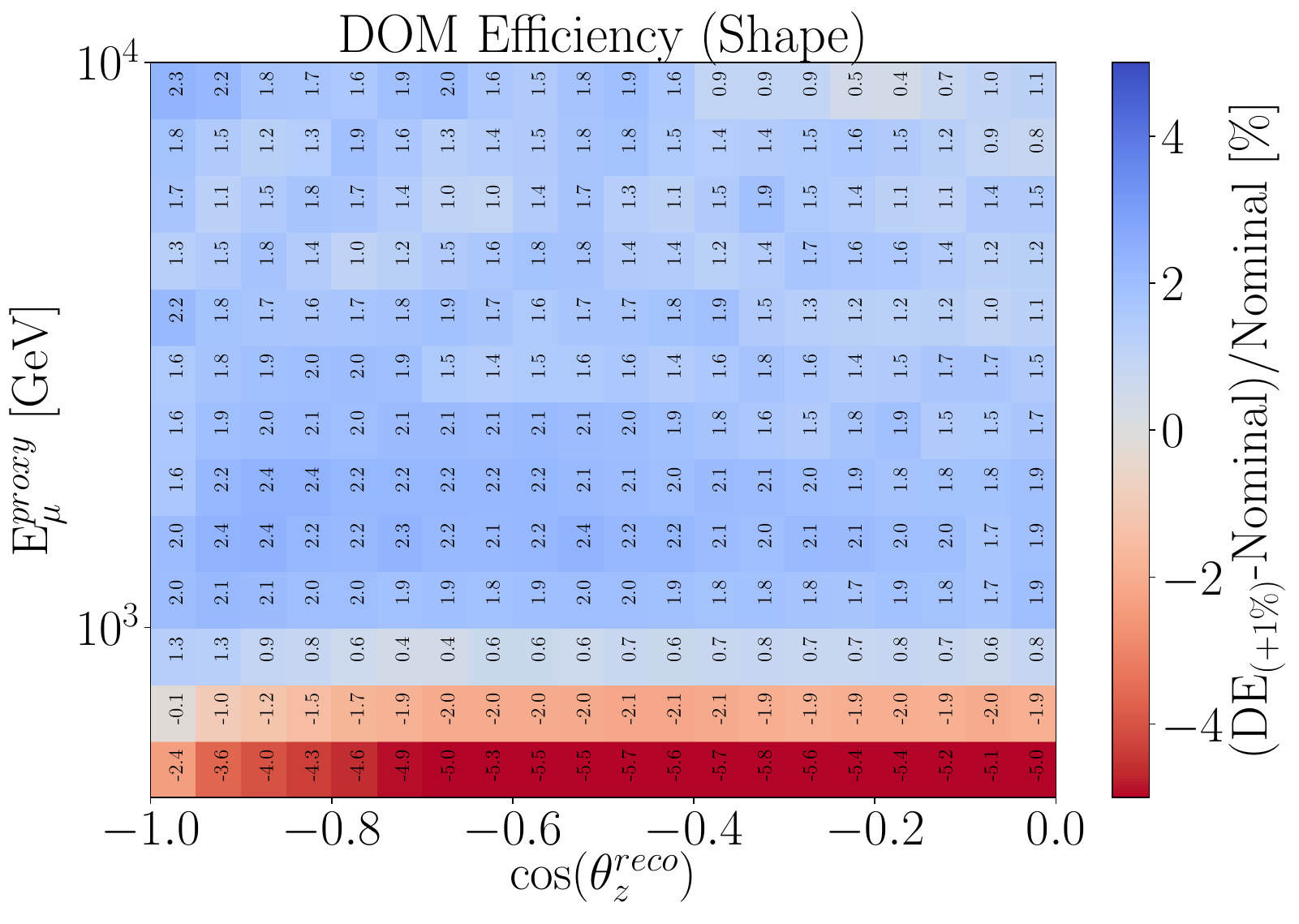} &
            \includegraphics[width=69.5mm]{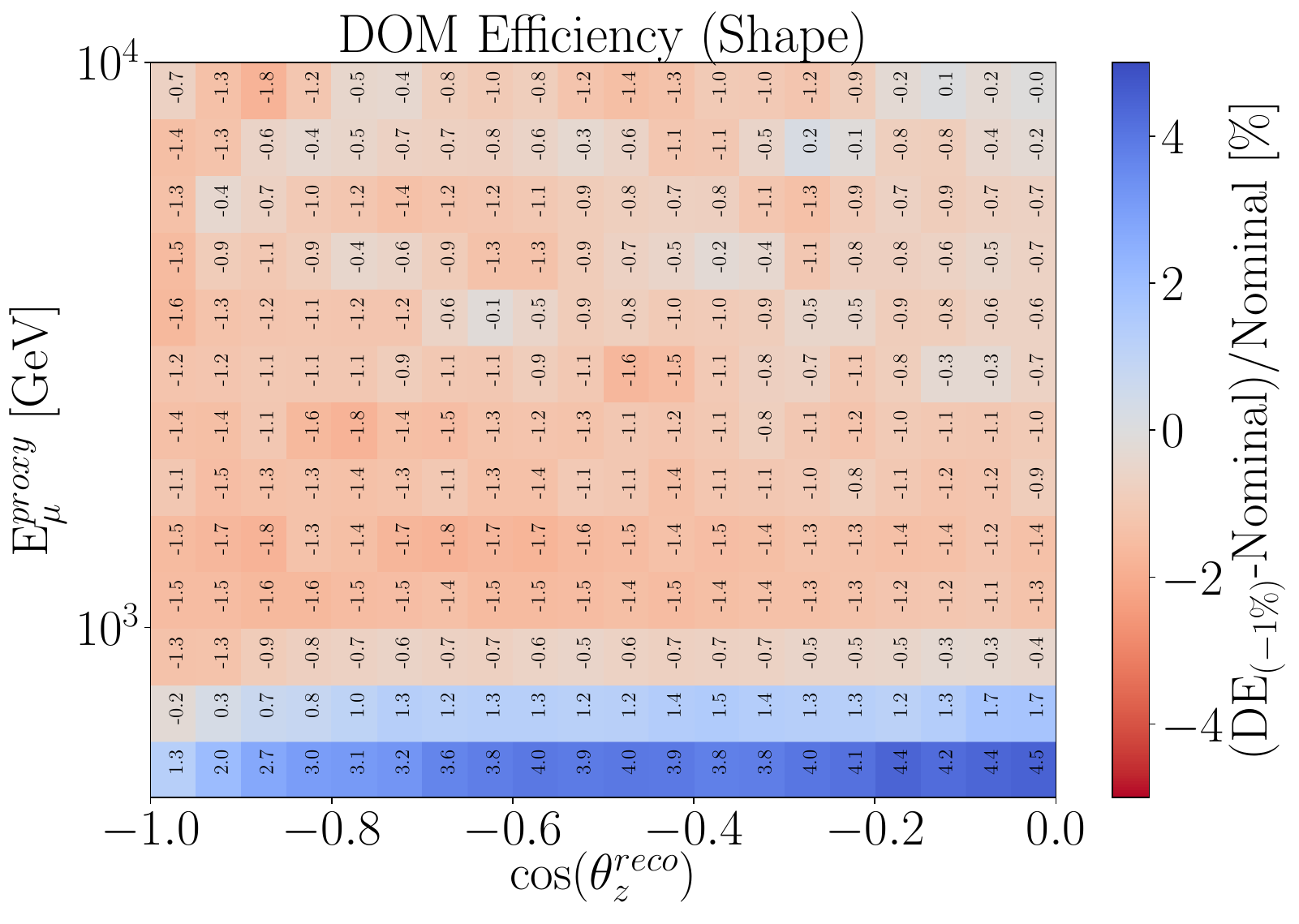} & 
            \includegraphics[width=69.5mm]{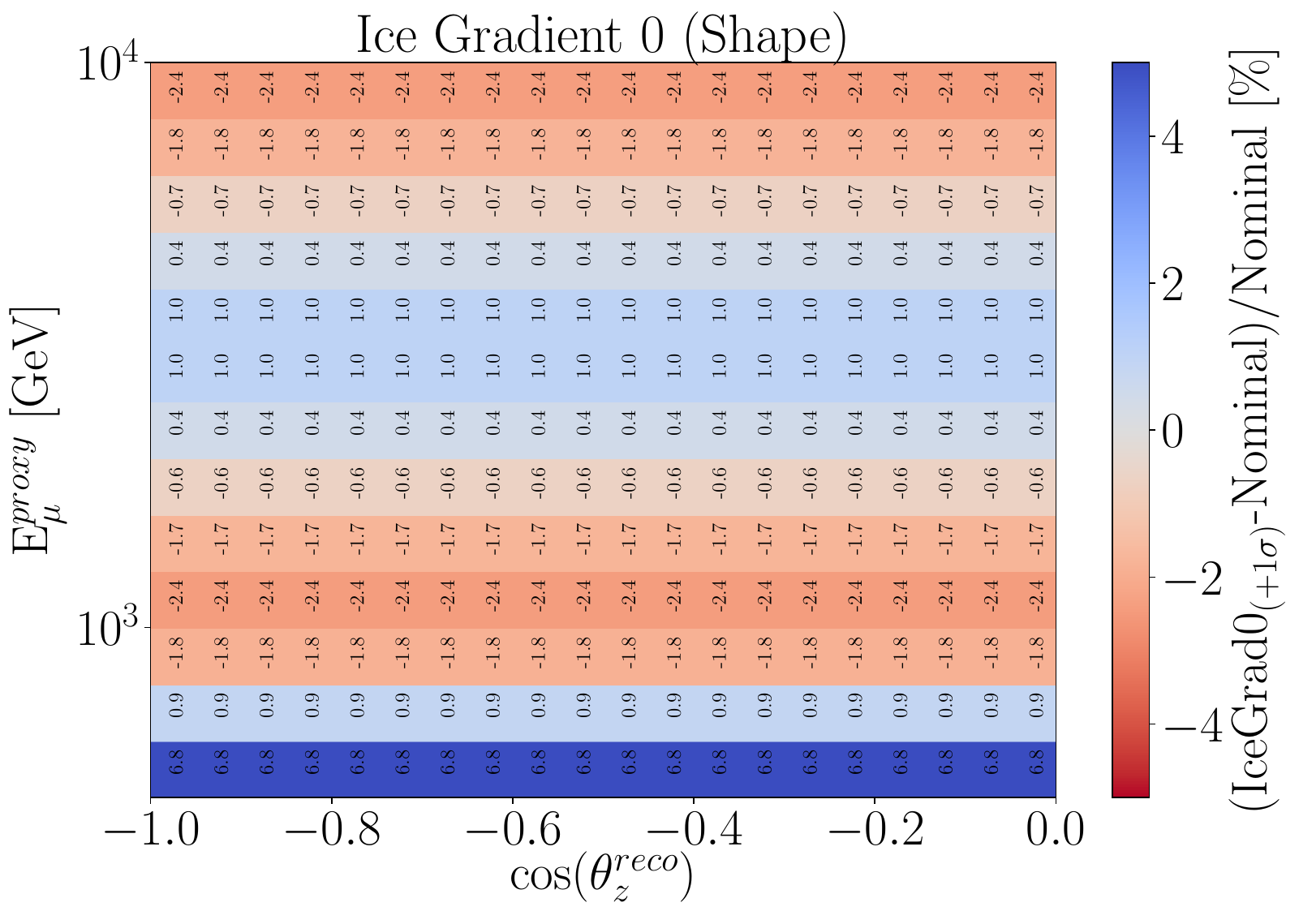}  \\
            ii & 
            \includegraphics[width=69.5mm]{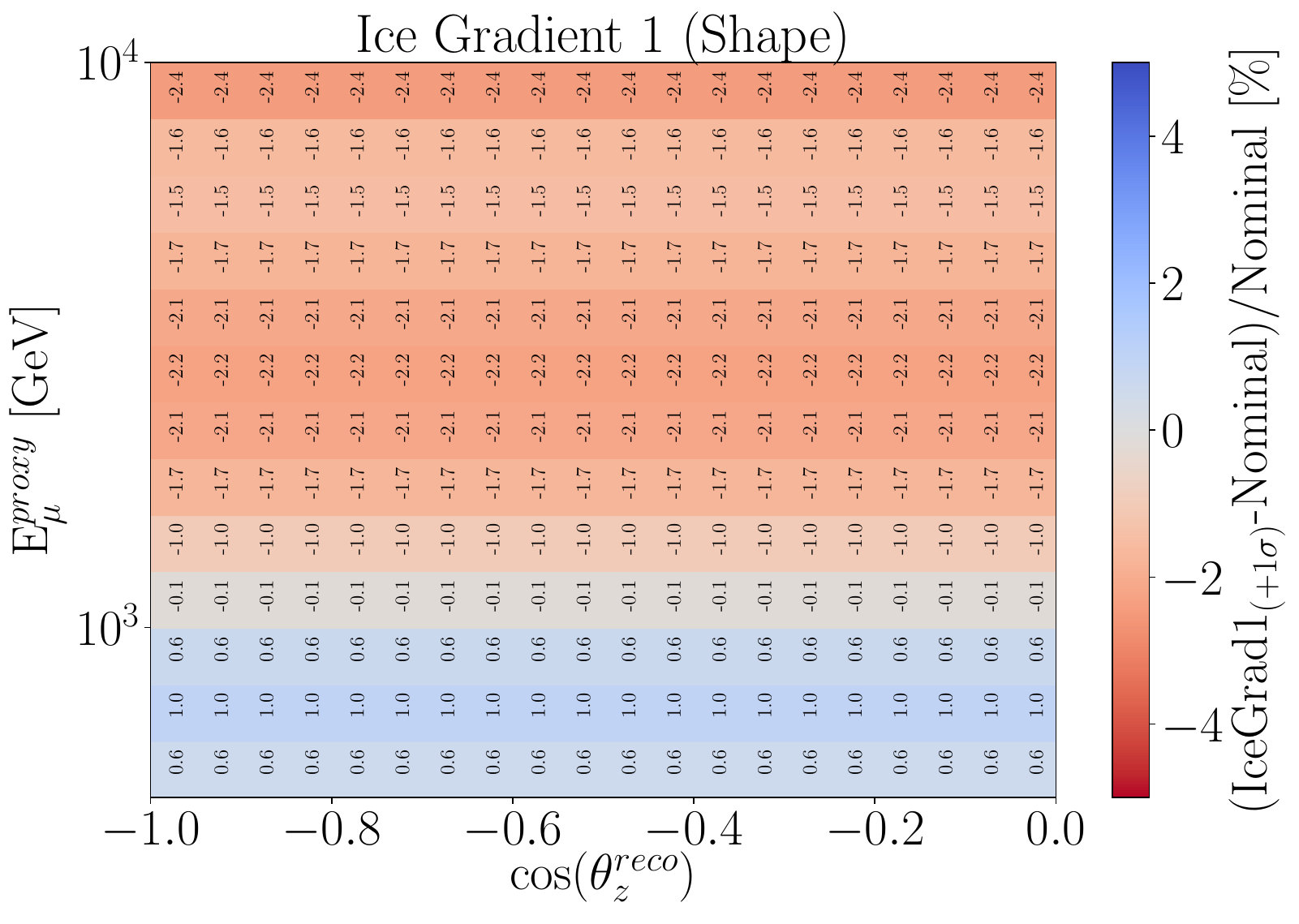} & 
            \includegraphics[width=69.5mm]{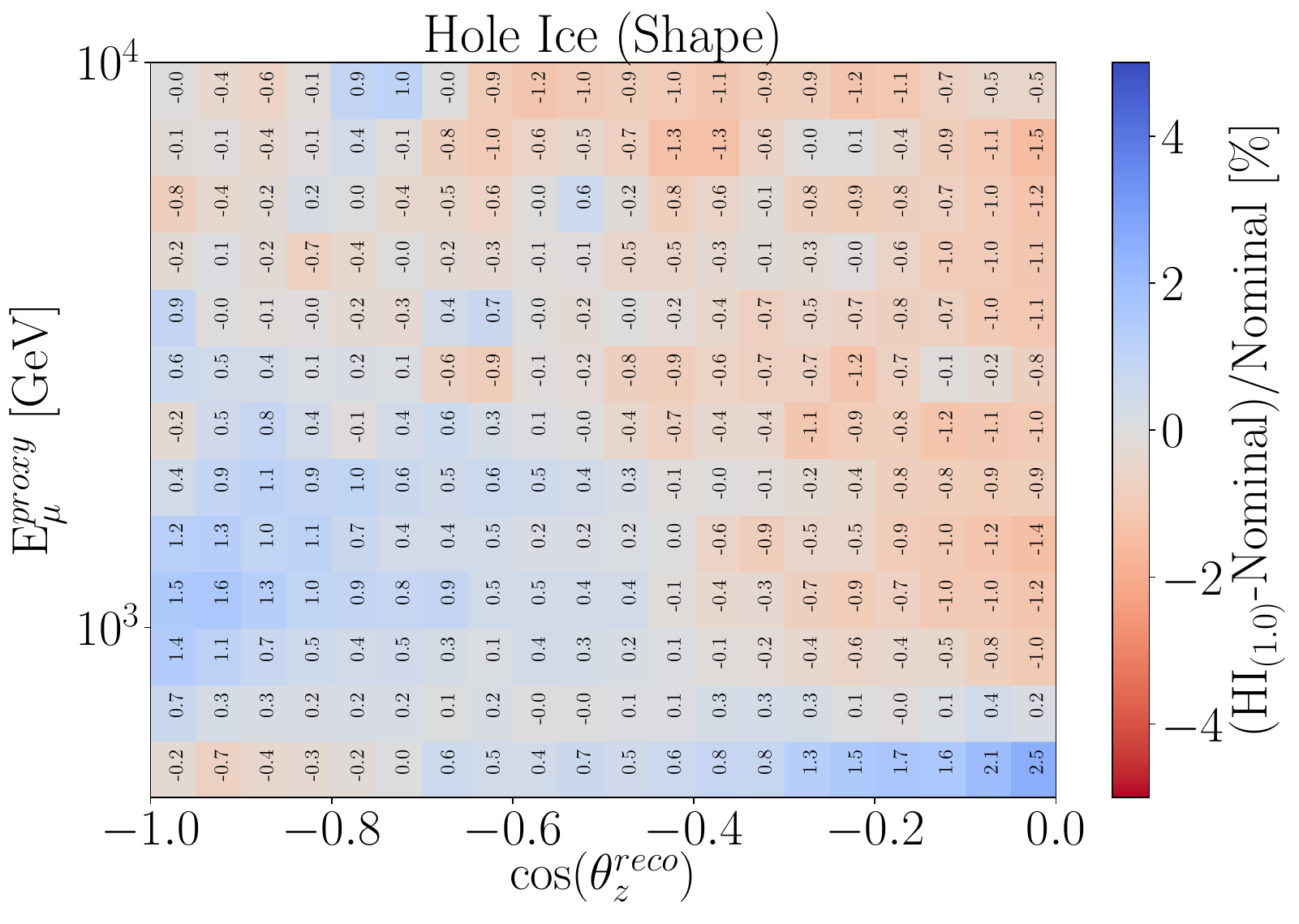} & 
            \includegraphics[width=69.5mm]{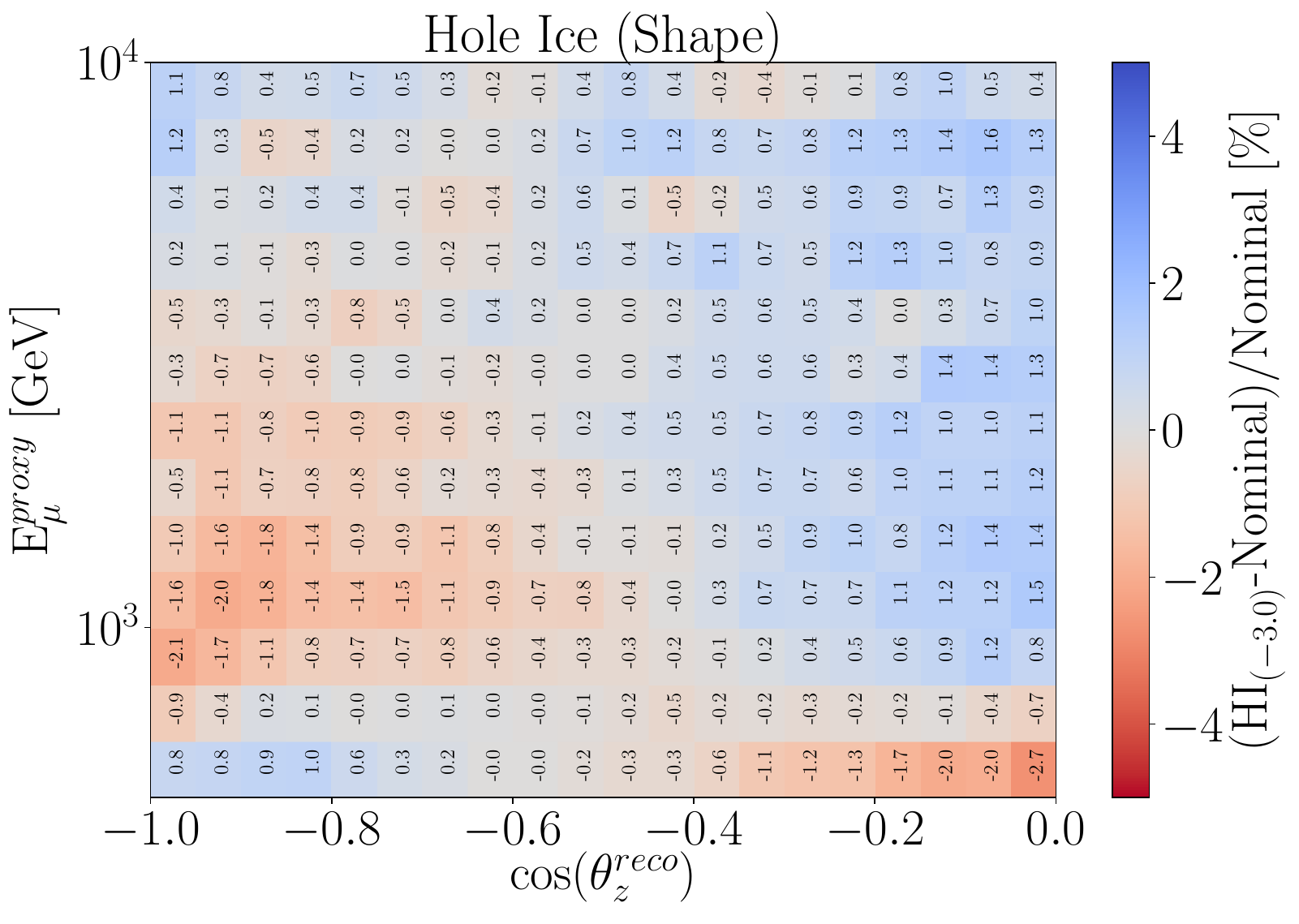}  \\
            iii & 
            \includegraphics[width=69.5mm]{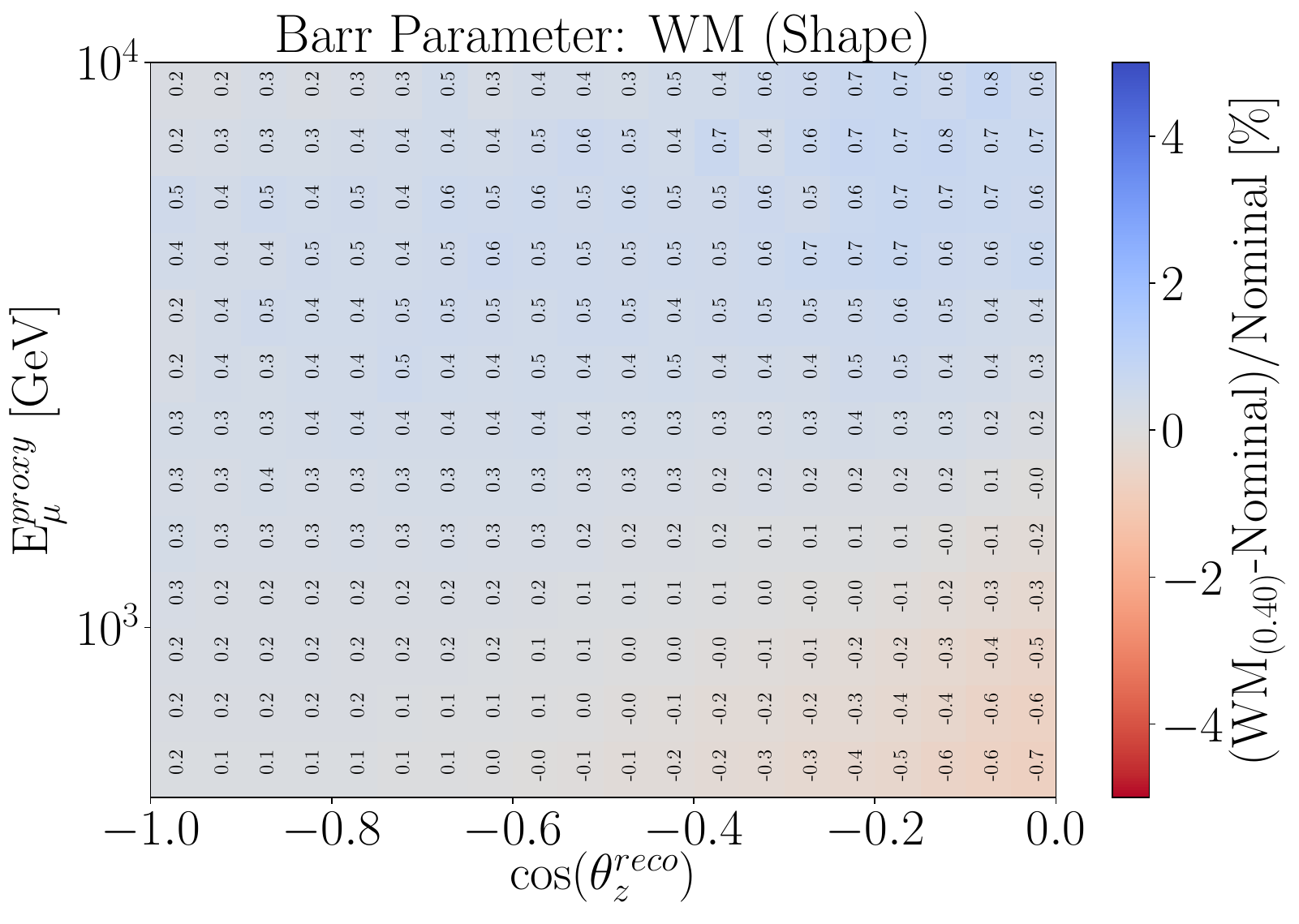} & 
            \includegraphics[width=69.5mm]{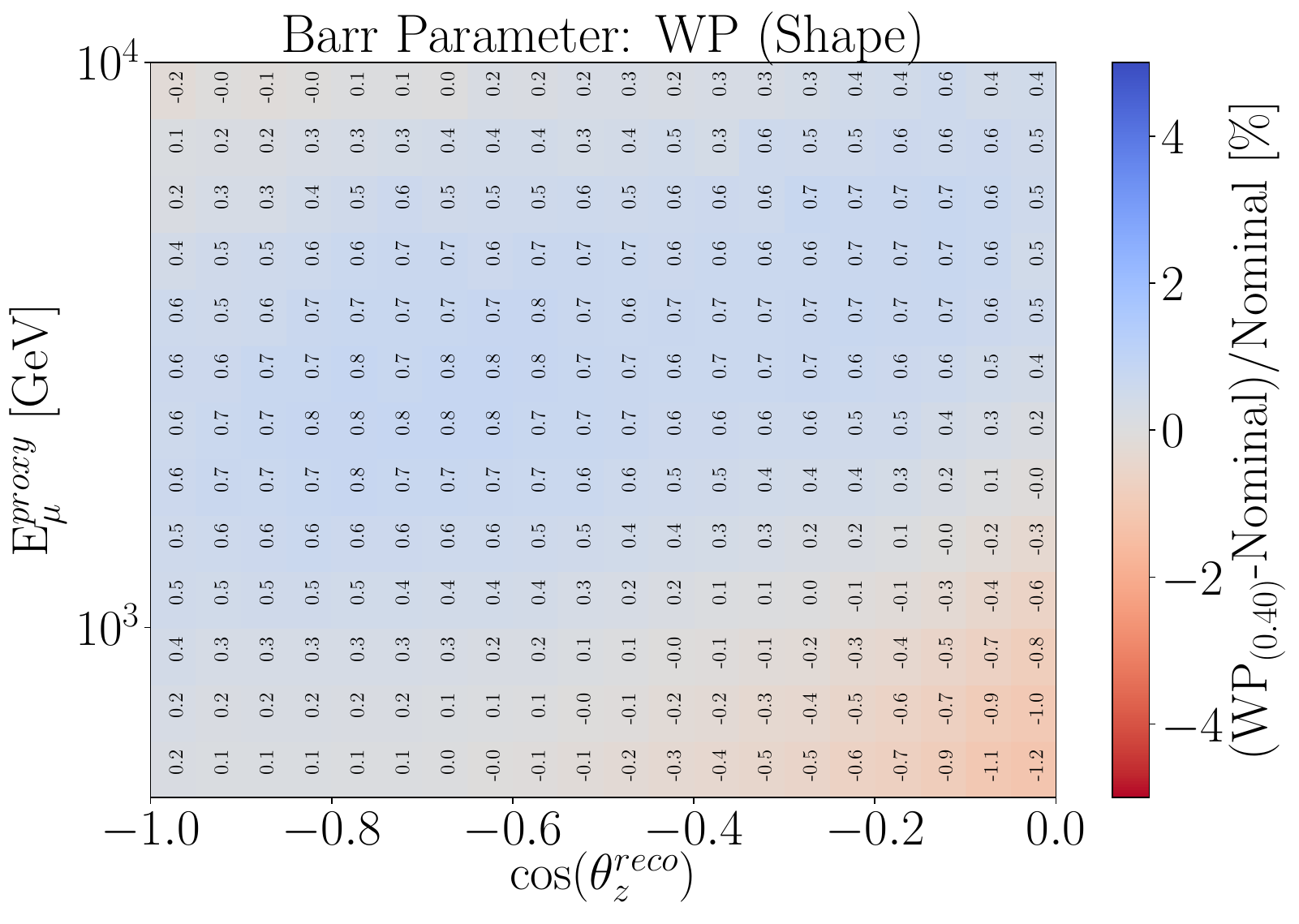} &  
            \includegraphics[width=69.5mm]{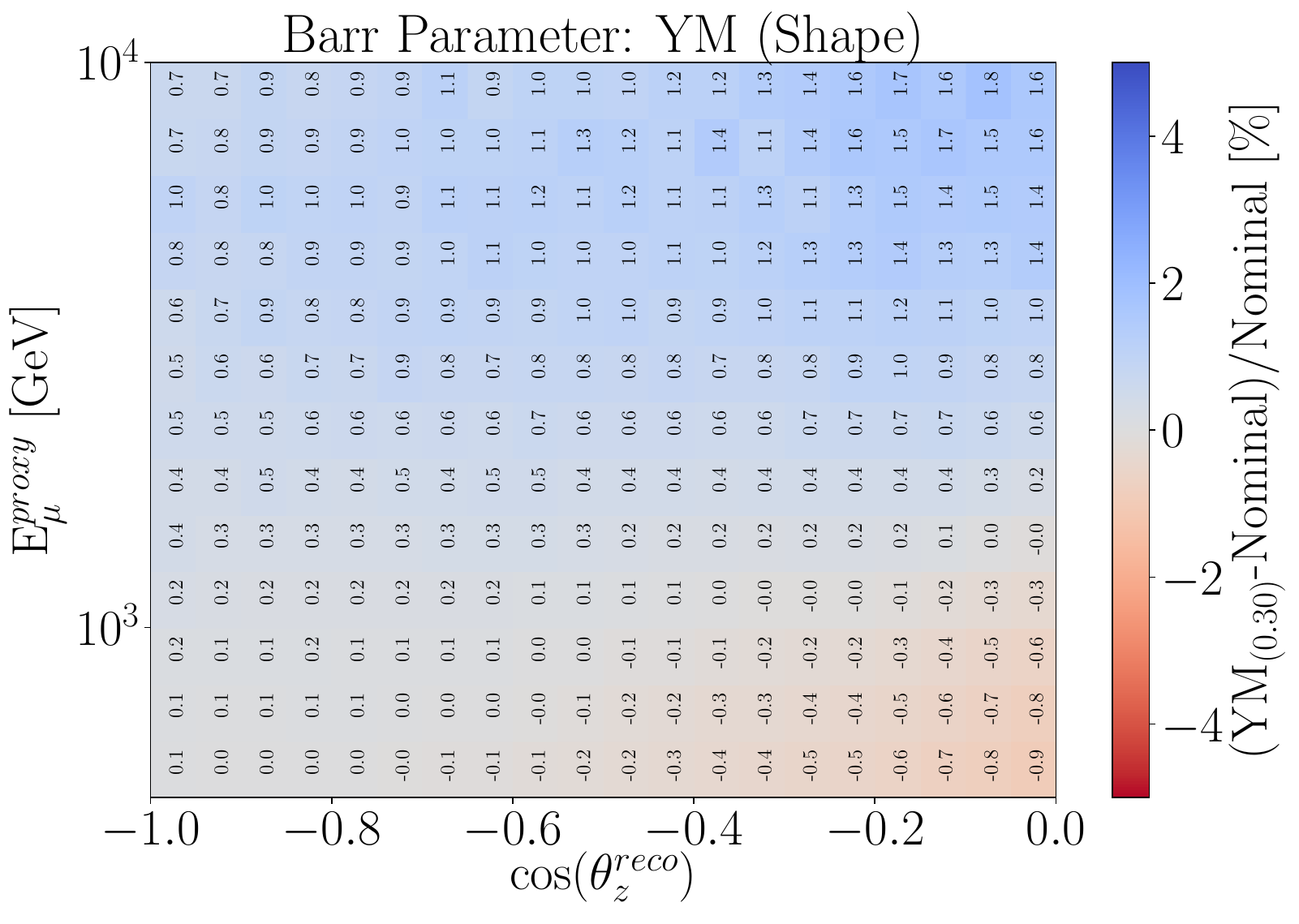}   \\
            \bottomrule
        \end{tabular}
        \caption{Effects of systematic nuisance parameters in analysis space (see text for details)}
        \label{tbl:syspanel1}
    \end{sidewaystable}
    
    \begin{sidewaystable}
        \centering
        Supplementary Material: Effect of Each Systematic Uncertainty (2)
        \begin{tabular}{cM{70mm}M{70mm}M{70mm}}
           \toprule
            Key. & a & b & c  \\
            \midrule
            iv & 
            \includegraphics[width=69.5mm]{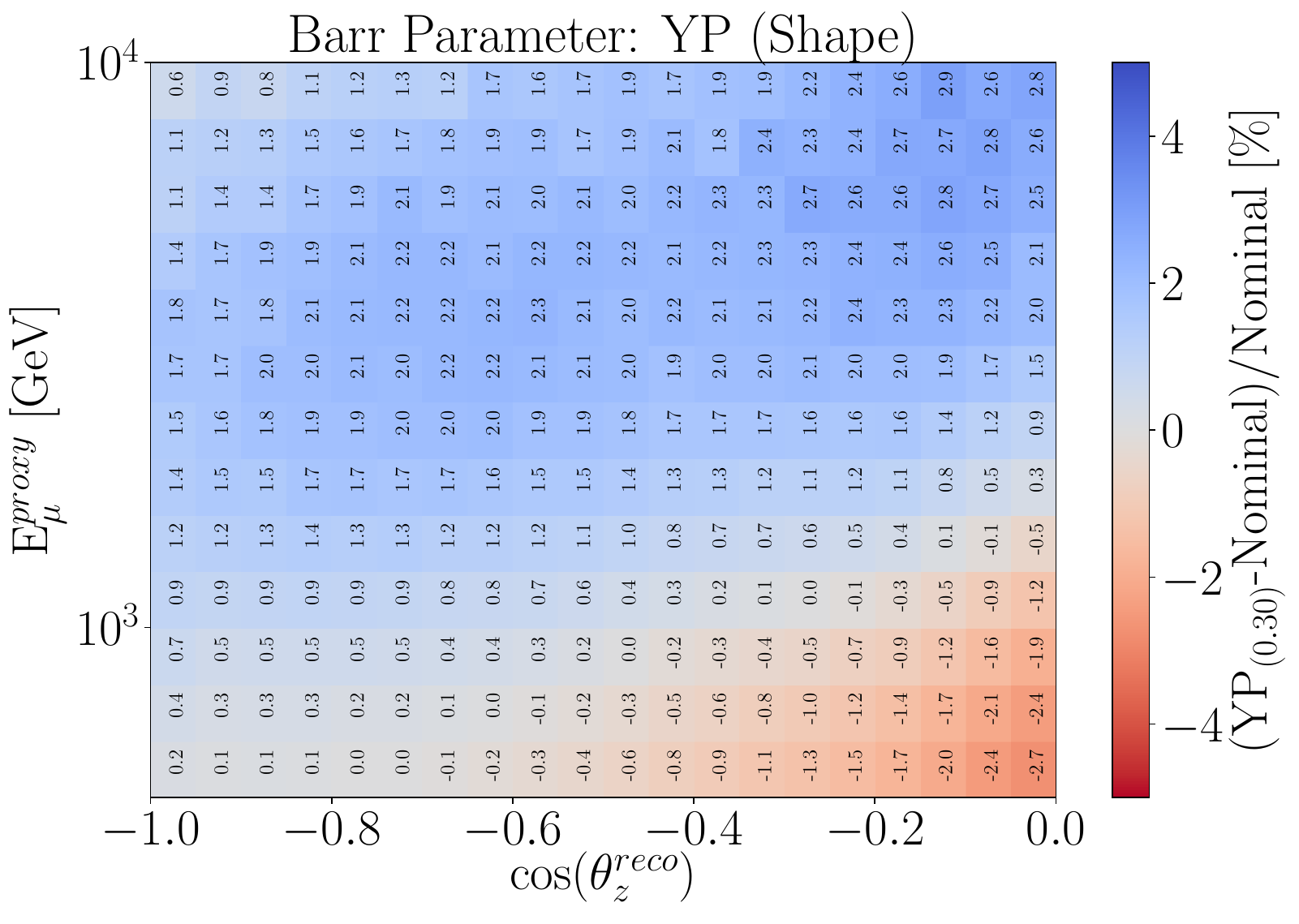} &
            \includegraphics[width=69.5mm]{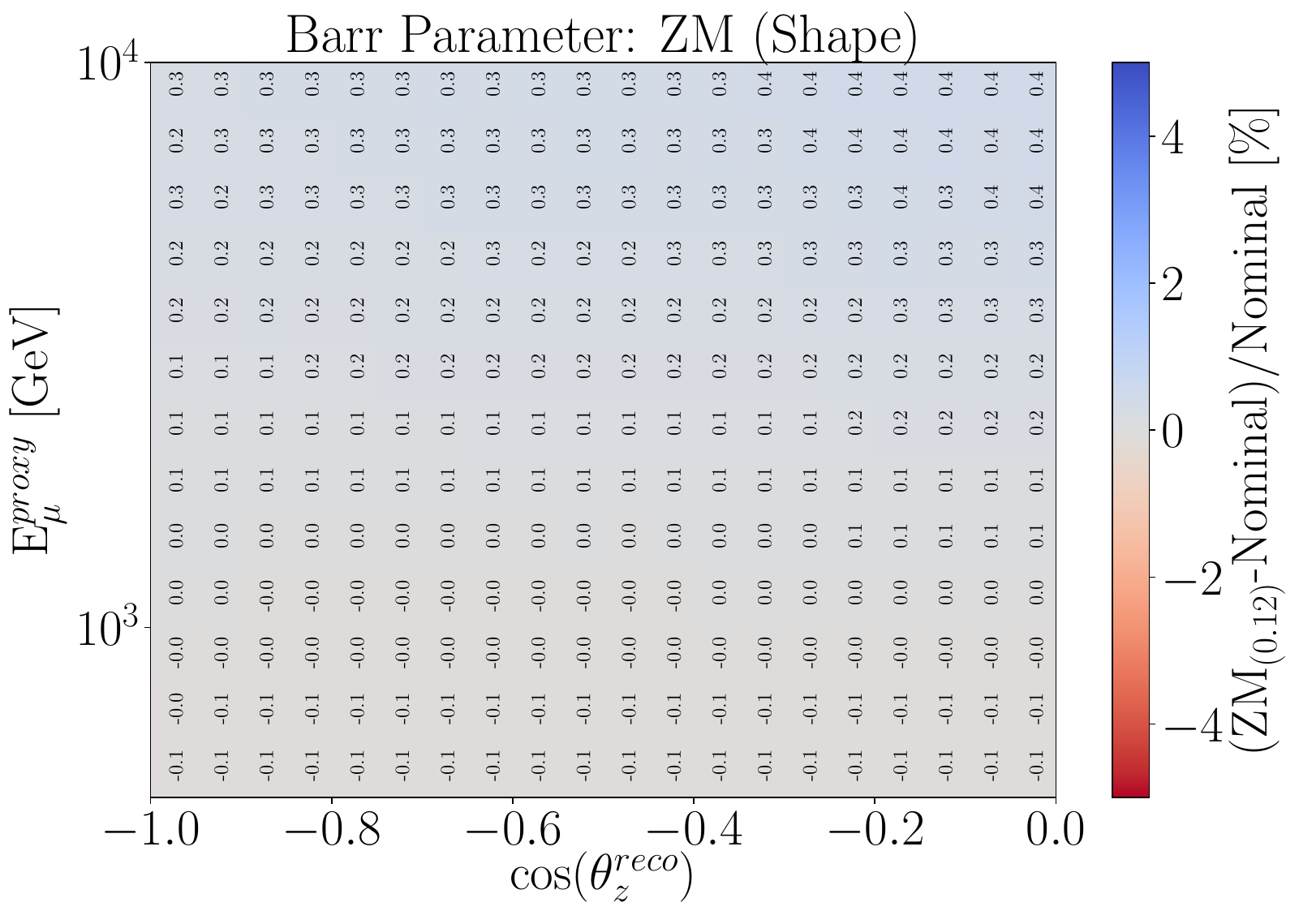} & 
            \includegraphics[width=69.5mm]{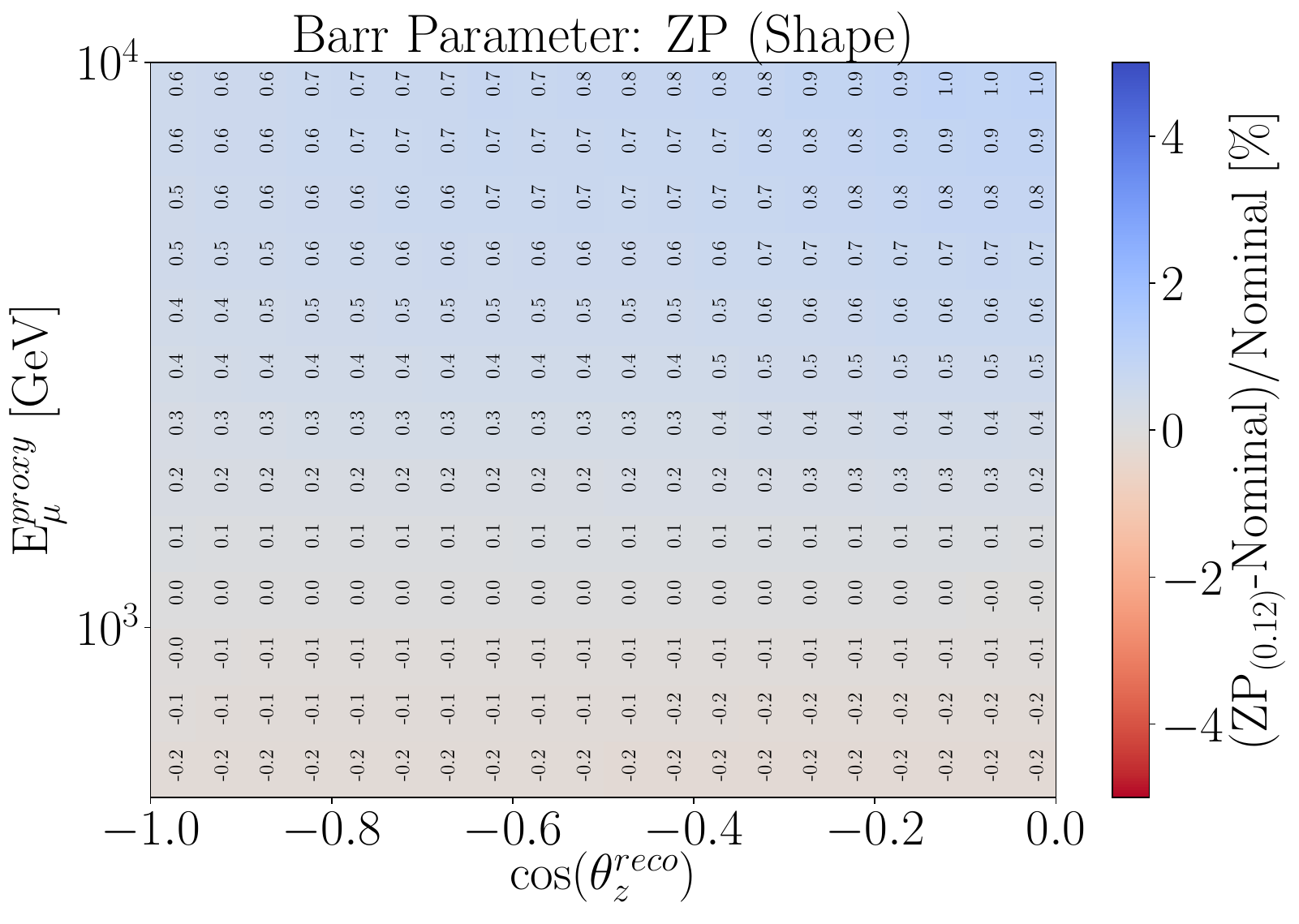}  \\
            v & 
            \includegraphics[width=69.5mm]{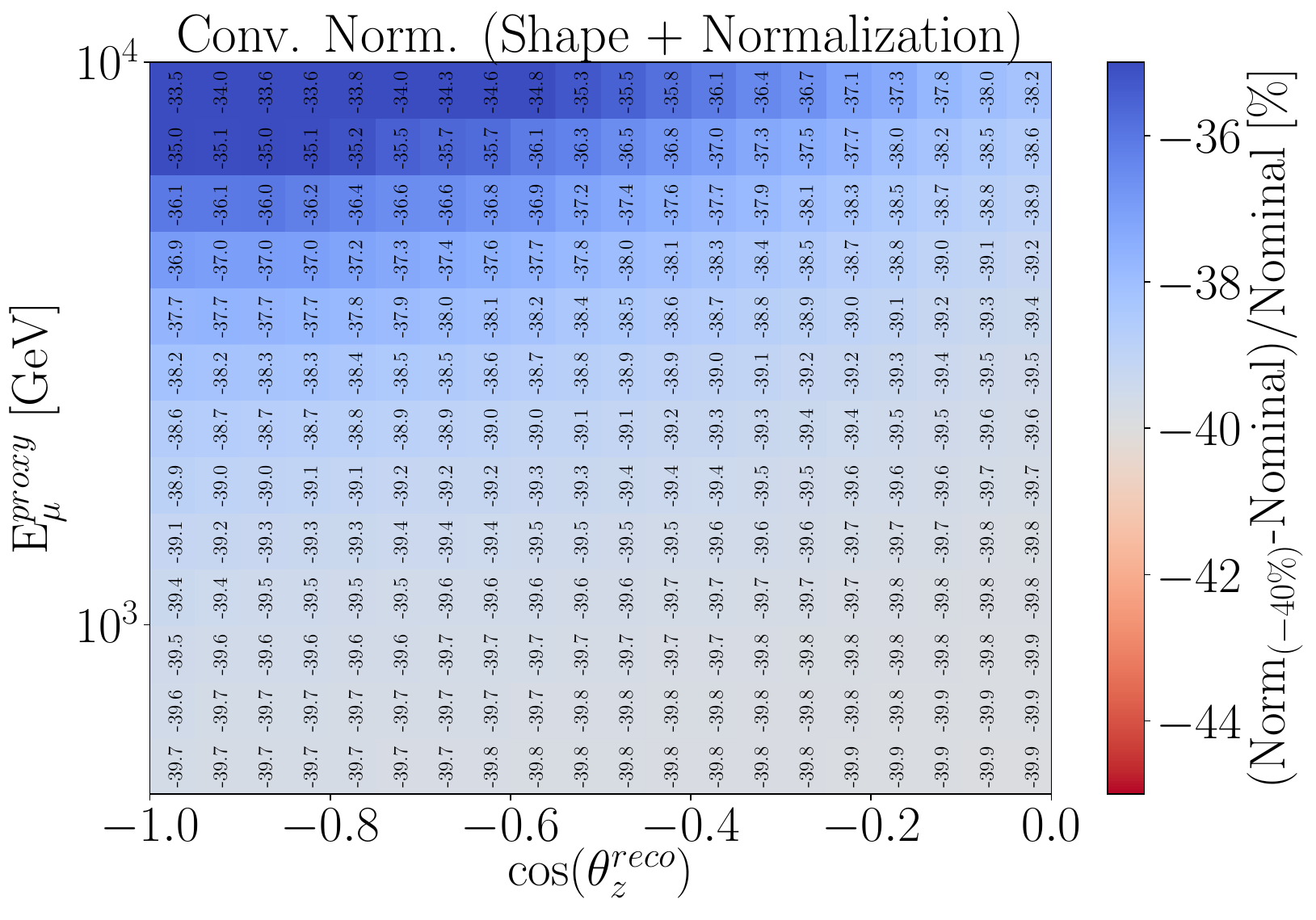} & 
            \includegraphics[width=69.5mm]{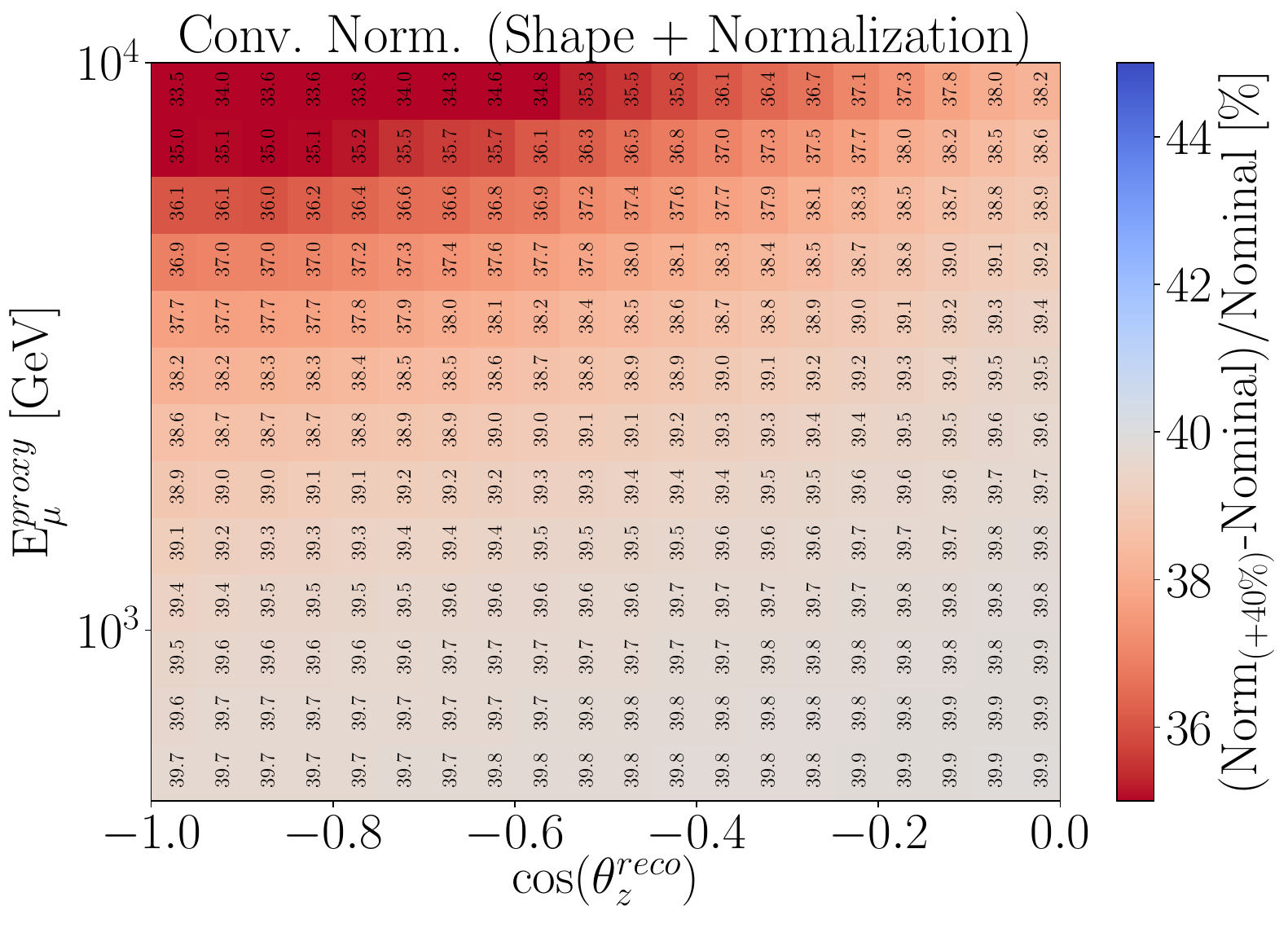} & 
            \includegraphics[width=69.5mm]{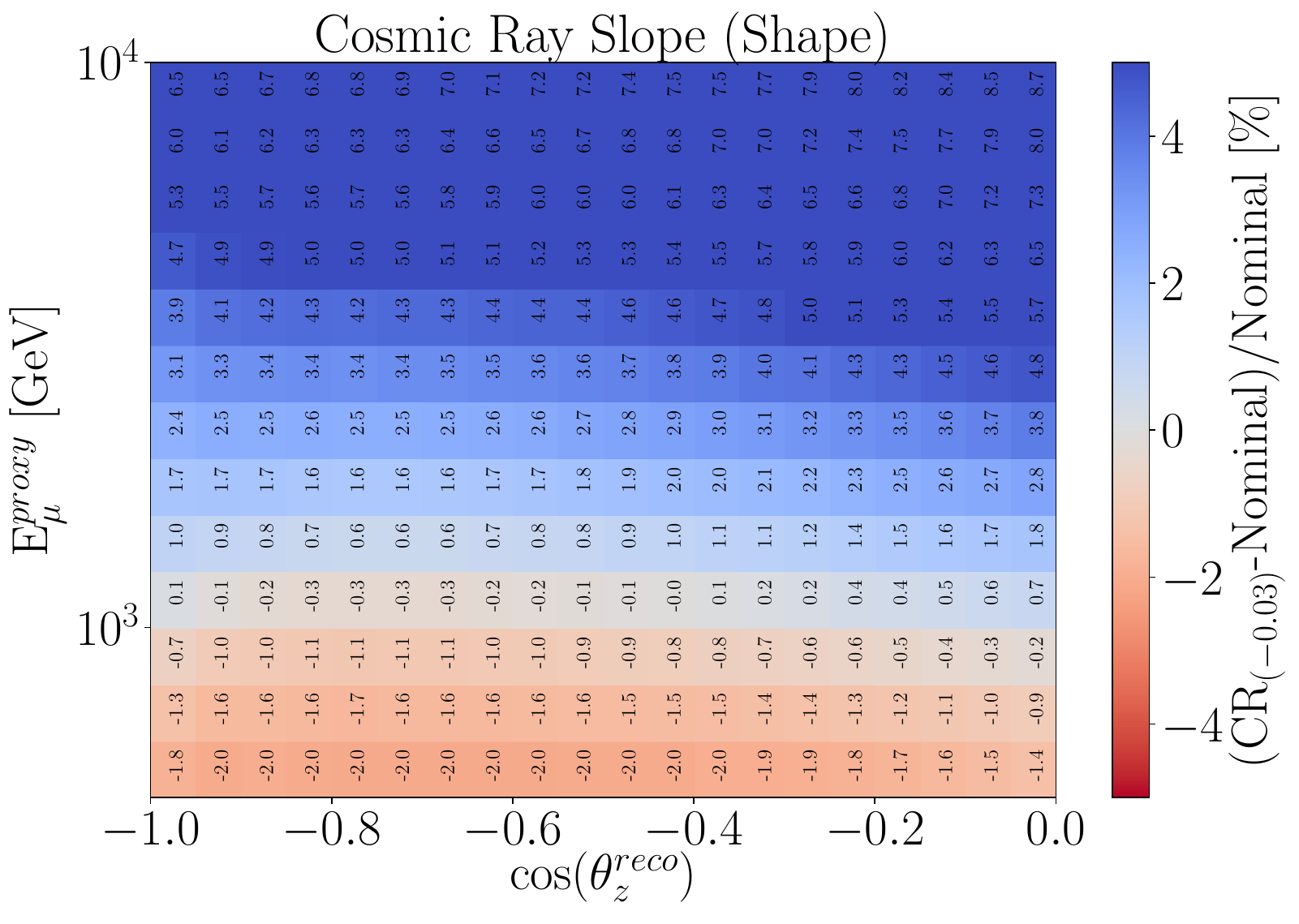}  \\
            vi & 
            \includegraphics[width=69.5mm]{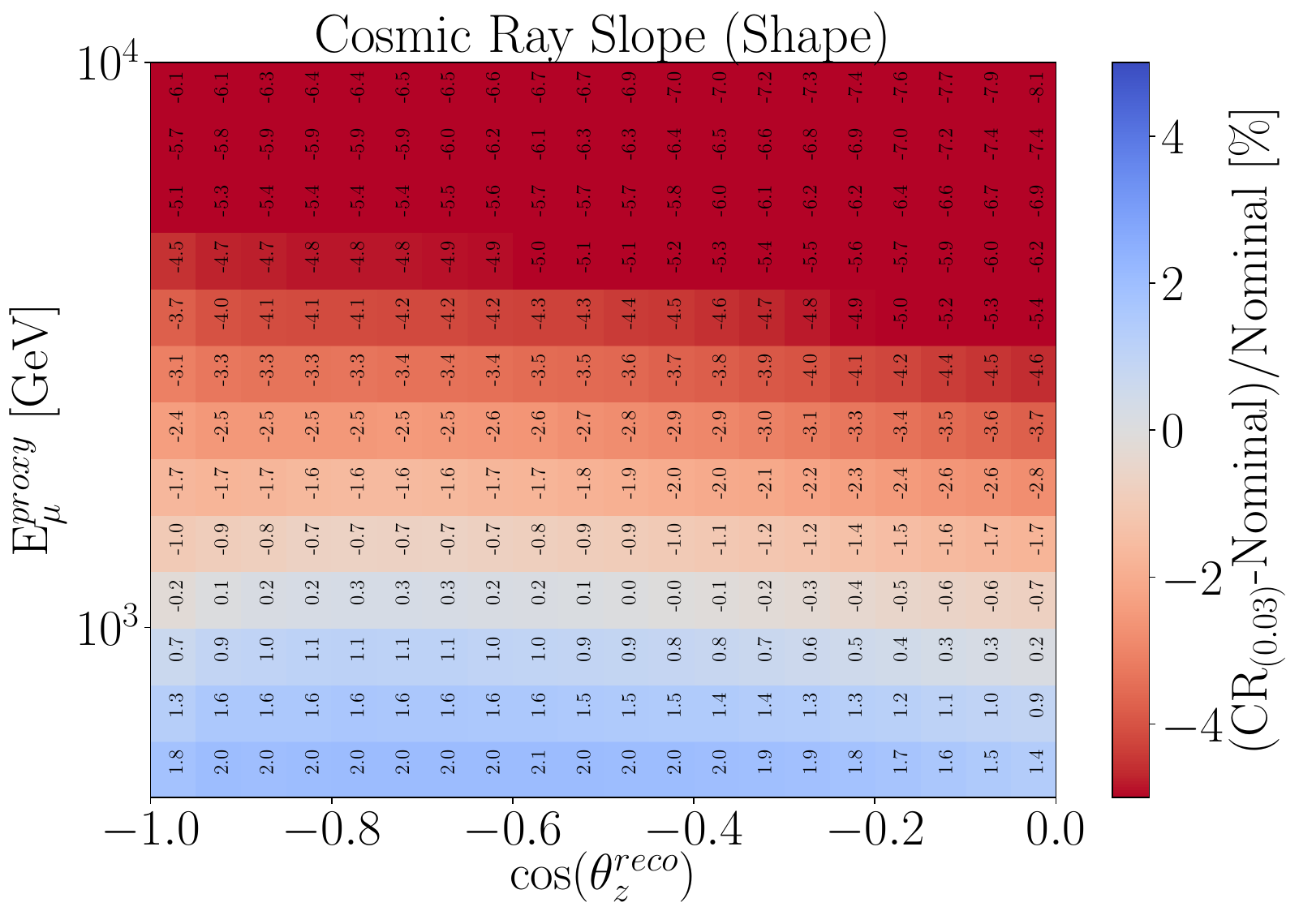} & 
            \includegraphics[width=69.5mm]{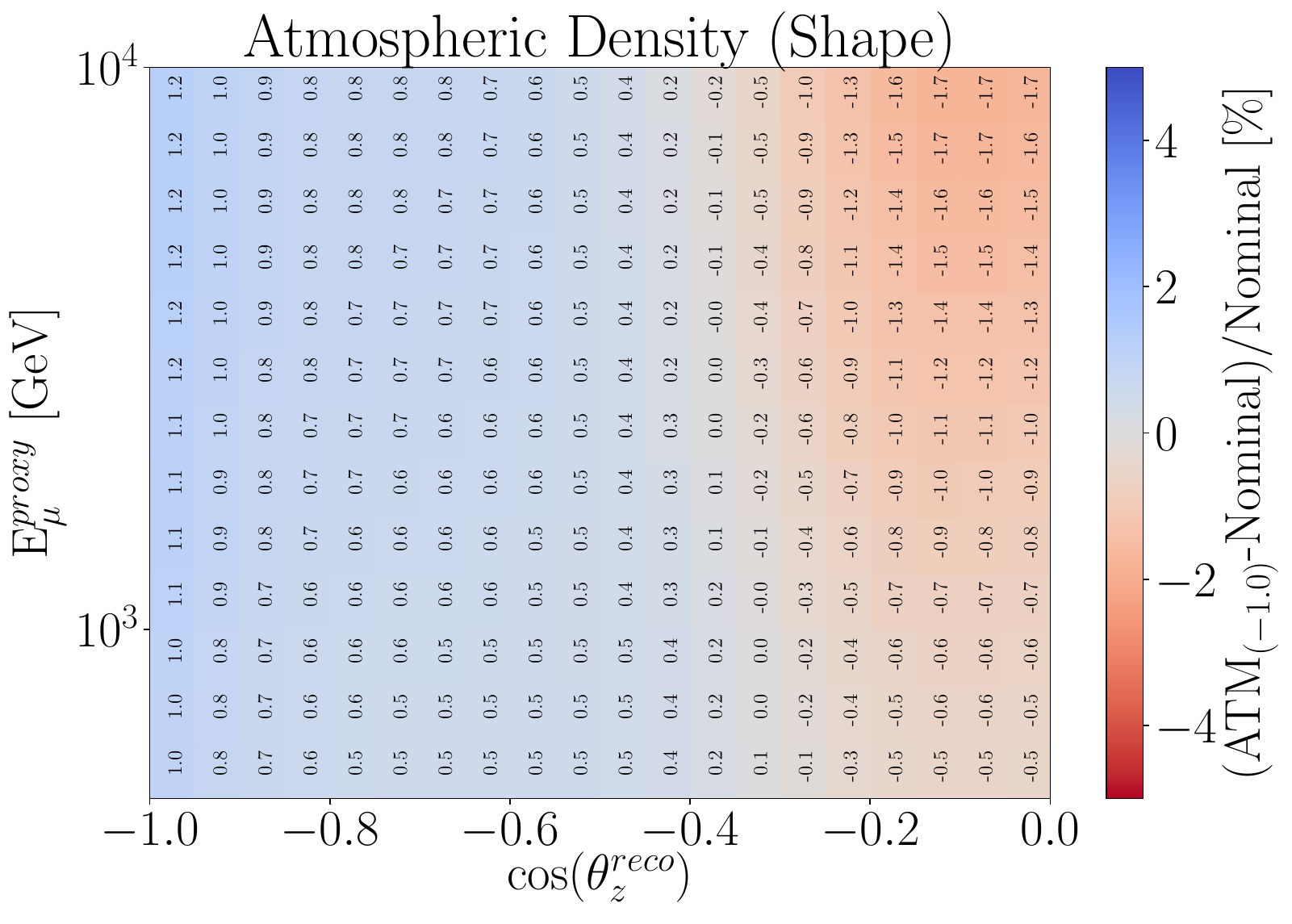} &  
            \includegraphics[width=69.5mm]{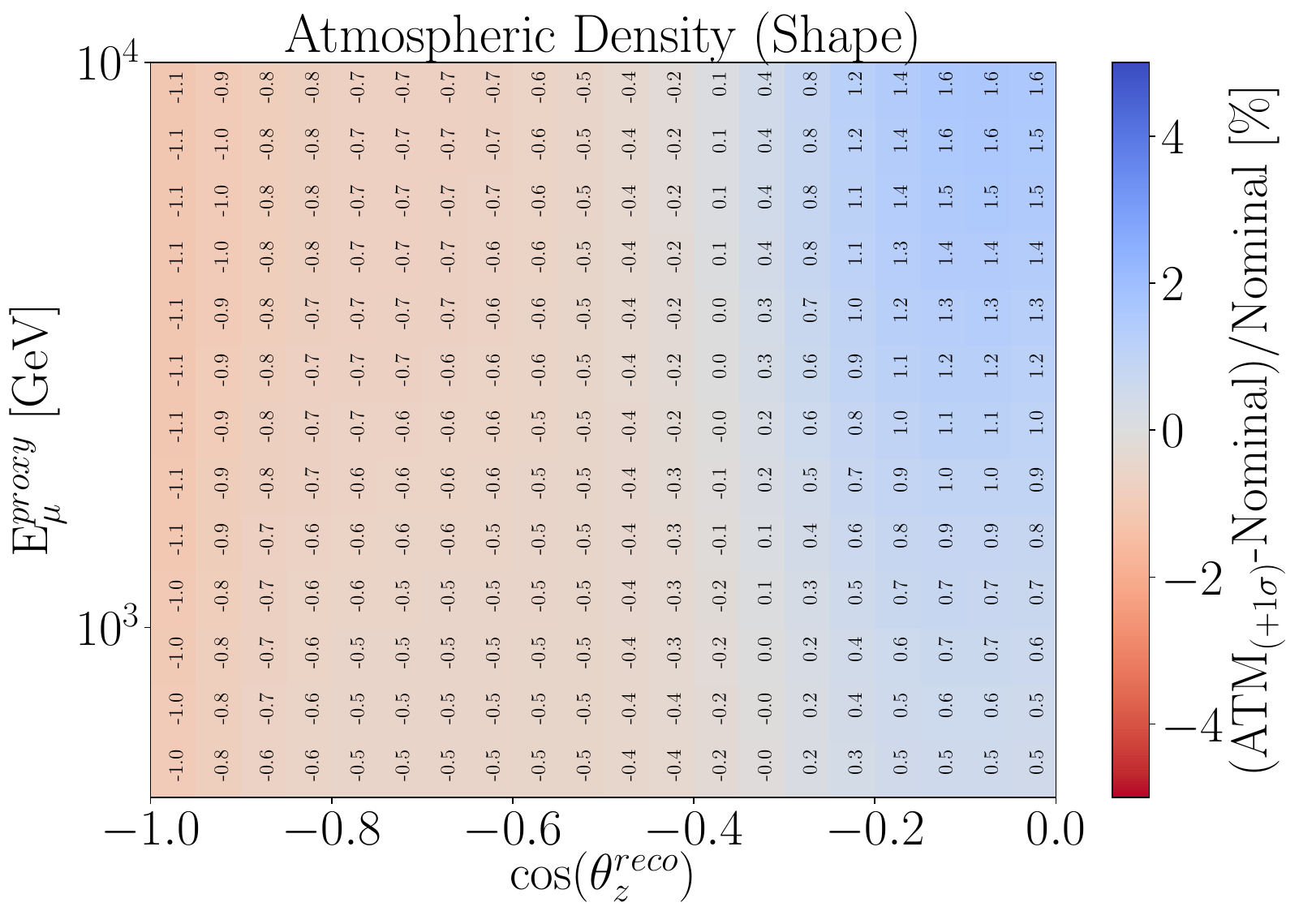}  \\
            \bottomrule
        \end{tabular}
        \caption{Effects of systematic nuisance parameters in analysis space (see text for details)}
        \label{tbl:syspanel2}
    \end{sidewaystable}

        \begin{sidewaystable}
        \centering
        Supplementary Material: Effect of Each Systematic Uncertainty (3)
        \begin{tabular}{cM{70mm}M{70mm}M{70mm}}
           \toprule
            Key. & a & b & c  \\
            \midrule
            vii & 
            \includegraphics[width=69.5mm]{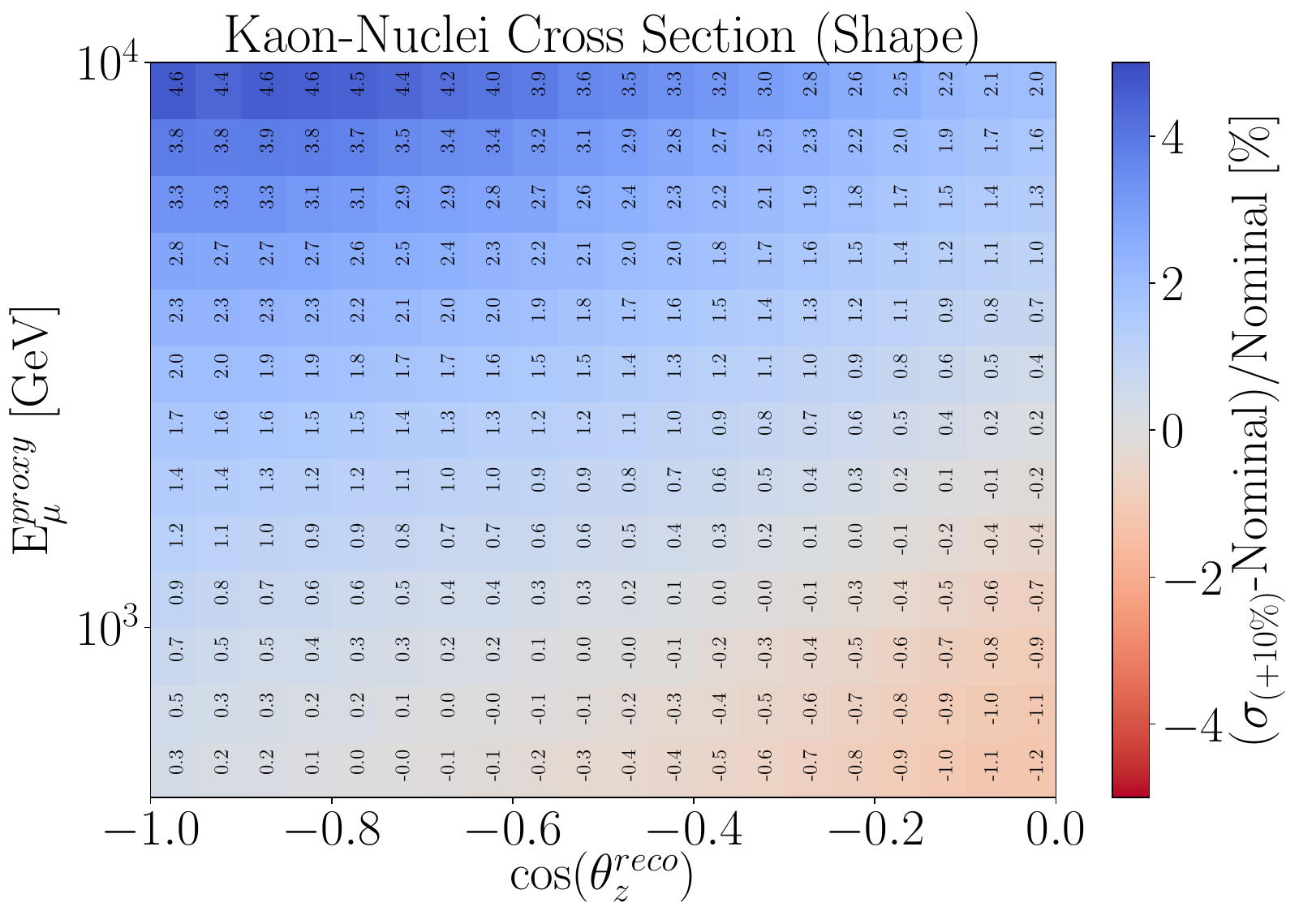} &
            \includegraphics[width=69.5mm]{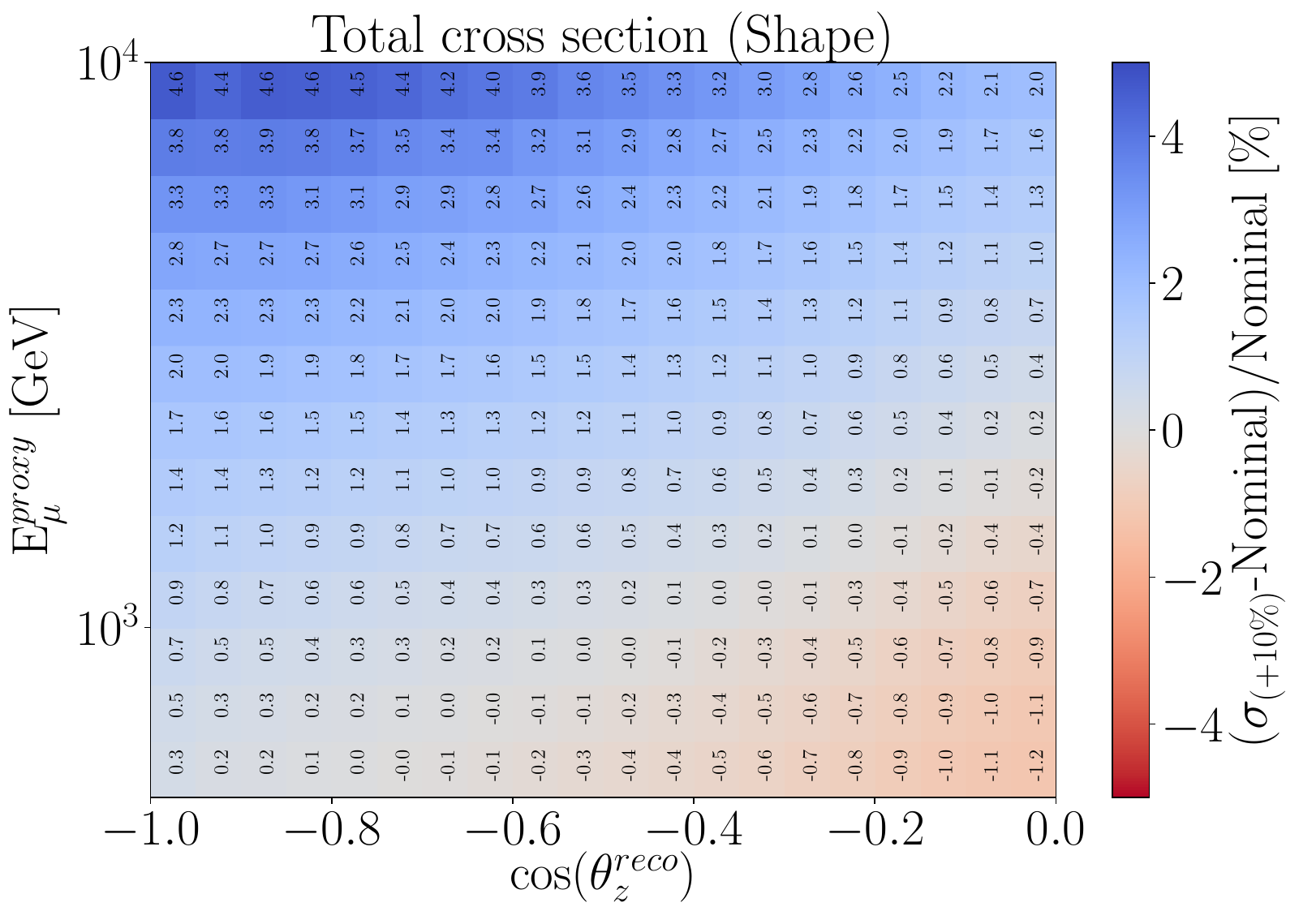} & 
            \includegraphics[width=69.5mm]{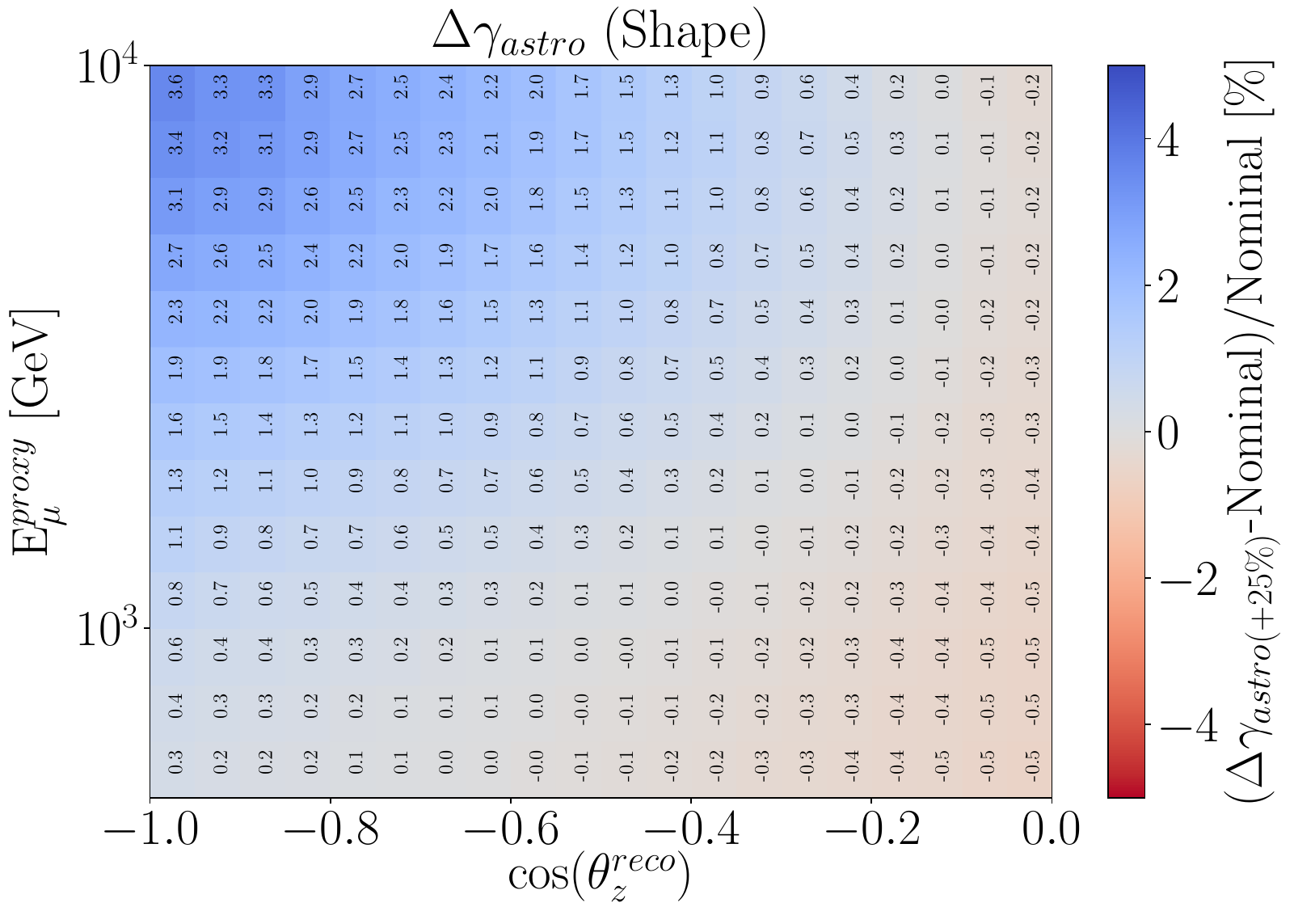}  \\
            viii & 
            \includegraphics[width=69.5mm]{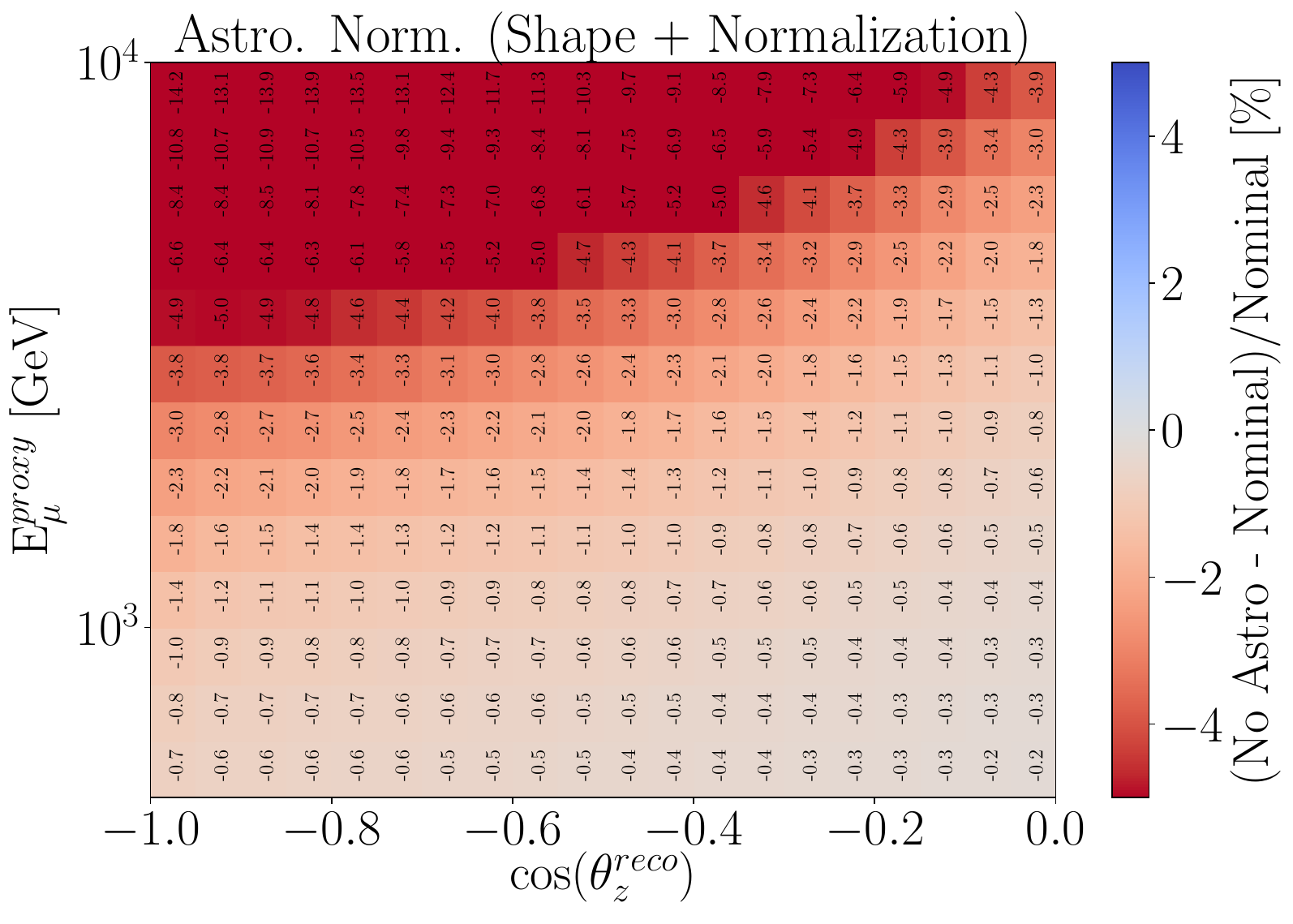} & 
            \includegraphics[width=69.5mm]{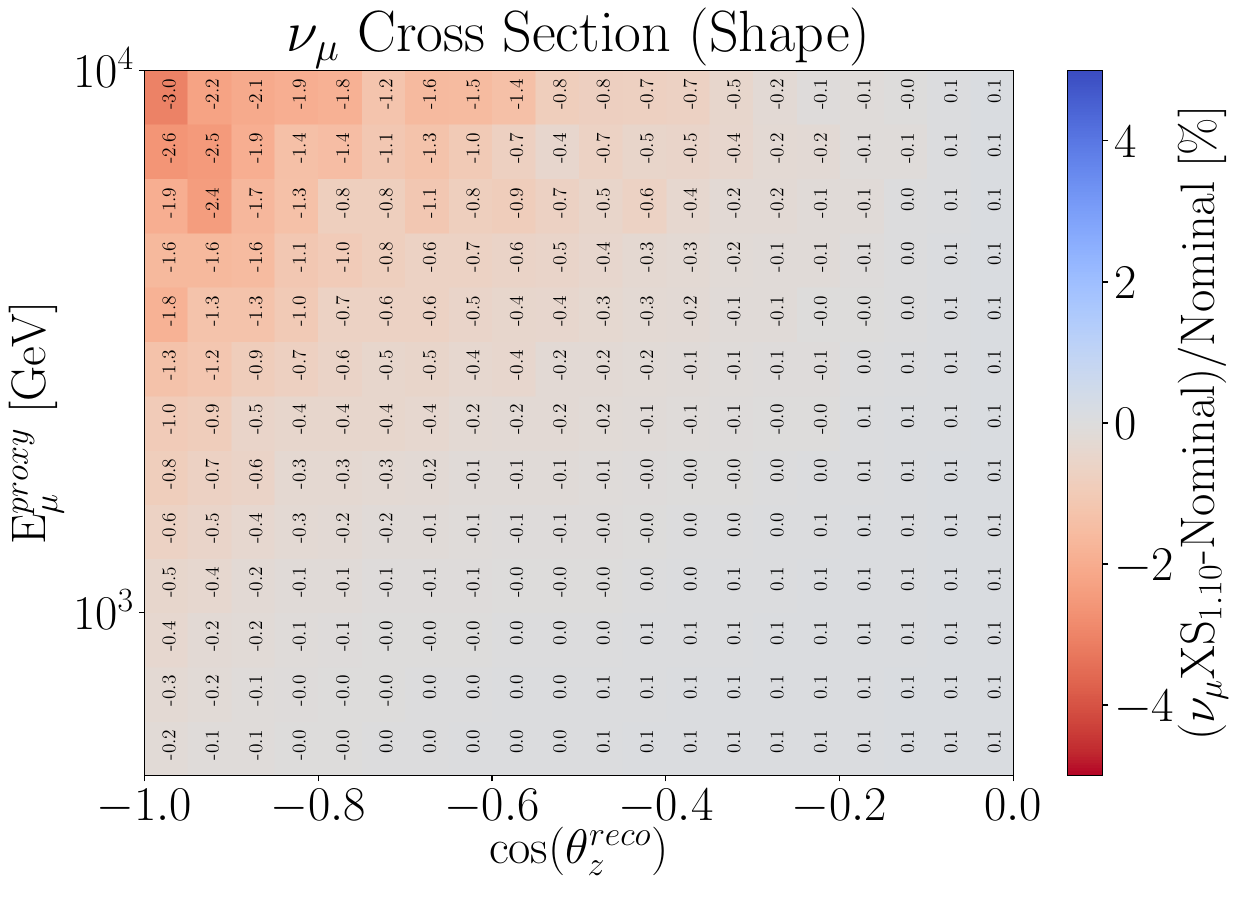} & 
            \includegraphics[width=69.5mm]{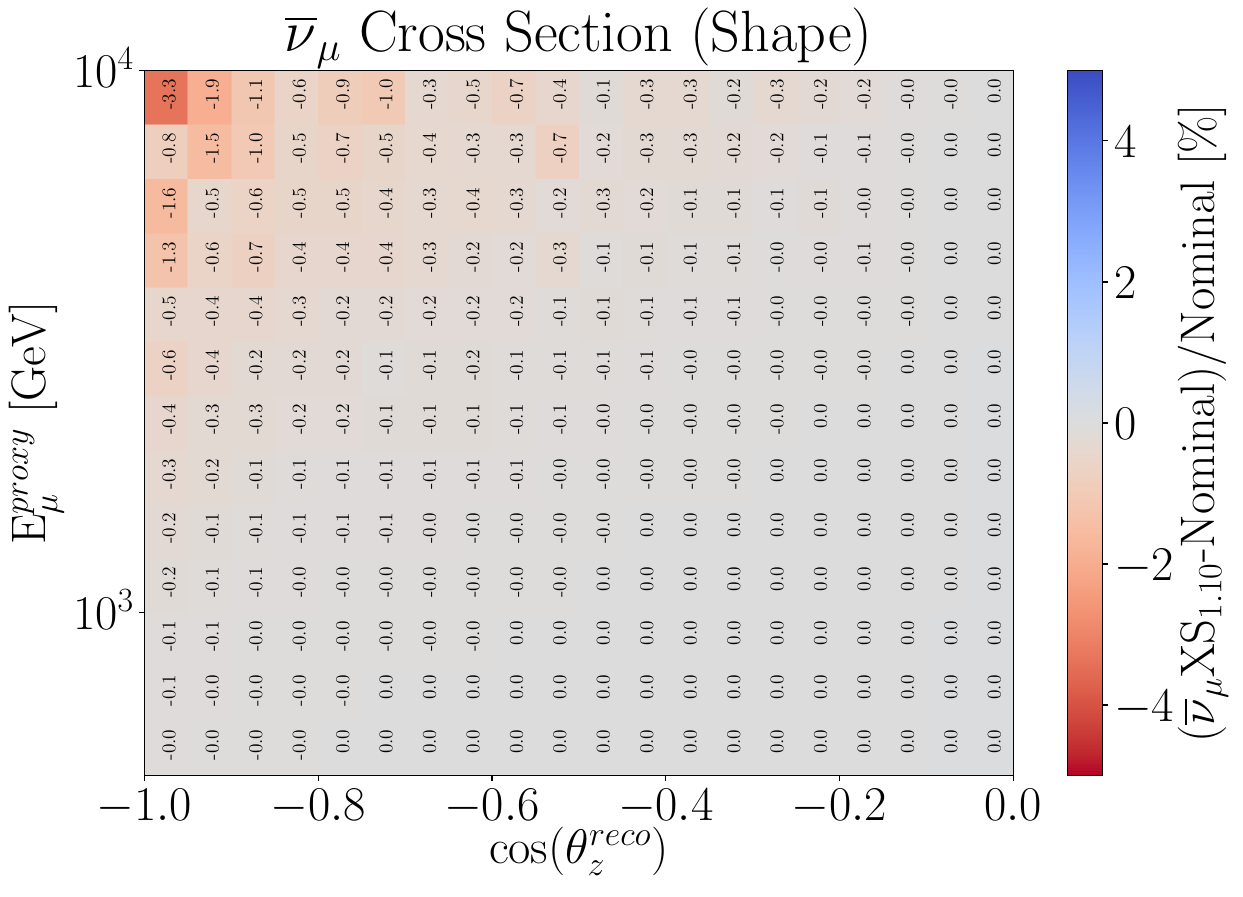}  \\
           
            \bottomrule
        \end{tabular}
        \caption{Effects of systematic nuisance parameters in analysis space (see text for details)}
        \label{tbl:syspanel3}
    \end{sidewaystable}

\section*{Supplementary figures}

\begin{figure*}[b!]
    \centering
    \begin{minipage}{0.24\textwidth}
        \centering
        \includegraphics[width=0.98\linewidth]{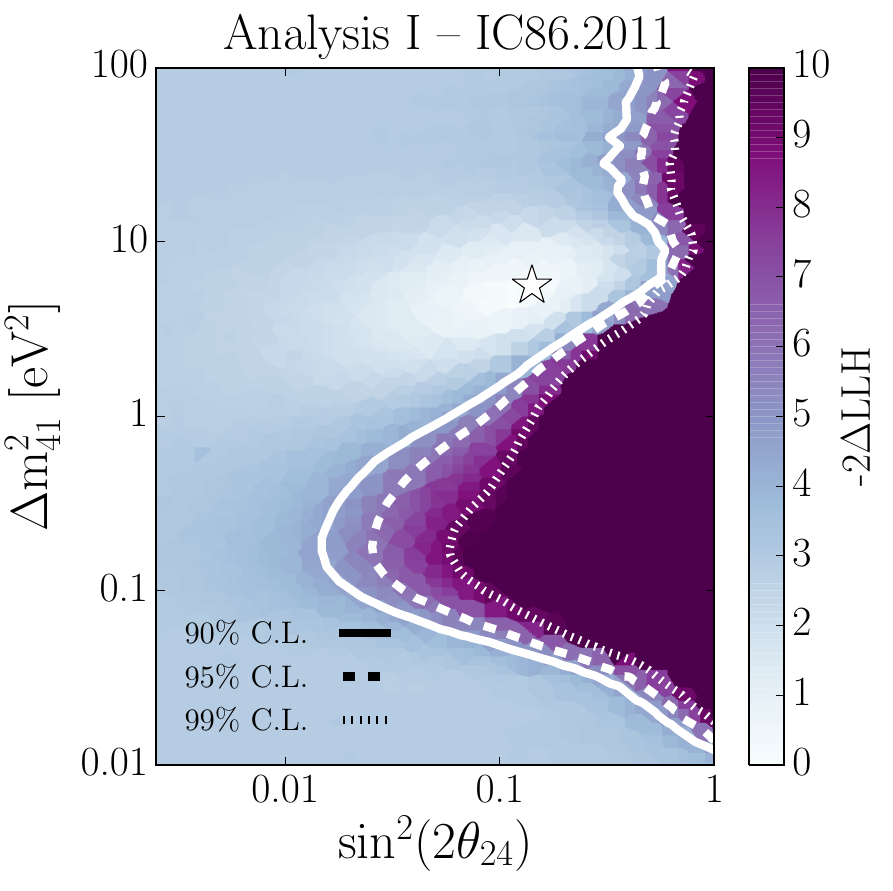}
    \end{minipage}%
    \begin{minipage}{0.24\textwidth}
        \centering
        \includegraphics[width=0.98\linewidth]{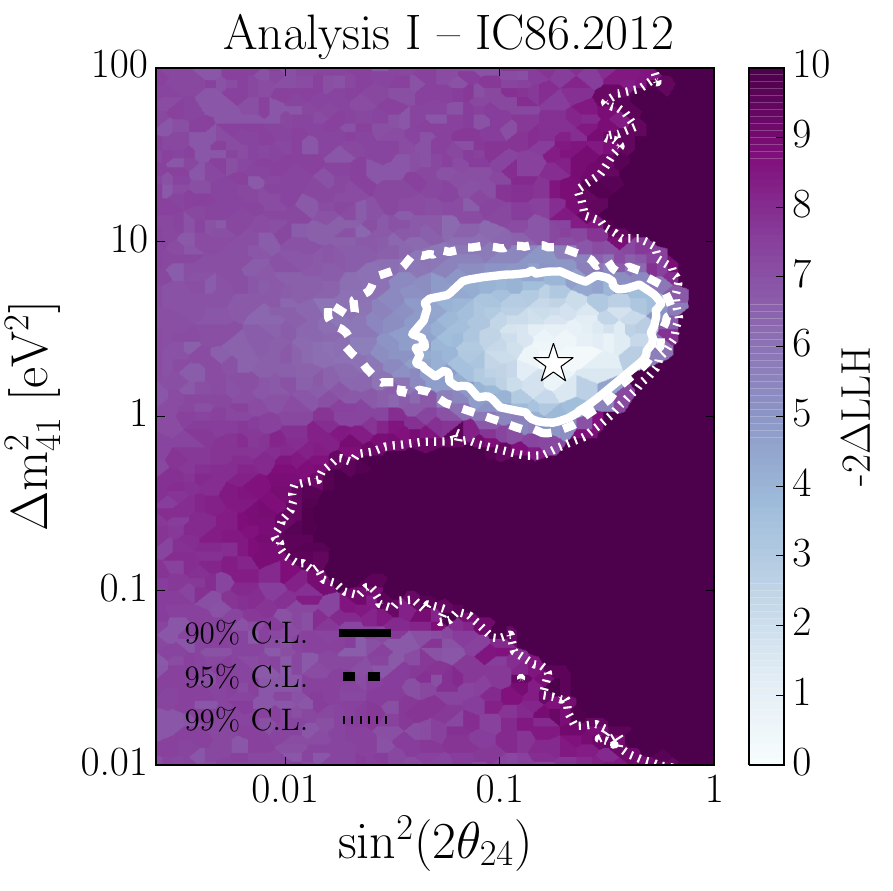}
    \end{minipage}
    \begin{minipage}{0.24\textwidth}
        \centering
        \includegraphics[width=0.98\linewidth]{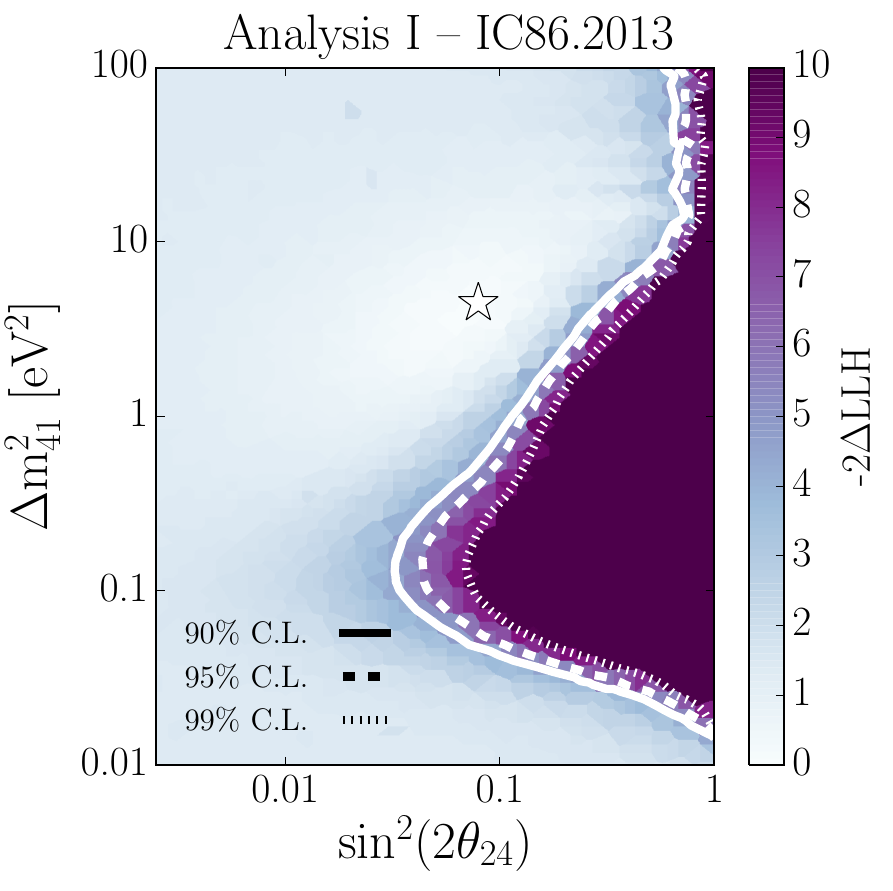}
    \end{minipage}%
    \begin{minipage}{0.24\textwidth}
        \centering
        \includegraphics[width=0.98\linewidth]{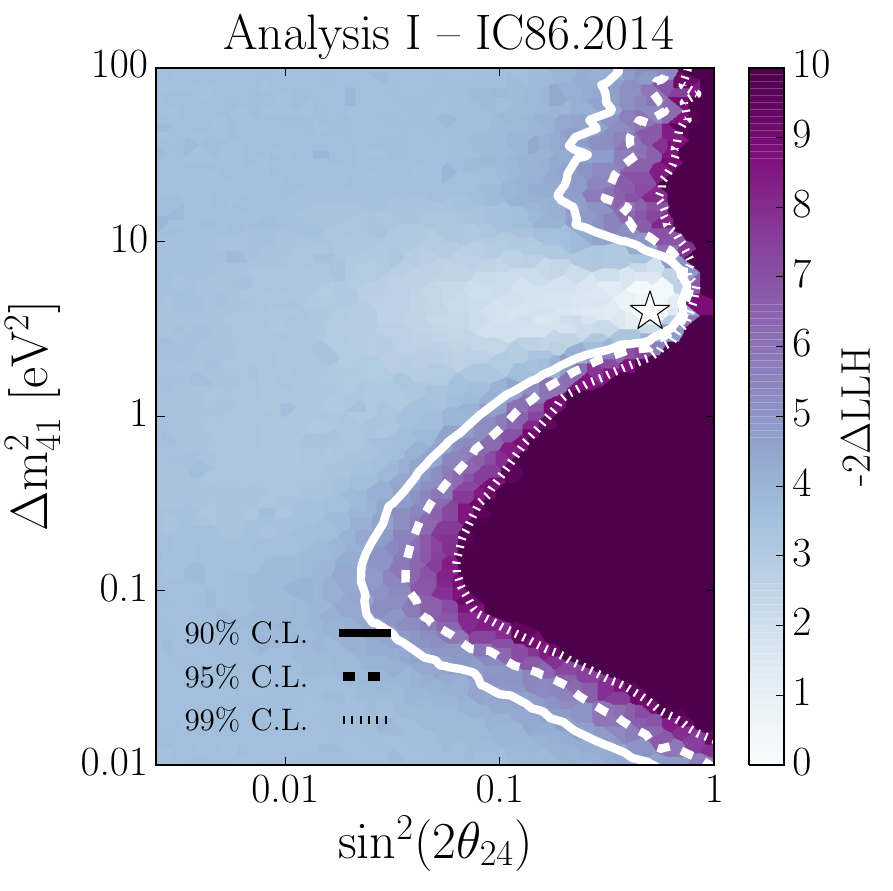}
    \end{minipage}
    \begin{minipage}{0.24\textwidth}
        \centering
        \includegraphics[width=0.98\linewidth]{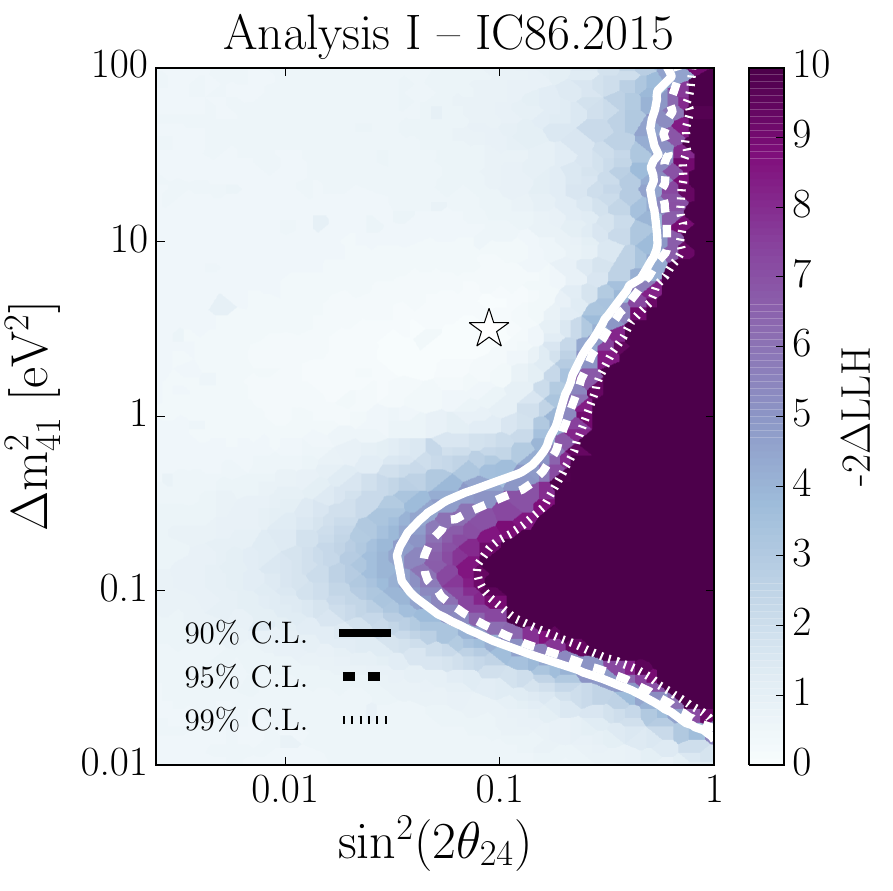}
    \end{minipage}%
    \begin{minipage}{0.24\textwidth}
        \centering
        \includegraphics[width=0.98\linewidth]{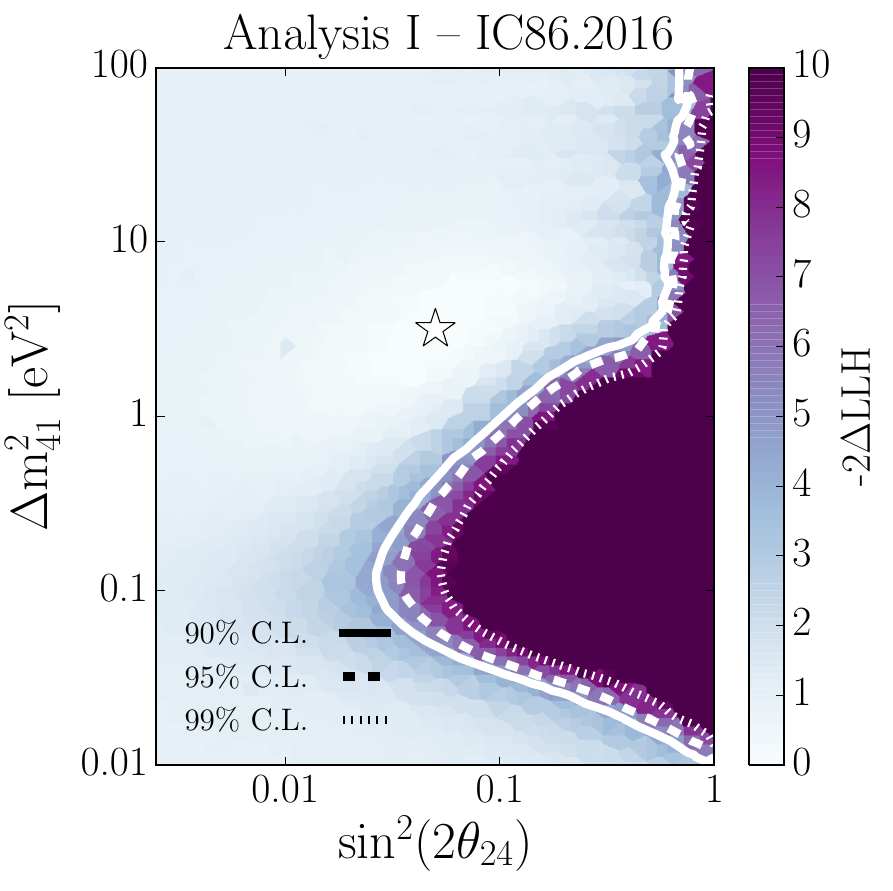}
    \end{minipage}
    \begin{minipage}{0.24\textwidth}
        \centering
        \includegraphics[width=0.98\linewidth]{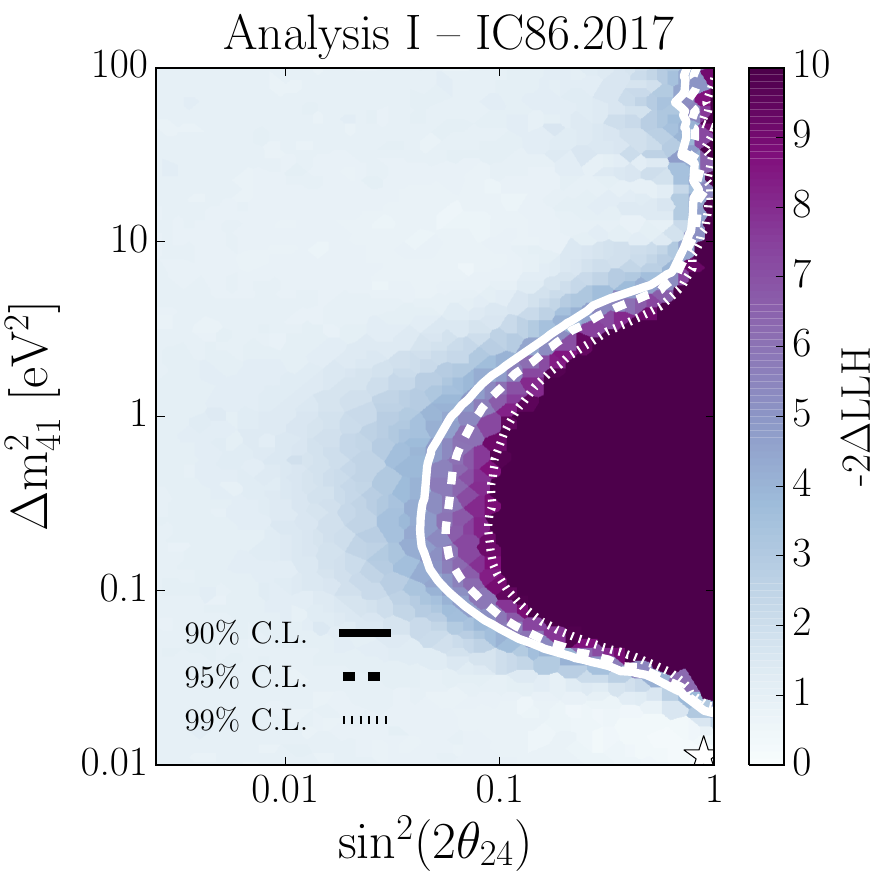}
    \end{minipage}%
    \begin{minipage}{0.24\textwidth}
        \centering
        \includegraphics[width=0.98\linewidth]{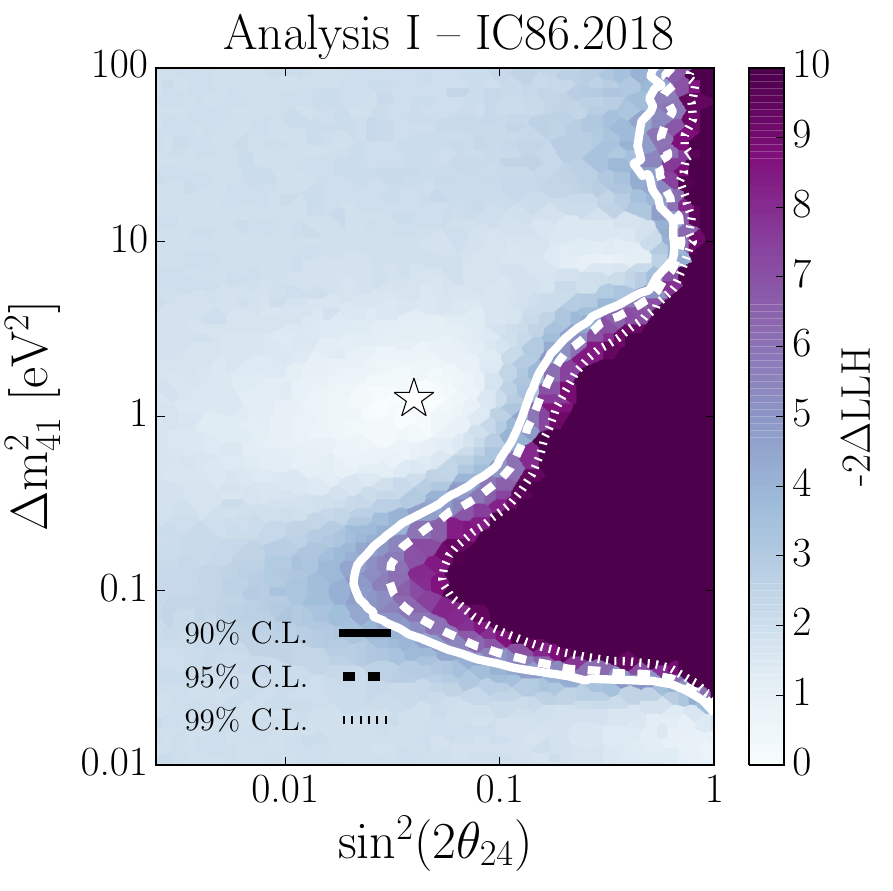}
    \end{minipage}
    
    \caption{\textbf{\textit{The test statistic distribution for Analysis I, determined independently for each IceCube season (IC86.2011 to IC86.2018).}}
    Each subplot shows the test statistic for each IceCube season. The reported confidence levels are calculated using Wilks' theorem. The white star represents the best fit point for each season.
    }
	\label{fig::yearlys1}
\end{figure*}

\begin{figure*}[b!]
    \centering
    \begin{minipage}{0.24\textwidth}
        \centering
        \includegraphics[width=0.98\linewidth]{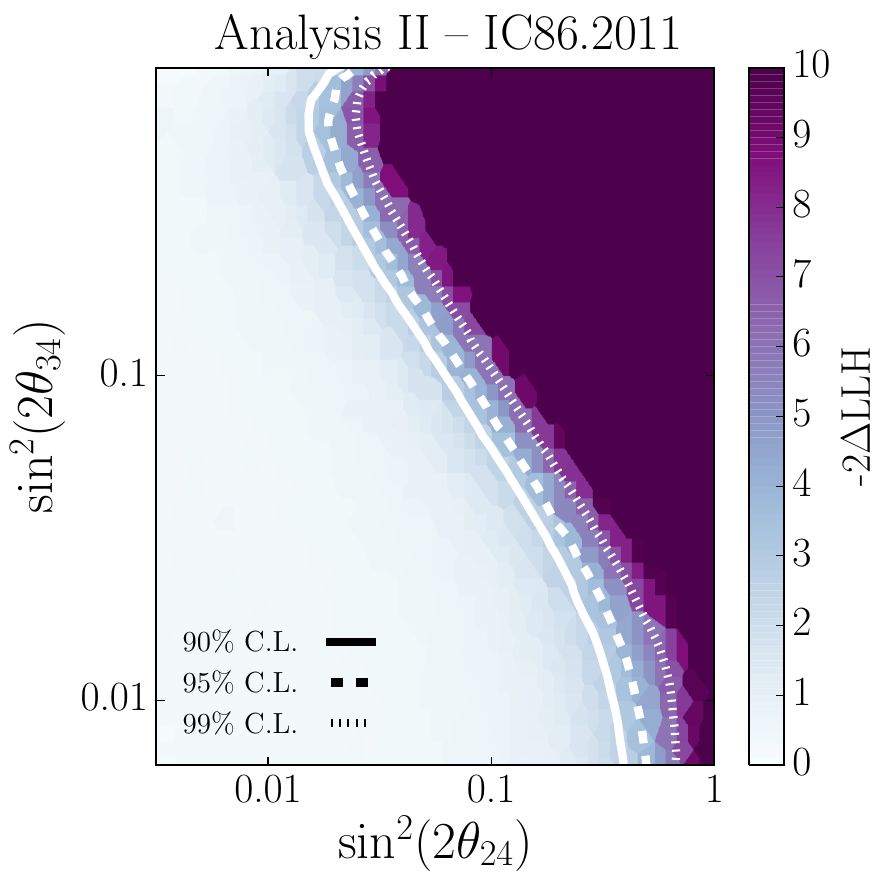}
    \end{minipage}%
    \begin{minipage}{0.24\textwidth}
        \centering
        \includegraphics[width=0.98\linewidth]{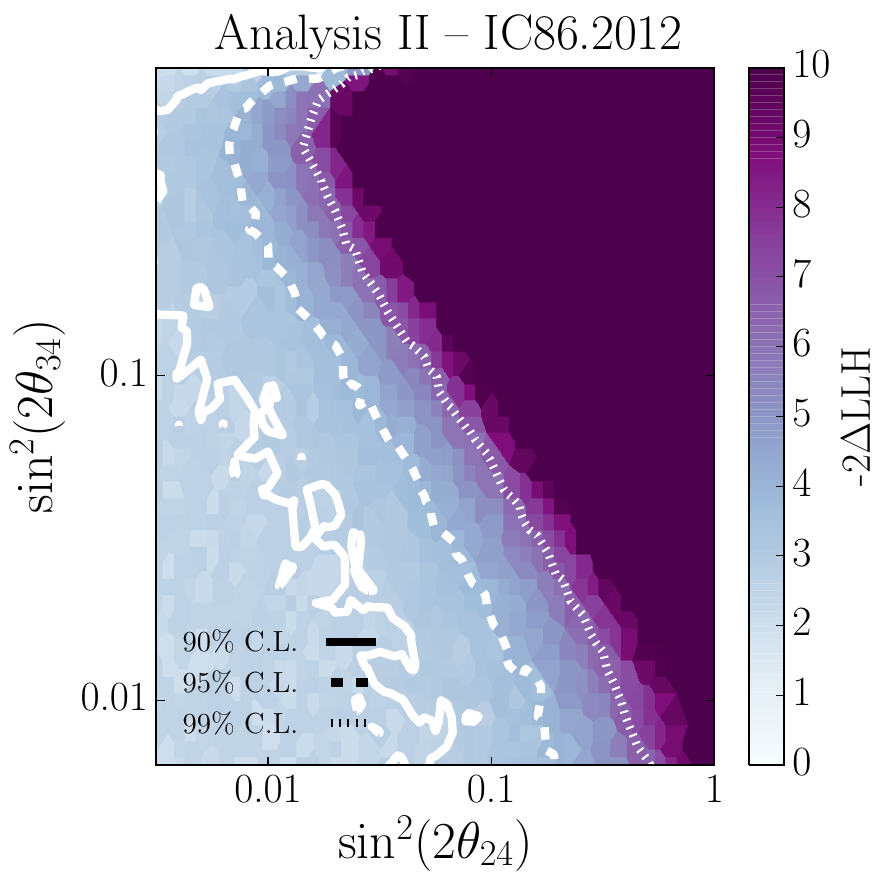}
    \end{minipage}
    \begin{minipage}{0.24\textwidth}
        \centering
        \includegraphics[width=0.98\linewidth]{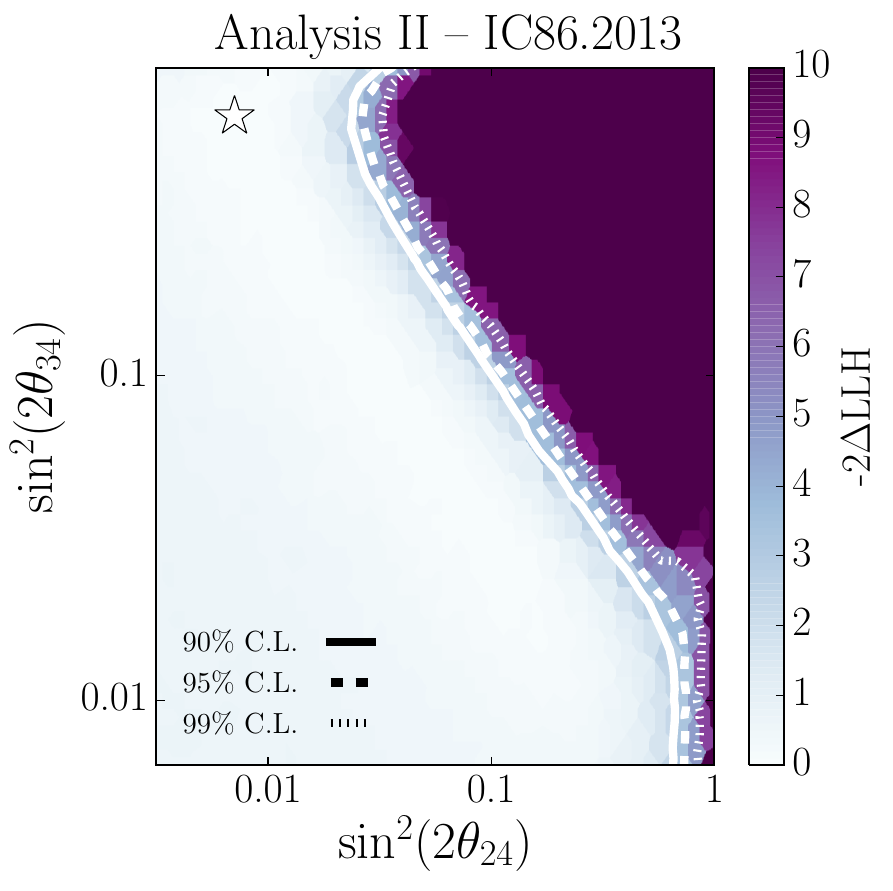}
    \end{minipage}%
    \begin{minipage}{0.24\textwidth}
        \centering
        \includegraphics[width=0.98\linewidth]{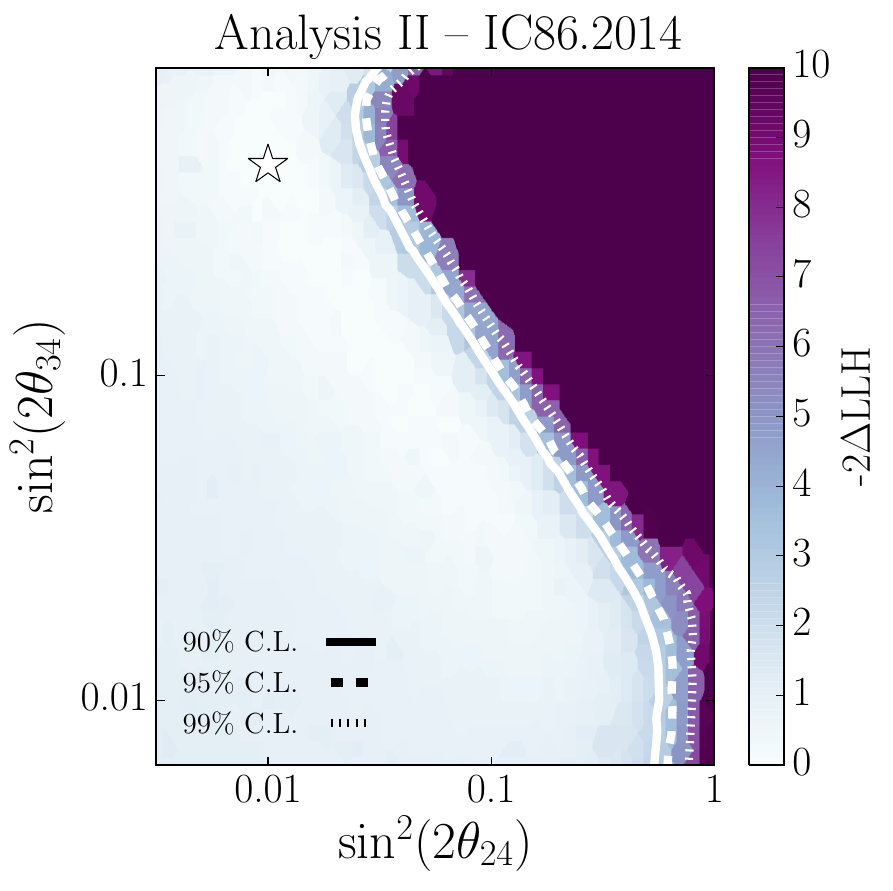}
    \end{minipage}
    \begin{minipage}{0.24\textwidth}
        \centering
        \includegraphics[width=0.98\linewidth]{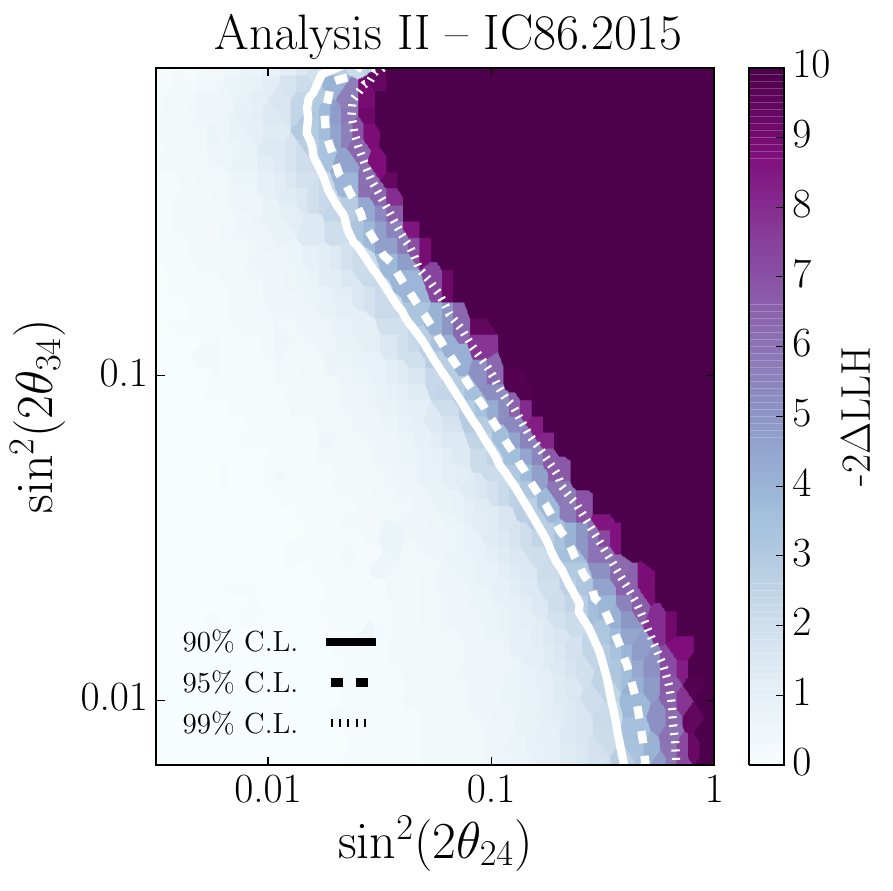}
    \end{minipage}%
    \begin{minipage}{0.24\textwidth}
        \centering
        \includegraphics[width=0.98\linewidth]{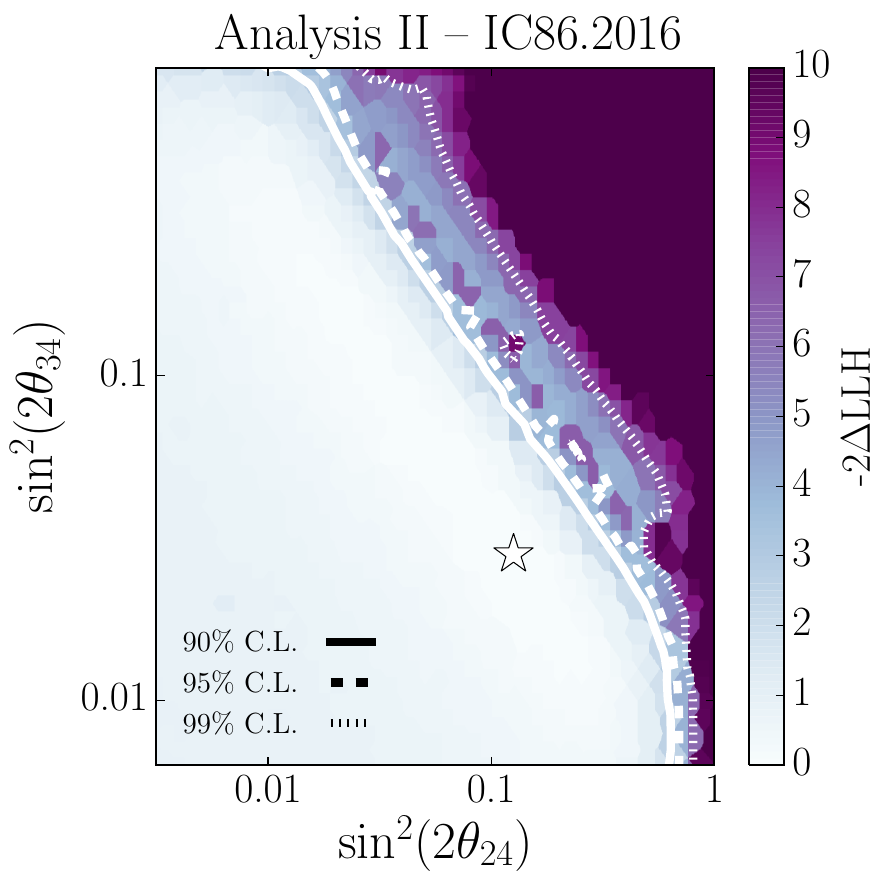}
    \end{minipage}
    \begin{minipage}{0.24\textwidth}
        \centering
        \includegraphics[width=0.98\linewidth]{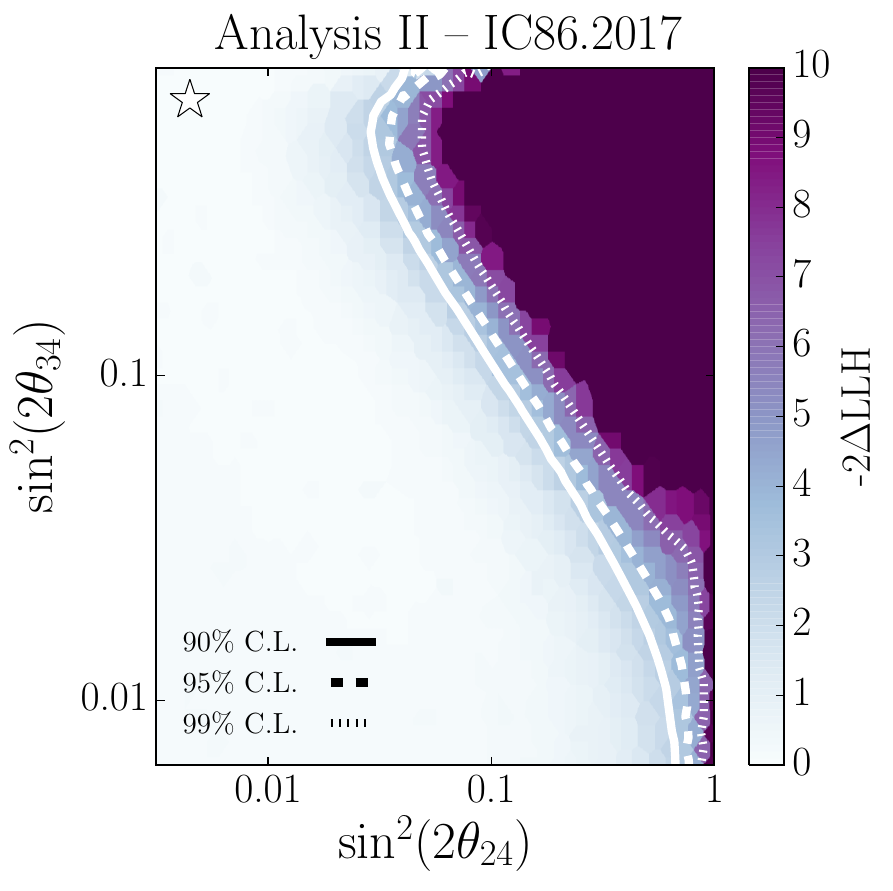}
    \end{minipage}%
    \begin{minipage}{0.24\textwidth}
        \centering
        \includegraphics[width=0.98\linewidth]{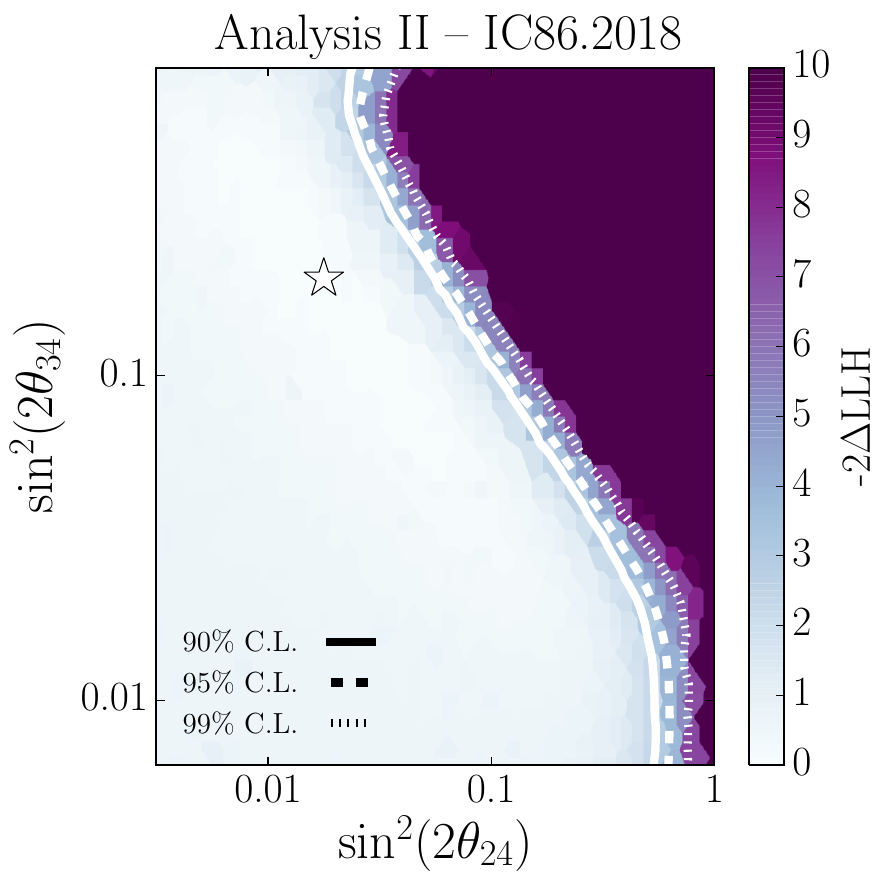}
    \end{minipage}
    
    \caption{\textbf{\textit{The test statistic distribution for Analysis II, determined independently for each IceCube season (IC86.2011 to IC86.2018).}}
    Each subplot shows the test statistic for each IceCube season. The reported confidence levels are calculated using Wilks' theorem. The white star represents the best fit point for each season.
    }
	\label{fig::yearlys2}
\end{figure*}

In this section we present, as supplementary material, analysis results for each year of IceCube data fit independently, showing consistency of the results (Suppl. Figs.~\ref{fig::yearlys1} and ~\ref{fig::yearlys2})  and the effects of varying each one of the nuisance parameters, as referred to in the main text (Suppl. Tables~\ref{tbl:syspanel1},\ref{tbl:syspanel2}, and \ref{tbl:syspanel3})

\end{document}